\documentclass[12pt]{cernart}
\usepackage{times}
\usepackage{epsfig}
\usepackage{amssymb}
\usepackage{cite}
\usepackage{colordvi}

\newcommand{\qx }{$q(x)$}
\newcommand{\Deqx }{$\Delta q(x)$}
\newcommand{\Deqtx }{$\Delta_T q(x)$}

\newcommand{\Deqt }{$\Delta_T q$}

\newcommand{\gevc }{GeV/$c$}
\renewcommand{\d }{\rm d}
\usepackage{graphicx}
\usepackage{rotating}
\usepackage{multirow}

\begin {document}

\begin{titlepage}
\docnum{CERN--PH--EP/2006--031}
\date{September 21, 2006}

\vspace{1cm}

\begin{center}
{\LARGE {\bf A new measurement of the Collins and Sivers  asymmetries
on a transversely polarised deuteron target}}

\vspace*{0.5cm}
\end{center}

\author{\large The COMPASS Collaboration}
\vspace{2cm}

\begin{abstract}
New high precision  measurements of the Collins and Sivers asymmetries of charged
hadrons produced in deep-inelastic scattering of muons
on a transversely  polarised $^6$LiD target are presented. 
The data were taken in 2003 and 2004 with the COMPASS spectrometer
using the muon beam of the CERN SPS at 160~\gevc .
Both the Collins and Sivers asymmetries turn out to be
 compatible with zero, 
within the present statistical errors, which are more than
a factor of 2 smaller than those of the published COMPASS results from the
2002 data.
The final results from the 2002, 2003 and 2004 runs are compared with
naive expectations and with existing model calculations.
\vfill
\submitted{(Submitted to Nuclear Physics B)}
\end{abstract}

\newpage
\begin{Authlist}
{\large ~~~~~~~~~~~~~~~~~~~The COMPASS Collaboration} \newline \newline 
%
%
E.S.~Ageev\Iref{protvino},
V.Yu.~Alexakhin\Iref{dubna},
Yu.~Alexandrov\Iref{moscowlpi},
G.D.~Alexeev\Iref{dubna},
M.~Alexeev\Iref{turin},
A.~Amoroso\Iref{turin},
B.~Bade{\l}ek\Iref{warsaw},
F.~Balestra\Iref{turin},
J.~Ball\Iref{saclay},
J.~Barth\Iref{bonnpi}
% ...

\end{Authlist}

\newpage
%
%
\Instfoot{bielefeld}{ Universit\"at Bielefeld, Fakult\"at f\"ur Physik, 33501 Bielefeld, Germany\Aref{g}}
\Instfoot{bochum}{ Universit\"at Bochum, Institut f\"ur Experimentalphysik, 44780 Bochum, Germany\Aref{g}}
\Instfoot{bonniskp}{ Universit\"at Bonn, Helmholtz-Institut f\"ur  Strahlen- und Kernphysik, 53115 Bonn, Germany\Aref{g}}
\Instfoot{bonnpi}{ Universit\"at Bonn, Physikalisches Institut, 53115 Bonn, Germany\Aref{g}}
\Instfoot{brno}{Institute of Scientific Instruments, AS CR, 61264 Brno, Czech Republic\Aref{h}}
\Instfoot{burdwan}{ Burdwan University, Burdwan 713104, India\Aref{j}}
\Instfoot{calcutta}{ Matrivani Institute of Experimental Research \& Education, Calcutta-700 030, India\Aref{k}}
\Instfoot{dubna}{ Joint Institute for Nuclear Research, 141980 Dubna, Moscow region, Russia}
\Instfoot{erlangen}{ Universit\"at Erlangen--N\"urnberg, Physikalisches Institut, 91054 Erlangen, Germany\Aref{g}}
\Instfoot{freiburg}{ Universit\"at Freiburg, Physikalisches Institut, 79104 Freiburg, Germany\Aref{g}}
\Instfoot{cern}{ CERN, 1211 Geneva 23, Switzerland}
\Instfoot{heidelberg}{ Universit\"at Heidelberg, Physikalisches Institut,  69120 Heidelberg, Germany\Aref{g}}
\Instfoot{helsinki}{ Helsinki University of Technology, Low Temperature Laboratory, 02015 HUT, Finland  and University of Helsinki, Helsinki Institute of  Physics, 00014 Helsinki, Finland}
\Instfoot{liberec}{Technical University in Liberec, 46117 Liberec, Czech Republic\Aref{h}}
\Instfoot{lisbon}{ LIP, 1000-149 Lisbon, Portugal\Aref{i}}
\Instfoot{mainz}{ Universit\"at Mainz, Institut f\"ur Kernphysik, 55099 Mainz, Germany\Aref{g}}
\Instfoot{miyazaki}{University of Miyazaki, Miyazaki 889-2192, Japan\Aref{l}}
\Instfoot{moscowlpi}{Lebedev Physical Institute, 119991 Moscow, Russia}
\Instfoot{munichlmu}{Ludwig-Maximilians-Universit\"at M\"unchen, Department f\"ur Physik, 80799 Munich, Germany\Aref{g}}
\Instfoot{munichtu}{Technische Universit\"at M\"unchen, Physik Department, 85748 Garching, Germany\Aref{g}}
\Instfoot{nagoya}{Nagoya University, 464 Nagoya, Japan\Aref{l}}
\Instfoot{praguecu}{Charles University, Faculty of Mathematics and Physics, 18000 Prague, Czech Republic\Aref{h}}
\Instfoot{praguectu}{Czech Technical University in Prague, 16636 Prague, Czech Republic\Aref{h}}
\Instfoot{protvino}{ State Research Center of the Russian Federation, Institute for High Energy Physics, 142281 Protvino, Russia}
\Instfoot{saclay}{ CEA DAPNIA/SPhN Saclay, 91191 Gif-sur-Yvette, France}
\Instfoot{telaviv}{ Tel Aviv University, School of Physics and Astronomy, 
              69978 Tel Aviv, Israel\Aref{m}}
\Instfoot{triestictp}{ INFN Trieste and ICTP--INFN MLab Laboratory, 34014 Trieste, Italy}
\Instfoot{triest}{ INFN Trieste and University of Trieste, Department of Physics, 34127 Trieste, Italy}
\Instfoot{turin}{ INFN Turin and University of Turin, Physics Department, 10125 Turin, Italy}
\Instfoot{warsaw}{ So{\l}tan Institute for Nuclear Studies and Warsaw University, 00-681 Warsaw, Poland\Aref{n} }
\Instfoot{warsawtu}{ Warsaw University of Technology, Institute of Radioelectronics, 00-665 Warsaw, Poland\Aref{o} }
\Anotfoot{a}{Also at IST, Universidade T\'ecnica de Lisboa, Lisbon, Portugal}
\Anotfoot{b}{Also at University of East Piedmont, 15100 Alessandria, Italy}
\Anotfoot{c}{deceased}
\Anotfoot{d}{On leave of absence from JINR Dubna}               
\Anotfoot{e}{Also at Chubu University, Kasugai, Aichi, 487-8501 Japan}
\Anotfoot{f}{Also at Yamagata University, Yamagata, 992-8510 Japan}
\Anotfoot{g}{Supported by the German Bundesministerium f\"ur Bildung und Forschung}
\Anotfoot{h}{Suppported by Czech Republic MEYS grants ME492 and LA242}
\Anotfoot{i}{Supported by the Portuguese FCT - Funda\c{c}\~ao para
               a Ci\^encia e Tecnologia grants POCTI/FNU/49501/2002 and POCTI/FNU/50192/2003}
\Anotfoot{j}{Supported by DST-FIST II grants, Govt. of India}
\Anotfoot{k}{Supported by  the Shailabala Biswas Education Trust}
\Anotfoot{l}{Supported by the Ministry of Education, Culture, Sports,
               Science and Technology, Japan; Daikou Foundation  and Yamada Foundation}
\Anotfoot{m}{Supported by the Israel Science Foundation, founded by the Israel Academy of Sciences and Humanities}
\Anotfoot{n}{Supported by KBN grants nr 621/E-78/SPUB-M/CERN/P-03/DZ 298 2000,
             nr 621/E-78/SPB/CERN/P-03/DWM 576/2003--2006,
             and by MNII reseach funds for 2005--2007}
\Anotfoot{o}{Supported by  KBN grant nr 134/E-365/SPUB-M/CERN/P-03/DZ299/2000}

%
\end{titlepage}

\newpage
\tableofcontents

\newpage

\section{An introduction to transverse spin physics}
\label{sec:introg}

\subsection{Historical introduction}
\label{sec:intro}

The importance of transverse spin effects at high energy
in hadronic physics was first suggested by the discovery in 1976 that
$\Lambda$ hyperons produced in $pN$ interactions exhibited 
an anomalously large transverse polarisation~\cite{Bunc76}. 
This effect could not be easily explained.
For a long time it was believed to be forbidden 
at leading twist in QCD~\cite{Kane78}, and
very little theoretical work was devoted to this field for more than a decade.
Nevertheless some important theoretical progress for the understanding of
single spin asymmetry (SSA) phenomena was done at that time~\cite{aefr}.

This situation changed in the nineties. 
After the first hints of large single transverse spin asymmetries 
in inclusive $\pi^0$ production in polarised pp scattering at 
CERN~\cite{Antille:1980th}
and in IHEP~\cite{ihep},
remarkably large asymmetries were found at 
Fermilab both for neutral and charged pions~\cite{E704a}. 
The
discovery of the EMC collaboration at CERN in the late eighties 
that the quark spin contributes
only a small fraction to the proton spin~\cite{EMC88}
caused a renewed interest
in the origin of the nucleon spin and proposals for new and versatile
experiments.
In parallel, intense theoretical activity was taking
place: the significance of the quark transversity distribution,
already introduced in 1979~\cite{RaSo79} to describe a quark in a
transversely polarised nucleon, was reappraised~\cite{ArMe90} in 1990, 
and its measurability via the Drell--Yan
process established. 
In 1991 a general scheme of all leading
twist and higher-twist parton distribution functions 
was worked out~\cite{JaJi91}, and in
1993 a way to measure
transversity in lepton nucleon polarised deep-inelastic scattering (DIS)
was suggested~\cite{Collins:1993kk}.
On the experimental
side, the RHIC-Spin Collaboration~\cite{RHIC} and
the HELP Collaboration~\cite{HELP} put forward the first proposals to 
measure transversity.
Today transversity is an
important part of the scientific programme of the HERMES experiment
at DESY, of the RHIC experiments at BNL, and of the 
COMPASS experiment at CERN, all presently taking data.
An experiment to measure transversity is being prepared at JLAB.
First results on a transversely polarised proton target have been published
recently by the HERMES Collaboration~\cite{Hermest} and, on a transversely 
polarised deuteron target by COMPASS~\cite{Alexakhin:2005iw}.

The COMPASS published results refer to the data taken in 2002.
Further data taking occurred in 2003 and 2004, and in this paper
results from these new data are presented, as well as a reanalysis of the
2002 data and the final results for the 2002--2004 data.

\subsection{The Collins mechanism}
\label{sec:collins}
To fully specify  the quark structure of the nucleon 
at the twist-two level, 
the transverse spin distributions \Deqtx\  must be added to the momentum
distributions \qx\ and the helicity distributions \Deqx ~\cite{JaJi91},
where $x$ is the Bjorken variable.
For a discussion on notation, see Ref.~\cite{BDR02}.
If the quarks are collinear with the parent nucleon (no intrinsic quark
transverse momentum $\vec k_T$), or after integration over $\vec k_T$,
these three distributions exhaust the information
on the internal dynamics of the nucleon.
More distributions are allowed admitting a finite 
$\vec k_T$, as we will see in Sections~\ref{sec:sivers} and~\ref{sec:colsiv},  
or at higher twist~\cite{BDR02,JaJi92,aram94,MuTa96}.

The distributions \Deqt\ are difficult to measure, since 
they are chirally odd and therefore absent in
inclusive DIS. They  may instead be
extracted from measurements of the single-spin asymmetries in cross-sections 
for semi-inclusive DIS (SIDIS)
of leptons on transversely polarised nucleons, in which 
a hadron is also detected in the final state.
In these processes the measurable asymmetry
is due to the combined effect 
of \Deqt\ and another chirally-odd function
which
describes the spin-dependent part of the hadronisation of a transversely 
polarised quark $q$ into a hadron $h$. 

The existence of an azimuthal asymmetry in transversely polarised
leptoproduction of spinless hadrons at leading twist, which depends on a
$T$-odd fragmentation function 
and arises from final-state
interaction effects, was predicted by Collins~\cite{Collins:1993kk} and is
now generally known as the Collins effect.
It is responsible for a left-right asymmetry in the fragmentation
of transversely polarised quarks.
In some models~\cite{Artru}, the Collins asymmetry is expected to be largest 
for the leading hadron in the current jet, i.e. the hadron with the
highest momentum.

\begin{figure}[t] %
\begin{center}
\includegraphics[width=0.7\textwidth]{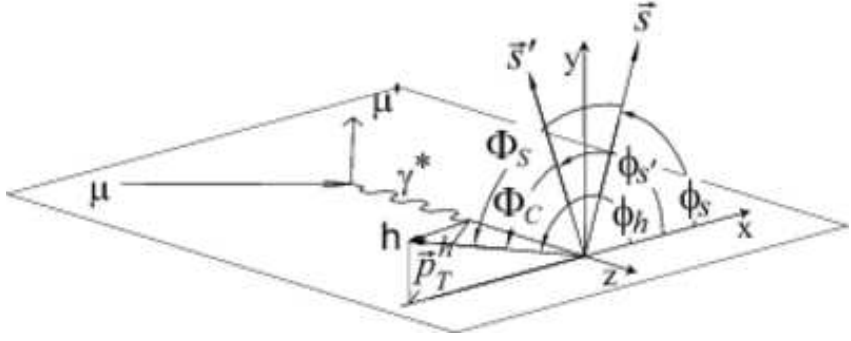}
\caption{Definition of the Collins and Sivers angles.
The vectors $\vec{p}_T^{\, h}, \, \vec{s}$ and $\vec{s}'$
are the hadron transverse momentum and spin of the initial and struck
quarks respectively.
}
\label{fig:angles}
\end{center}
\end{figure}

Assuming the detected hadron to be spinless, 
and a collinear quark distribution in the nucleon, at leading twist the
SIDIS cross-section can be written as
\begin{eqnarray}
 \hspace{-1em}
\frac{d\sigma}{d x \, d y \, d z \, d^2p_T^{\, h} \, d\phi_{S}} &=& 
  \frac{2\alpha_\mathrm{em}^2s}{Q^4} \, \sum_q e_q^2 \, \cdot 
  \left\{
    \frac12 \left[ 1 + ( 1 - y)^2 \right]
    \cdot x \cdot  q(x) \cdot  D_q^h(z, {p}_T^{\, h}) \, + \right.
\nonumber \\
&& \hspace{-4em}   +
  (1- y)  \left.  \, 
  | \vec{S}_\perp | \, \sin(\phi_h - \phi_{s'})
  \cdot x \cdot  \Delta_T q (x) \cdot  \Delta_T^0 D_q^h (z, {p}_T^{\, h})
  \right\} 
  \label{eq:sicross}
\end{eqnarray}
where Bjorken $x$ is $Q^2 / [ 2 M (E_{l}-E_{l'}) ]$,
$y$ is the fractional energy of the virtual photon,
$Q^2$ the photon virtuality,
$M$ the target nucleon mass,
and $z= E_h /(E_{l}-E_{l'})$ the fraction of available energy carried by 
the hadron.
The energies $E_h$, $E_{l}$, and $E_{l'}$ are the energies of the hadron, 
the incoming lepton,
 and the scattered lepton respectively in the target rest frame system.
The hadron transverse momentum $\vec{p}_T^{\, h}$ is evaluated with respect to 
the virtual photon direction.
Referring to Fig.~\ref{fig:angles},  $\phi_h$ and $\phi_{s'}$ are the
azimuthal angles of the hadron and of the struck quark spin 
in a coordinate 
system in which the z-axis is the virtual photon direction, and the x-z plane 
is the lepton scattering plane with positive x-direction along the scattered 
lepton transverse momentum.
$\vec{S}_\perp$ is the target spin normal to the virtual photon direction
and $\phi_{S}$ is its azimuthal angle with respect to the
lepton scattering plane.

The $\vec{p}_T^{\, h}$-dependent fragmentation function 
can be obtained by
investigating the fragmentation of a polarised quark
$q$ into a hadron $h$, and is expected to be of the form
\begin{equation}
D_{T\,q}^{\; \; \; h}(z, \vec{p}_T^{\, h}) = D_q^h(z, {p}_T^{\, h}) + 
    \Delta_T^0 D_q^h(z, {p}_T^{\, h}) \cdot \sin(\phi_h - \phi_{s'}) .
\label{eq:collfun}
\end{equation}
where $\Delta_T^0 D_q^h(z,{p}_T^{\, h})$ is the $T$-odd part of 
the fragmentation function,
responsible for the left-right asymmetry in the fragmentation of the 
transversely polarised quark.

The ``Collins angle'' $\Phi_C$ was originally defined 
in~\cite{Collins:1993kk} as
the angle between the transverse momentum of the outgoing hadron and
the transverse spin vector of the fragmenting quark, i.e.,
\begin{equation}
  \sin\Phi_C =
  \frac{( \vec{p}_T^{\, h} \times \vec{q}){\cdot}\vec{s'}}
       {| \vec{p}_T^{\, h} \times \vec{q} | \, | \vec{s'} |} \, .
  \label{eq:defac2}
\end{equation}
or
\begin{equation}
  \Phi_C = \phi_h - \phi_{s'} \, .
  \label{eq:defac1}
\end{equation}
Since, as dictated by QED, the directions of the
final and initial quark spins are related to each other by
$\phi_{s'} = \pi - \phi_s$ ,
equation (\ref{eq:defac1}) becomes
\begin{equation}
\Phi_C=\phi_h+\phi_S-\pi \, .
  \label{eq:defac3}
\end{equation}

By comparing the cross-sections on oppositely polarised
target nucleons one obtains from expression~(\ref{eq:sicross}) the transverse 
single spin asymmetry
\begin{equation}
A_T^h  \equiv 
  \frac{\d\sigma(\vec{S}_\perp) - \d\sigma(-\vec{S}_\perp)}
       {\d\sigma(\vec{S}_\perp) + \d\sigma(-\vec{S}_\perp)}
 =  | \vec{S}_\perp |  \cdot D_{NN}  \cdot A_{Coll}   \cdot \sin \Phi_C
  \label{eq:acoll}
\end{equation}
where the ``Collins asymmetry'' is
\begin{equation}
A_{Coll} = \frac {\sum_q e_q^2 \cdot \Delta_T q(x) \cdot 
                  \Delta_T^0 D_q^h(z, {p}_T^{\, h}}
{\sum_q e_q^2 \cdot q(x) \cdot D_q^h(z, {p}_T^{\, h})} \, ,
\label{eq:collass}
\end{equation}
and 
\begin{equation}
D_{NN} = \frac{1-y}{1-y+y^2/2} 
\label{eq:dnn}
\end{equation}
is the transverse spin transfer coefficient from the initial to 
the struck quark~\cite{BDR02}.

\subsection{The Sivers mechanism}
\label{sec:sivers}
An entirely different mechanism was suggested by Sivers~\cite{Sivers} 
as a possible cause of the transverse spin effects observed in pp
scattering. 
This mechanism could also be responsible for a spin asymmetry in the 
cross-section of SIDIS of leptons on transversely polarised nucleons.
Sivers conjecture was the possible existence of a correlation
between the transverse momentum $\vec{k}_T$ of an unpolarised quark
in a transversely polarised nucleon and the nucleon polarisation vector, 
i.e. that the quark distribution $q(x)$
in expression~(\ref{eq:sicross}) could be written as
\begin{equation}
q_{T}(x,\vec{k}_T)= q(x,{k}_T) + | \vec{S}_\perp | \cdot 
            \Delta_0^T q(x,{k}_T)\cdot \sin \, \Phi_{ S}
  \label{eq:fsiv}
\end{equation}
where the ``Sivers angle''
\begin{equation}
\Phi_{ S}= \phi_q -\phi_S
  \label{eq:angsiv}
\end{equation}
is the relative azimuthal angle between the quark transverse momentum
$\vec{k}_T$ and the target spin $\vec{S}_\perp$.

Under the assumption that the hadron produced in
the fragmentation and the fragmenting quark are collinear, i.e. that
all the hadron transverse momentum originates from the intrinsic 
transverse momentum of the quark in the nucleon
($\vec{p}_T^{\, h} = z \vec{k}_T$), the Sivers angle,
shown in Fig.~\ref{fig:angles}, becomes
\begin{equation}
\Phi_{ S}= \phi_h -\phi_S 
  \label{eq:angsiv1}
\end{equation}
 and the SIDIS cross-section in leading order QCD is given by
\begin{eqnarray}
\lefteqn{\frac{d\sigma}{d x \, d y \, d z \, d^2 {p}_T^{\, h} \, d \phi_S} =
  \frac{2\alpha_\mathrm{em}^2s}{Q^4} \sum_q e_q^2 \cdot 
    \frac12 \left[ 1 + ( 1 - y)^2 \right] \cdot x  \cdot }
\nonumber \\
   && \hspace{6em} \null \cdot
 \left[ q(x, {p}_T^{\, h}/z)
  +  | \vec{S}_\perp | \sin(\Phi_S)\Delta_0^T q(x, {p}_T^{\, h}/z) \right]
 \cdot D_q^h(z) 
  \, .
  \label{eq:sicross1}
\end{eqnarray}

Comparing the cross-sections on oppositely polarised target nucleons,
the transverse spin asymmetry of expression~(\ref{eq:acoll}) becomes
\begin{equation}
A_T^h  \equiv 
  \frac{\d\sigma(\vec{S}_\perp) - \d\sigma(-\vec{S}_\perp)}
       {\d\sigma(\vec{S}_\perp) + \d\sigma(-\vec{S}_\perp)}
=
  | \vec{S}_\perp |  \cdot  A_{Siv}   \cdot \sin \Phi_S 
  \label{eq:asiv}
\end{equation}
where the ``Sivers asymmetry'' 
\begin{equation}
A_{Siv} =  \frac {\sum_q e_q^2 \cdot \Delta_0^T q(x, {p}_T^{\, h}/z) 
         \cdot D^h_q(z)}
{\sum_q e_q^2 \cdot q(x, {p}_T^{\, h}/z) \cdot D_q^h(z)} 
\label{eq:sivass}
\end{equation}
could be revealed as a $\sin \Phi_{S}$ modulation
in the number of produced hadrons.

The existence of the Sivers function requires final/initial state 
interaction, and an interference between different helicity
Fock states.
In the absence of interactions the Sivers function would vanish
by time-reversal invariance of QCD (see e.g. Ref.~\cite{MuTa96})
and indeed
it was believed for several years that the Sivers function is zero.
Recently it was shown 
however~\cite{Brodsky:2002pr,Collins:2002kn,Belitsky:2002sm} 
that these interactions 
are represented naturally by the gauge link that is required
for a gauge invariant definition of a transverse momentum dependent
(TMD) parton distribution, thus the Sivers function has become a 
very important piece in the most fundamental issues of QCD.

\subsection{More general formalism}
\label{sec:colsiv}

A combined treatment of the the Collins and Sivers mechanisms
requires a more general approach and the introduction of
TMD distributions and fragmentation 
functions~\cite{aram94,MuTa96,colsiv}.
Convolution integrals appear in the SIDIS cross-section, whose expression 
becomes
\begin{eqnarray}
d\sigma & \sim &\sum_q e_q^2 \cdot 
  \left\{\frac12 \left[ 1 + ( 1 - y)^2 \right] \, x \, 
    \left[ q \otimes D_q^h +  
    | \vec{S}_\perp | \sin(\Phi_S)\Delta_0^T q \otimes D_q^h \right]  + \right.
\nonumber \\
&&      + (1- y)   \, \left. 
  | \vec{S}_\perp | \, \sin(\Phi_C) 
  \, x \, \Delta_T q \otimes  \Delta_T^0 D_q^h
  \right\} \, .
  \label{eq:sicrosst}
\end{eqnarray}
In the general expression of the SIDIS cross-section 
at leading order QCD other terms related to different single and
double spin azimuthal asymmetries appear. 
Here they are neglected since they are beyond the scope of this paper.
The symbol $\otimes$, which replaces the products in Eq.~(\ref{eq:sicross}) and
(\ref{eq:sicross1}), indicates the convolution integral
\begin{equation}
DF \otimes FF = \int d^2{k}_T \,  \, 
DF(x,\vec{k}_T) \cdot FF(z,\vec{p}_T^{\, h}- z\vec{k}_T)
\end{equation}
where $DF$ and $FF$ are generic TMD distributions and fragmentation 
functions respectively.

The transverse spin asymmetry is then given by
\begin{eqnarray}
A_T^h & \equiv &
  \frac{\d\sigma(\vec{S}_\perp) - \d\sigma(-\vec{S}_\perp)}
       {\d\sigma(\vec{S}_\perp) + \d\sigma(-\vec{S}_\perp)}
  \nonumber
\\
  &=&
  | \vec{S}_\perp |  \cdot D_{NN}  \ A_{Coll}   \cdot \sin \Phi_C +
  | \vec{S}_\perp |  \cdot A_{Siv}   \cdot \sin \Phi_S 
  \label{eq:acollsiv}
\end{eqnarray}
where the Collins and the Sivers asymmetries are still given
by Eq.~(\ref{eq:collass}) and  (\ref{eq:sivass}) when replacing
the products of the distribution and fragmentation functions with
the corresponding convolutions.
In the usual assumption of Gaussian distributions for the parton and the hadron
transverse momenta in the $DF$ and in the $FF$, the only effect of the
convolution integral in Eq. ~(\ref{eq:collass}) and  (\ref{eq:sivass})
is the presence of a factor which depends on $< { \vec{k}_T}^{\, \; 2}>$ and 
$<\vec{p}_T^{\, h \,\,2}>$
(see f.i.~\cite{Collins:2005ie,efre}).

Since the Collins and Sivers terms in the transverse spin asymmetry
depend on the two independent angles $\Phi_C$ and $\Phi_S$,
measuring SIDIS on a transversely polarised target allows the 
Collins and the Sivers effects to be disentangled and the
two asymmetries can separately be extracted from the data.

\section{The COMPASS experiment}
\label{sec:compass}

\subsection{Physics objectives}
\label{sec:cphys}
The COMPASS experiment was proposed to CERN in 1996 to investigate 
hadron structure and hadron spectroscopy
by carrying on a number of key measurements using both hadron 
($\pi$, K and protons) and muon high-energy
beams.
Apart from a short pilot run in 2004 with 190~\gevc\ pions
to measure the pion polarisability via Primakoff scattering on high-Z targets,
the experiment in so far has focused on the investigation of the
spin structure of the nucleon using a 160 \gevc\ $\mu^+$ beam 
and a polarised deuteron target.
In the longitudinal target spin mode, the main goal of the experiment
is the measurement of $\Delta G/G$~\cite{Ageev:2005pq}, the polarisation 
of the gluons
in a longitudinally polarised nucleon, but very precise
$A_1^d$ data are also collected~\cite{Ageev:2005gh}.
In about 20\% of the running time, the target polarisation was
set along the vertical direction, orthogonal to the beam axis,
and transverse spin effects were measured, which are the subject 
of this paper.

\subsection{The experimental set-up}
\label{sec:setup}

The COMPASS spectrometer~\cite{spectro} has been set up in the CERN SPS
North Area. 

\begin{figure*}[b] %
\includegraphics[width=\textwidth]{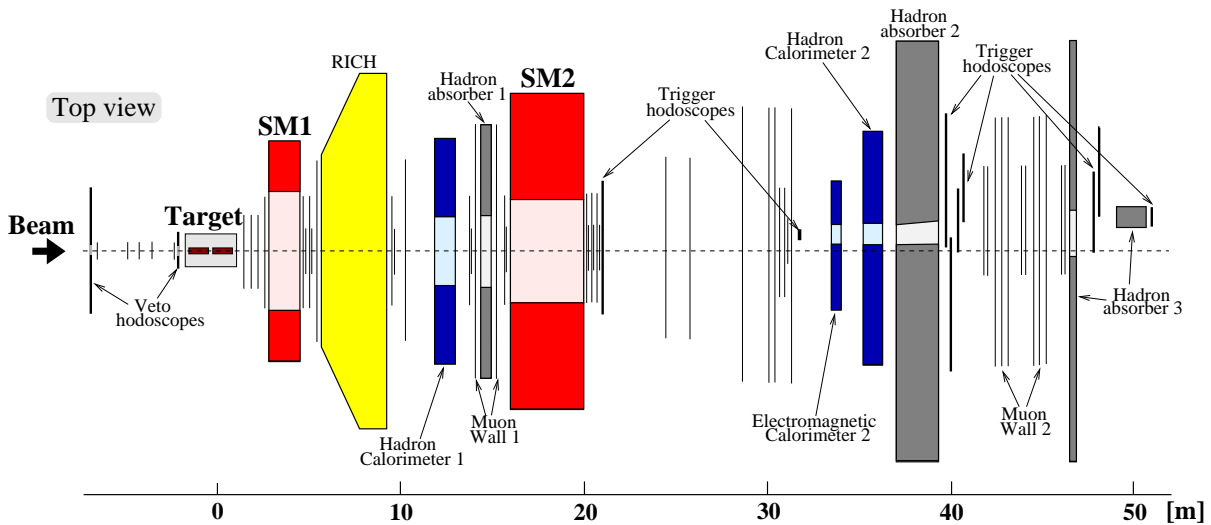}
\caption{Top view of the lay-out of the spectrometer for 
the COMPASS experiment in 2003.
The labels and the arrows refer to the major components of the
trigger and PID systems.
The thin vertical lines represent the tracking detectors.}
\label{fig:setup}
\end{figure*}

The layout of the spectrometer which was on the floor in 2003 is shown 
in Fig.~\ref{fig:setup}.
To combine large geometrical acceptance and broad dynamical range, 
the spectrometer comprises two magnetic stages, and is made up
of a variety of tracking detectors, a fast RICH, two hadronic
calorimeters, and provides muon identification via filtering 
through thick absorbers.

The first stage is centred around the spectrometer 
magnet SM1 (1~Tm bending power)
and the design acceptance is about $\pm 200$~mrad in both planes,
to fully contain the hadrons of the current jet.
The second stage uses the spectrometer magnet SM2 (operated at a bending power
of 4.4 Tm),
located 18~m downstream from the target, and the acceptance is $\pm$50
and $\pm$25~mrad in the horizontal and vertical plane respectively.

 The design of detector
components, electronics and data acquisition system allows to
handle beam rates up to 10$^8$ muons/s and about 5$\cdot$10$^7$
hadrons/s. 
The triggering system and the tracking system of COMPASS have been designed to
stand the associated rate of secondaries, and use state-of-the-art detectors.
Also, fast front-end electronics, multi-buffering, and a large
and fast storage of events are essential.

The experiment has been run at a muon energy of 160~GeV.
The muons originate from the decay of $\pi$ and K mesons produced by 
the 400~GeV proton beam on a primary beryllium target.
The $\mu^+$ beam is naturally polarised by the weak decay mechanism,
and the beam polarisation is about -80\%.
The beam polarisation contributes to the 
transverse spin dependent part of the cross-section
only by higher twist effects, which are not considered in the leading-order 
analysis of this paper.
The  $\mu^+$ intensity is $2\cdot10^8$ per 
spill of 4.8~s with a cycle time of $16.8$~s.  
The beam profile presents a Gaussian core and a large non-Gaussian 
tail due to halo muons.
The beam has a nominal energy of 160~GeV and is focused at the target 
centre, with a  spread of 7~mm (r.m.s.) 
and a momentum spread of $\sigma_p/p=0.05$ for the Gaussian core. 
The momentum of each muon is measured upstream of the experimental 
area in a beam momentum station  consisting of
several planes of scintillator strips  with 
3 dipole magnets (30~Tm total bending power) in between. 
The precision of the momentum determination is typically 
$\Delta p/p \leq 0.003$. 

Due to the major problems and delay in the construction of the new
large-acceptance COMPASS polarised target (PT) magnet, the experiment has 
utilised in so far the magnet from the SMC experiment, which has 
a similar design and identical magnetic properties, 
but a smaller bore (26.5~cm diameter).
The resulting angular acceptance is reduced, going from $\pm$170~mrad
at the downstream end to $\pm$69~mrad at the upstream end of the
target.
The polarised target system~\cite{Ctarget} consists of two oppositely 
polarised target cells (upstream $u$, downstream $d$),
60~cm long each and 3~cm diameter, so that data are collected 
simultaneously for the 
two target spin orientations.
The PT magnet can provide both a solenoidal field (2.5 T) and a dipole field
for adiabatic spin rotation (up to 0.5 T) and for the transversity 
measurements (set at 0.42 T).
Correspondingly, the
target polarisation can then be oriented either longitudinally
or transversely to the beam direction.
The target is cooled to temperatures below 100 mK by a $^3$He-$^4$He
dilution refrigerator.
When operated in the frozen spin mode, temperatures of $\sim$50 mK
are reached.
The polarisation was lost at
rate of 0.4 - 1.0\%/day in the 0.42 T dipole field and 0.05 - 0.1\%/day in
2.5 T solenoid field.
The use of two different target materials, NH$_3$ as proton target and
$^6$LiD as deuteron target, is foreseen. 
Polarisations of 90\%~\cite{Adams:1999qy} and 50\%~\cite{spectro} have 
been reached, respectively.
In so far only $^6$LiD has been used as target: its 
favourable dilution factor of $\sim$0.4
is of the utmost importance for the measurement
of $\Delta G$.
The dynamical nuclear polarisation system can polarise the target only
at 2.5~T, i.e. in the solenoidal field.
To run in the transverse polarisation mode, the spins are frozen and rotated 
adiabatically by first lowering the longitudinal field to 0.5~T, then rotating
the magnetic field to the vertical direction with the help of the dipole coils.
The polarisation values are obtained from corresponding measurements 
done at the beginning and
at the end of each data taking period, with the polarised target field 
set back 
at 2.5 T, and take into account the relaxation time of the target polarisation.

To match the expected particle flux in the various locations
along the spectrometer, COMPASS uses very different tracking detectors.
The small area trackers consist of several
stations of scintillating fibres, silicon detectors, micromegas
chambers~\cite{microo} and gaseous chambers using the 
GEM-technique~\cite{gems}.
Large area tracking devices are made from gaseous detectors
(Drift Chambers, Straw tubes~\cite{straw}, and MWPC's)
placed around the two spectrometer magnets. 

Muons are identified in large-area Iarocci-like tubes and drift 
tubes downstream of muon absorbers.
Hadrons are detected by two large  iron-scintillator sampling calorimeters, 
installed in front of the absorbers
and shielded to avoid electromagnetic contamination.
The charged particles identification relies on the RICH 
technology~\cite{rich1}.
In this paper we have not utilised the information of the RICH, and give 
results for non-identified hadron asymmetries only.
The asymmetries for RICH-identified hadrons (pions and kaons) will be 
the subject of a separate paper.

The trigger~\cite{Bernet:2005yy} is formed by a combination
of signals indicating the presence of a scattered
muon at a given angle or in a given energy range. In
most DIS events ($Q^2 > 1 $ (\gevc )$^2$), the scattered muon
is identified by coincidence signals in the trigger hodoscopes,
that  measure the projection of the scattering angle in the
non-bending plane and check its compatibility with the target
position
Several veto counters installed upstream of the target are used
to avoid triggers due to halo muons. 
In addition to this
inclusive trigger mode, several semi-inclusive
triggers select events fulfilling requirements
based on the muon energy loss and on the presence of
a hadron signal in the calorimeters. 
The acceptance is further extended toward high $Q^2$ values
by the addition of a standalone calorimetric trigger
in which no condition is set for the scattered muon.

A complete description of the spectrometer can be found in Ref.~\cite{spectro}.

\subsection{Data taking and off-line system}
\label{sec:datataking}

\begin{figure}[tbp]
  \begin{center}
    \includegraphics[width=.8\textwidth]{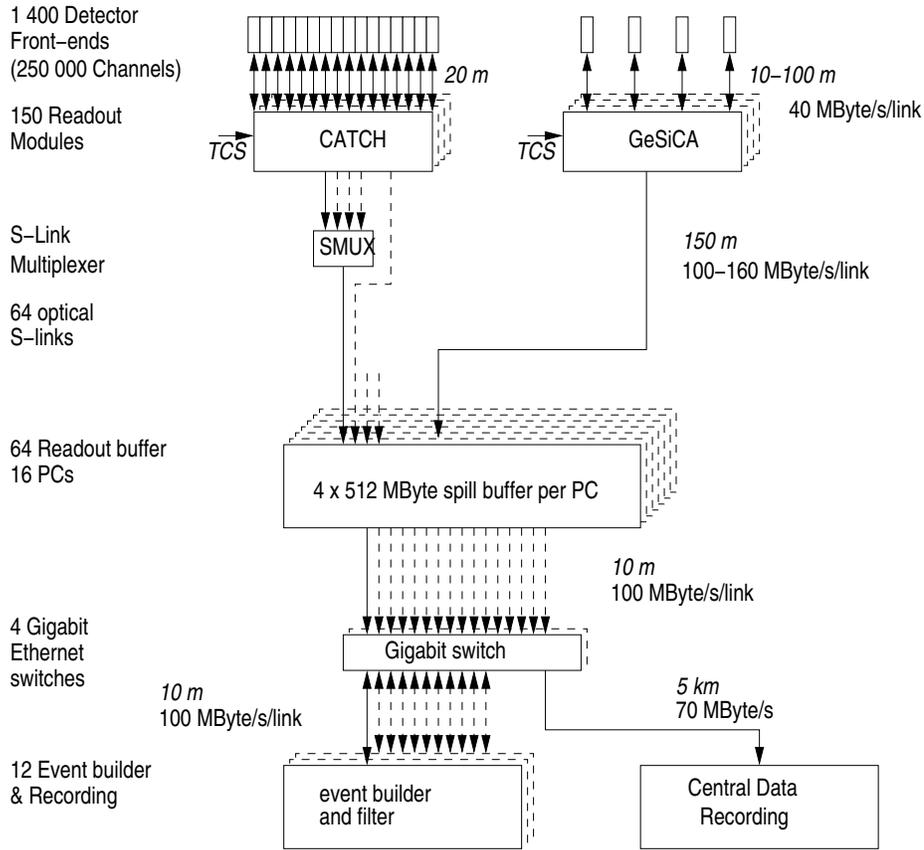}
  \end{center}
  \caption{General architecture of the DAQ system.}
  \label{daqoverview}
\end{figure}

The data acquisition system~\cite{daq} was designed to read the
250\,000 detector channels with an event rate of up to 100~kHz
and with essentially no dead-time. 
This required a full custom design of the readout electronics (mounted
directly on the detectors) and of the readout-driver modules
(named CATCH and GeSiCA), which
perform local event building and trigger distribution to the front-end
boards (Fig.~\ref{daqoverview}). The trigger control system (TCS)
performs trigger distribution and synchronisation of the
time-to-digital converters to better than 50~ps. 
The data collected during the 4.8~s spills is stored in 32~GByte buffers. 
The event building is performed on high performance Linux PCs connected to 
the read out system via Gigabit Ethernet. The data are grouped in ``runs'' 
corresponding to about 100 consecutive spills in 2003 and 200 in 2004.
General information on the run is extracted and stored in
the meta-data tables of an Oracle database. 
The files are then sent to the Central Data Recording 
facility in the CERN computer centre in 
parallel multiple streams over a dedicated optical fibre 
network at an average speed of 70 MBytes/s.

\begin{figure}[tbp]
  \begin{center}
    \includegraphics[width=.8\textwidth]{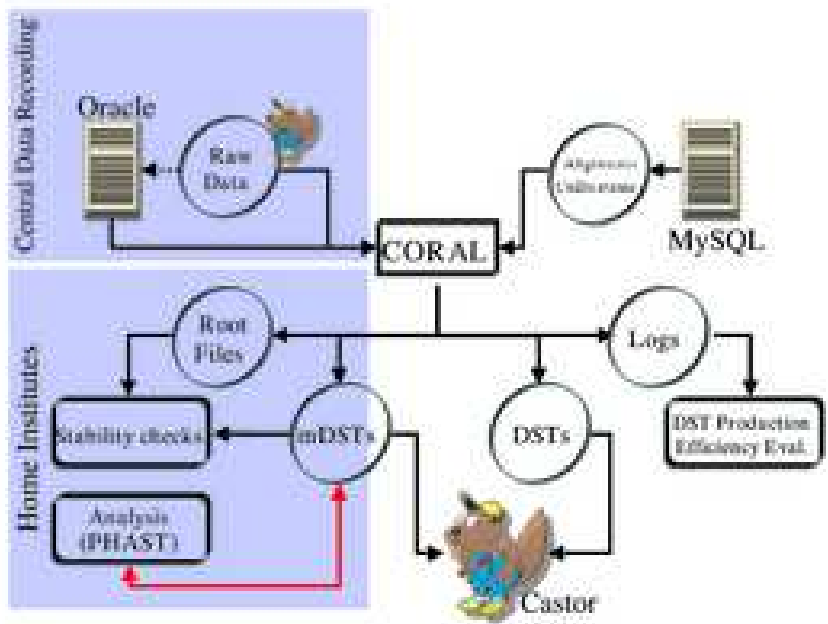}
  \end{center}
  \caption{Schematic view of the off-line system and reconstruction and 
analysis flow.}
  \label{fig:offline}
\end{figure}

Once at the computer centre, the files are registered in the name space of 
CASTOR (the CERN hierarchical storage management system): from this moment 
onward, CASTOR controls thoroughly the events data handling (copy to tape, 
managing of the disk space), while an Oracle RAC database system is in charge 
of translating high-level requests of data into file requests.
The huge amount of data of about 350 TBytes/year is reconstructed at the CERN 
computer centre, requiring a computing power of about 200k SPECint2000.
The event reconstruction is carried on by CORAL, the COMPASS reconstruction 
and analysis framework, a fully object-oriented program with a modular
architecture written in C++, which provides interfaces for the event 
reconstruction algorithms and insulation layers to access the data and for 
external pluggable packages~\cite{COMPASSsoftw}.
The reconstruction is carried out
in parallel, by some 600 jobs running on the CERN batch system.
CORAL decodes the data, reconstructs 
tracks 
and vertexes, and performs the particle identification making use of the 
alignment and calibration data describing the apparatus, which are stored as 
time/version-dependent information in a MySql database.
The reconstructed data is output in a proprietary format to files called Data 
Summary Tapes (DSTs), and in ROOT~\cite{root} format to mini-DSTs, which 
are selectively 
filtered out from the DSTs during production and turn out to have a 
size of about 1\% of the original RAW data. 
DSTs and miniDSTs are stored also centrally on tape, under CASTOR.
The physics analysis is performed on the mini-DSTs, replicated
in the different institutes,
by means of PHAST, the COMPASS framework for the final data 
analysis~\cite{COMPASSsoftw}. 
Fig.~\ref{fig:offline} depicts the reconstruction and analysis system to 
which the data flow after they are stored centrally.
For the simulations, the Monte Carlo program 
COMGeant~\cite{COMPASSsoftw}, based on
GEANT~3 and LEPTO~6.5.1, has been written and used
extensively.

In the first three years of data taking, from 2002 to 2004, in which the
experiment benefited of about $12 \cdot 10^5$ spills, COMPASS
collected 30 billion events, corresponding to a total data sample of more 
than 1 PByte.
Thanks to continuous work on alignment and improvement of the 
reconstruction code, these data have been processed several times.
About 20\% of these data have been taken in the 
transverse target spin mode.

\subsection{Principle of the measurement}

As explained in Section~\ref{sec:intro}, 
the Collins and the Sivers asymmetries can be estimated from the SIDIS data 
by measuring, in a given kinematical bin defined by $x, \, z$,
or $p_T^h$, the number of events $N^+$ and $N^-$ collected on oppositely
polarised target nucleons, and fitting the measured asymmetries 
$(N^+ - N^-)/(N^+ - N^-)$ either with
a $\sin \Phi_C$ or a $\sin \Phi_S$ dependence.
Since the spectrometer acceptance is somewhat different for the two target 
halves, the cross-section asymmetries cannot be obtained from a direct
comparison of the number of events collected in the $u-$ and $d-$cell.
It is necessary to compare the number of events collected in each cell
with the two target spin orientations, and the two-cell system allows
to reduce the systematic effects.

The orientation of the target polarisation cannot be flipped fast
in the COMPASS target system.
The holding field is 0.42 T, and transverse to the beam, thus it is out
of question to flip the target spin by inverting the magnetic
field because the acceptance would change.
Polarisation reversal is done by exchanging the microwave frequencies
for $u-$ and $d-$cells,
and to achieve $\sim$50\% polarisation one needs two days.
Therefore, once the target is polarised, data are taken for
several days (typically 5) before a polarisation reversal is done,
and to evaluate an asymmetry one compares the number of events
collected almost a week apart.
For this reason, the transverse spin measurement usually was 
carried out at the end of the run, when  the spectrometer was 
fully operational and functioning smoothly, and great care was taken
not to intervene on any detector to avoid changes in efficiency.

The typical cycle (a data taking period) consisted therefore 
of 5 days of measurement in one configuration (yielding f.i. $N_u^+$
and $N_d^-$), 2 days to change the target polarisation, and 5 more
days of data taking in the opposite configuration
(yielding $N_u^-$
and $N_d^+$).
At this point, one can estimate the cross-section asymmetry.
A straightforward procedure was used for the published 2002 data
analysis~\cite{Alexakhin:2005iw}, namely to derive one asymmetry from
$N_u^+$ and $N_u^-$, a second asymmetry from $N_d^+$ and $N_d^-$,
check their consistency and then average the two.
A different method, which uses at once the four numbers and
is less sensitive to differences in acceptance, has been used in this analysis,
and is described in Section~\ref{subs:rpm}.

In this paper final results are presented for all the data
collected with the deuteron target in the three years 2002--2003--2004.
Data were collected during two periods in 2002, one  period in 2003,
and during two  periods in 2004.
The focus is on the data collected in 2003 and 2004, and
most of the next section will be devoted to the analysis of these data,
which is slightly different from the analysis of the 2002 data
(already published), thanks to various improvements in the analysis code.
The new chain of analysis has also been used to reanalyse the 2002 data:
the results are perfectly compatible, as will be shown in 
Section~\ref{sec:meas}.
Also, the new 2003--2004 data are in excellent  agreement with the
2002 data.
At the end, overall results will be given for the whole
data set, 2002--2004.

\section{Data analysis I - event reconstruction and selection}
\subsection{Event reconstruction}

In the event reconstruction, a track reconstructed before
the target is assumed to be an incoming muon if
it is reconstructed in the scintillating fibres and silicons
upstream of the target, its momentum is reconstructed in the BMS, and
the track time is within 3 standard deviations of BMS and trigger time.
If several valid BMS-tracks are compatible with the
track time, a backtracking algorithm is used
to resolve ambiguities.

The scattered muon candidates are defined as the positively charged
outgoing tracks with a momentum larger than 1 \gevc , going through SM1, and
their extrapolation at the entrance and at the exit of the target
is within 5~cm from the target axis. 
In addition, for
all triggers based on the hodoscope information, the track must be
compatible with the hodoscope hits as given in the trigger matrix.
In the case of a calorimetric trigger, a minimal number of hits is required 
in the muon walls and the amount of material traversed in the spectrometer
must be larger than 66 and 74 radiation lengths for tracks reconstructed  
in the first and in the second spectrometer stage respectively.

The muon interaction point (the so-called ``primary vertex'') is 
defined by one beam particle and one scattered muon. If
in the event there are more than one beam particles and/or  scattered muons,
several vertexes may be reconstructed.
The distance of closest approach between the scattered muon and beam
track must be less than 10~$\sigma$ (about 2--3~mm).
The position of the vertex along the spectrometer axis 
z$_{vtx}$~\footnote{
the used reference system is defined to have a right handed frame with the
z axis along  the spectrometer axis (i.e. the nominal beam direction) 
the y axis in the vertical direction pointing upwards, and the origin
at the centre of the $u$-cell.}
is given by the average of the distance of
closest approach of all tracks in the vertex, while in the orthogonal plane
the coordinates x$_{vtx}$ and y$_{vtx}$ are defined from the beam track at 
z$_{vtx}$.
The tracks belonging to the vertex are selected using a Kalman fit.
The ``best primary vertex'', used in the following steps of the analysis,
 is defined as the one with the maximum 
number of tracks and, if the number of tracks is the same, the
one with smaller vertex $\chi^2$.

Only the events with at least one primary vertex reconstructed
with at least one more outgoing track,
and $Q^2 > 1$ (\gevc )$^2$, which are about 1\% of the initial raw event 
sample, are written on miniDST and used in the physics analysis.

The event reconstruction was almost the same for the 2003
and 2004 data, and is very similar to that used for the 2002 data.
Each year an improvement in the ratio of reconstructed events
over the useful beam was achieved, due both to additional
detectors in the spectrometer and to the better tuned
reconstruction and analysis software.

\subsection{Data quality checks}

To monitor the performances of the apparatus 
before and after the target spin reversal (i.e. on the two ``sub-periods'')
in each data taking period, the time stability of many distributions
has been checked dividing the
whole sample into runs or clusters of neighbouring runs.

Using the histograms produced during the event reconstruction,
the detector performance stabilities were scrutinised looking at 
the stability of the shape of the hit distributions in the about 360 detector 
planes.
The time stability of the detector and reconstruction efficiencies 
was checked looking at:\\
- the number of clusters per plane and per event,\\
- the mean number of tracks per event,\\
- the mean number of track segments in the different spectrometer regions per 
event,\\
- the mean number of primary vertexes per event,\\
- the mean number of secondary vertexes per event.\\
Using the miniDST events,
the stability was checked monitoring run per run\\
- the number of reconstructed $K^0$ per primary vertex,\\
- the reconstructed $K^0$ mass distribution,\\
- the energy measured in the two hadronic 
calorimeters $HCAL1$ and $HCAL2$,\\
- the x$_{vtx}$ and y$_{vtx}$ distributions in the two cells,\\
- the vertex $\chi^2$ distribution.\\
Also, the time stability of the distributions of several kinematical 
observables (like $x$, $Q^2$, $y$, the azimuthal angles and the momenta of 
the scattered muon and of the hadrons) was  investigated in detail.

\begin{figure}[bt]
\begin{center}
\includegraphics[width=.8\textwidth]{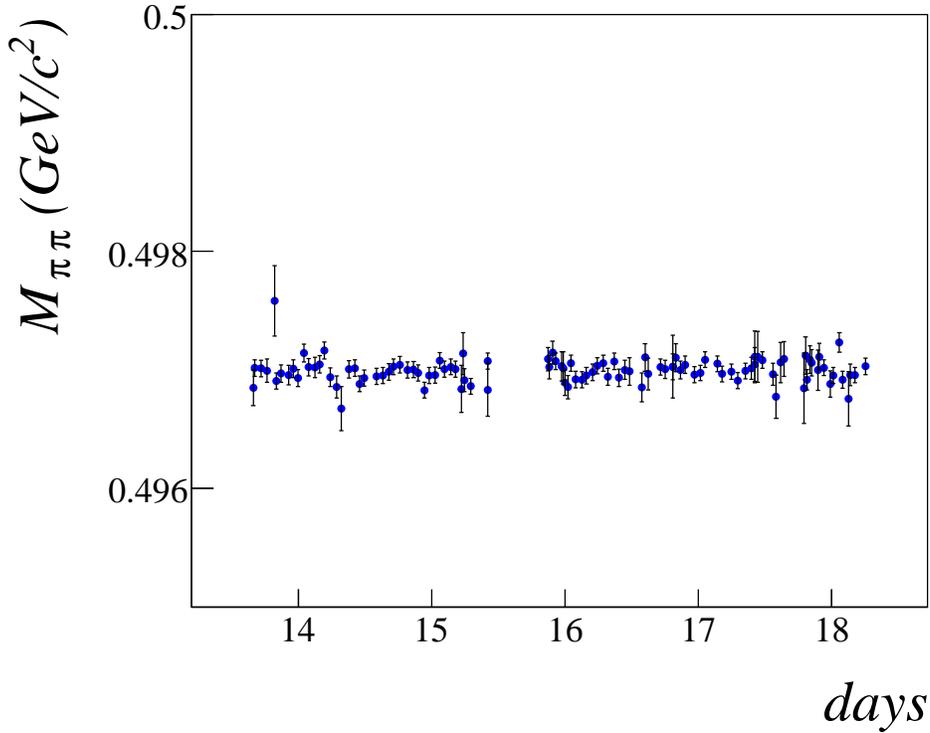} 
\end{center}
\caption{$\pi \pi$ invariant mass in the $K^0$ region as a function of time 
from August 13 to August 19, 2004. }
\label{fig:k0}
\end{figure}

As an example, the mean $\pi \pi$ invariant mass in the $K^0$ region 
from the data collected in the first sub-period of the 2004 run is shown 
in Fig.~\ref{fig:k0} as a function of time.

Runs showing some instability were not used for the physics analysis.
The runs rejected by this criterion are 28 (over a total
of 458) for the 2003 data, and 44 (over 462) for the 2004 data,
corresponding to about 5\% and 4\% of the initial raw event sample.

\subsection{DIS events selection}
To better define the DIS events, more refined cuts were applied.
In particular:
\begin{itemize}
\item[1.] 
primary vertex: if more primary vertexes were reconstructed 
in one event, the best  primary vertex
was selected. \\
The vertex had then to be inside the target.
A radial cut $r_{vtx} < 1.3$~cm was  applied on the distance of the  
vertex from the target axis.
The event was then accepted only if z$_{vtx}$ was inside one of the
two target cells (-100~cm $<$ z$_{vtx} <$ -40~cm or 
-30~cm $<$ z$_{vtx} <$ 30~cm), and 
assigned correspondingly to the \textit{u}- or to the \textit{d}-cell.\\
The distribution of the primary vertex z-coordinate for the final sample
is shown in Fig.~\ref{fig:zvtx}. 
The increase in the number of events with z$_{vtx}$ is
due to the increase in geometrical acceptance going from the 
upstream to the downstream end of the target, as explained in
Section~\ref{sec:setup}, and is very well reproduced by the
Monte Carlo simulation.
The two target cells are clearly separated.
One can also notice the continuum of events produced on the helium bath, 
as well as a sample of events produced in 
the Al window, at z$=50$~cm, which seals the vessel
of the PT magnet.
\begin{figure}[bt]
\begin{center}
\includegraphics[width=.8\textwidth]{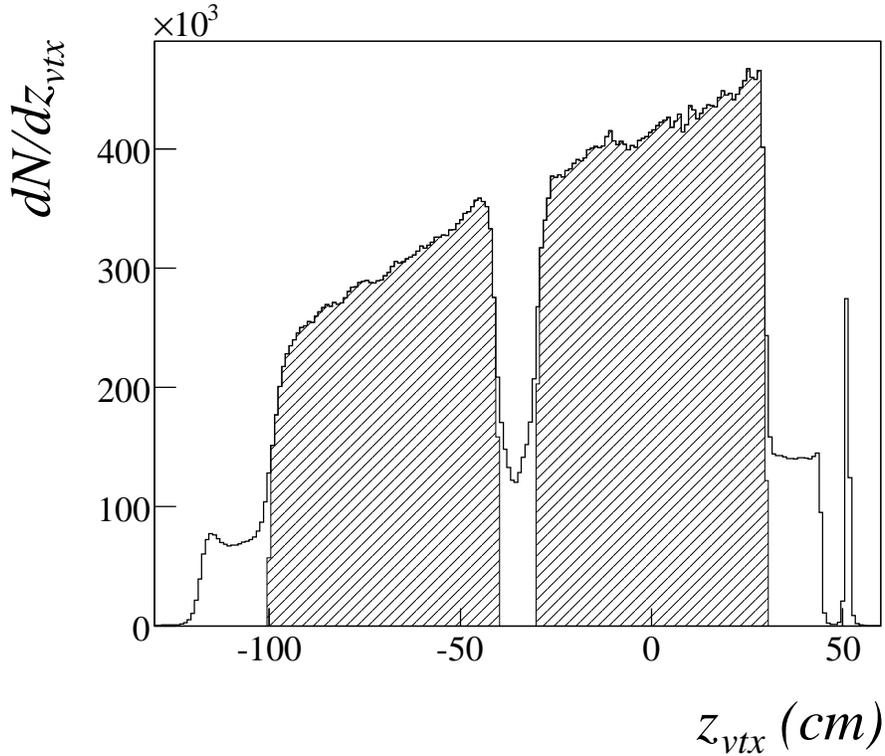} 
\end{center}
\caption{Distribution of the primary vertex z-coordinate for the final sample.
The dashed histogram indicates the events selected by the z$_{vtx}$ cut.}
\label{fig:zvtx}
\end{figure}
\item[2.] 
incoming muon: the cut $p_{beam} < 200 $ \gevc\ was applied
on the momentum of the incoming particle.
A check was also performed on the quality of the beam track, 
and a safe cut on $\chi^2/\nu$ was applied.\\
To ensure an identical beam intensity in both target
cells, i.e. a nearly identical luminosity, the beam track projection 
at the entrance and at the exit of the target region (e.g. at
z$=-100$~cm and z$=30$~cm)
had to be inside the target region as defined for the
primary vertex.
The momentum distribution for the final data sample is shown in
Fig.~\ref{fig:beam}.
\begin{figure}[bt]
\begin{center}
\includegraphics[width=.8\textwidth]{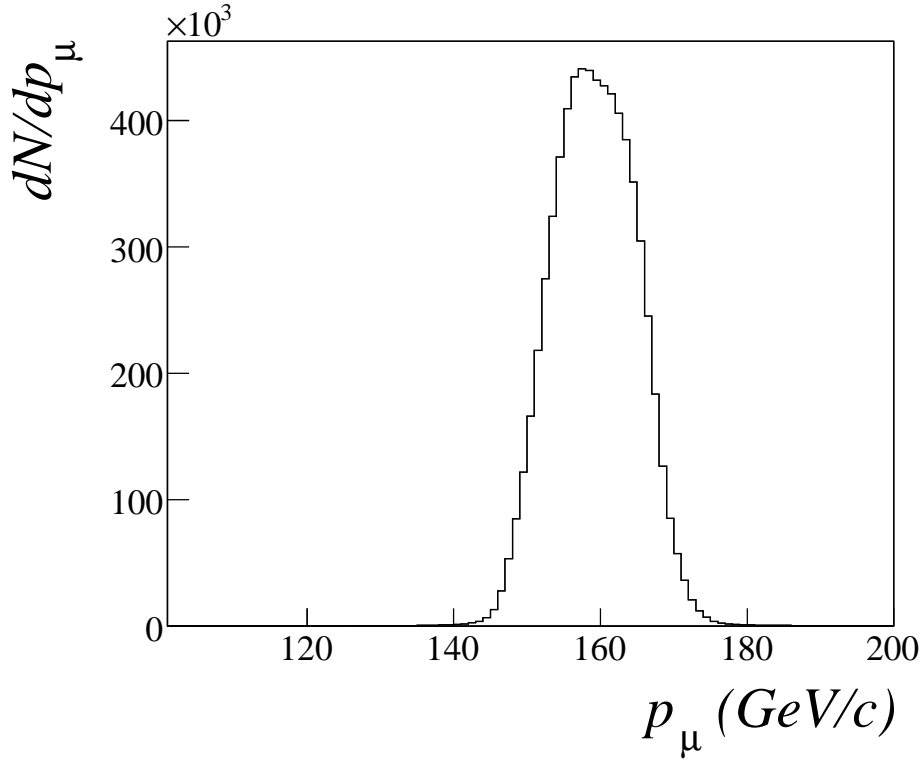} 
\end{center}
\caption{Momentum distribution of the reconstructed incoming muons
for the final sample of events.}
\label{fig:beam}
\end{figure}
\item[3.] 
scattered muon ($\mu^{\prime}$):
as well as for the incident muon, a suitable cut was applied
on the $\chi^2/\nu$ of all reconstructed tracks from the vertex, including the 
scattered muon candidate.
Other outgoing tracks were flagged as muon candidate if they had hits
in one of the two Muon Walls.
To achieve a clean muon identification, the amount of material 
traversed in the spectrometer had to be larger than 30 radiation lengths.
Only events with one and only one muon candidate entered the following 
steps of the analysis.
\end{itemize}
In addition, standard DIS cuts $Q^2 > 1$~(\gevc )$^2$, 
mass of the final hadronic state
$W > 5$~\gevc $^2$, and $0.1 < y < 0.9$ were
applied.
All these cuts reduced the number of miniDST events by a further 45\%.

\subsection{Hadron identification}
All the outgoing particles not flagged as $\mu^{\prime}$ candidates
were assumed to be hadrons.

Only particles with the last measured coordinate after the first spectrometer
magnet were used, to reject tracks  reconstructed in the fringe field
on SM1 which have a poorer momentum resolution.

The
following requirements had to be satisfied in order
to identify  the particle with a hadron:
\begin{itemize}
\item[1.] 
the amount of material traversed in the spectrometer had to be smaller 
than 10 radiation lengths;
\item[2.] the particle must not  give a signal in any of the two hadronic 
calorimeters $HCAL1$ and $HCAL2$.
\item[3.] if the particle gave a signal in one calorimeter, the measured energy
had to be $E_{HCAL1}\,>$ 5 (2003 data) or 4 (2004 data) GeV, 
or $E_{HCAL2}\,>$ 5 (2003 and 2004 data) GeV. The correlation
between the energy measured in HCAL and that measured by
the spectrometer is shown in Fig.~\ref{fig:hcal}.
\item[4.] the transverse momentum of the particle with respect to the virtual 
photon direction had to be larger than 0.1 \gevc .\\
The first requirement reduced the muon contamination, the second and the third
the muon and the electron contamination. 
The last cut was introduced to assure a good resolution in the measured 
azimuthal angle.
\begin{figure}[bt]
\begin{center}
\includegraphics[width=0.45\textwidth]{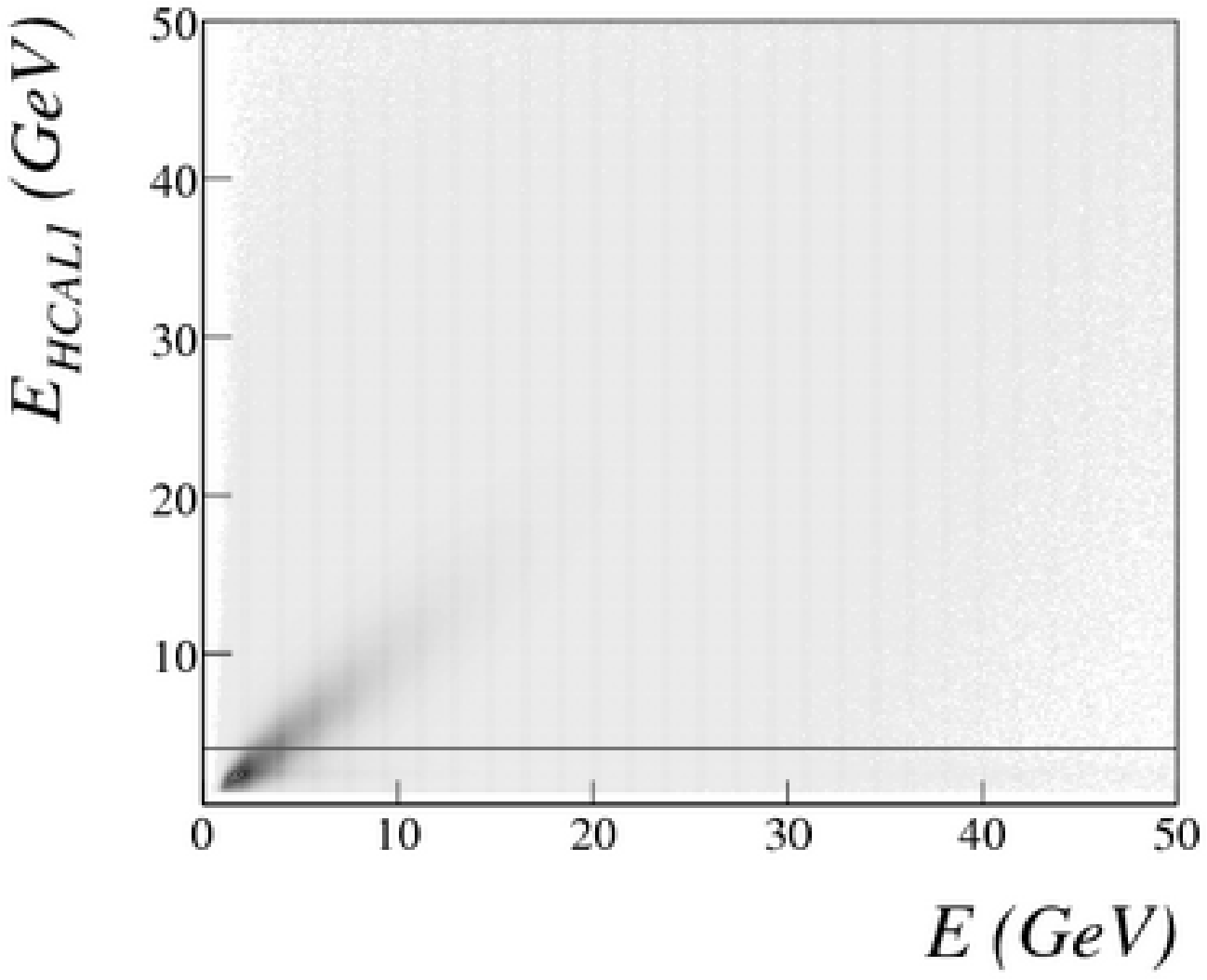} 
\includegraphics[width=0.45\textwidth]{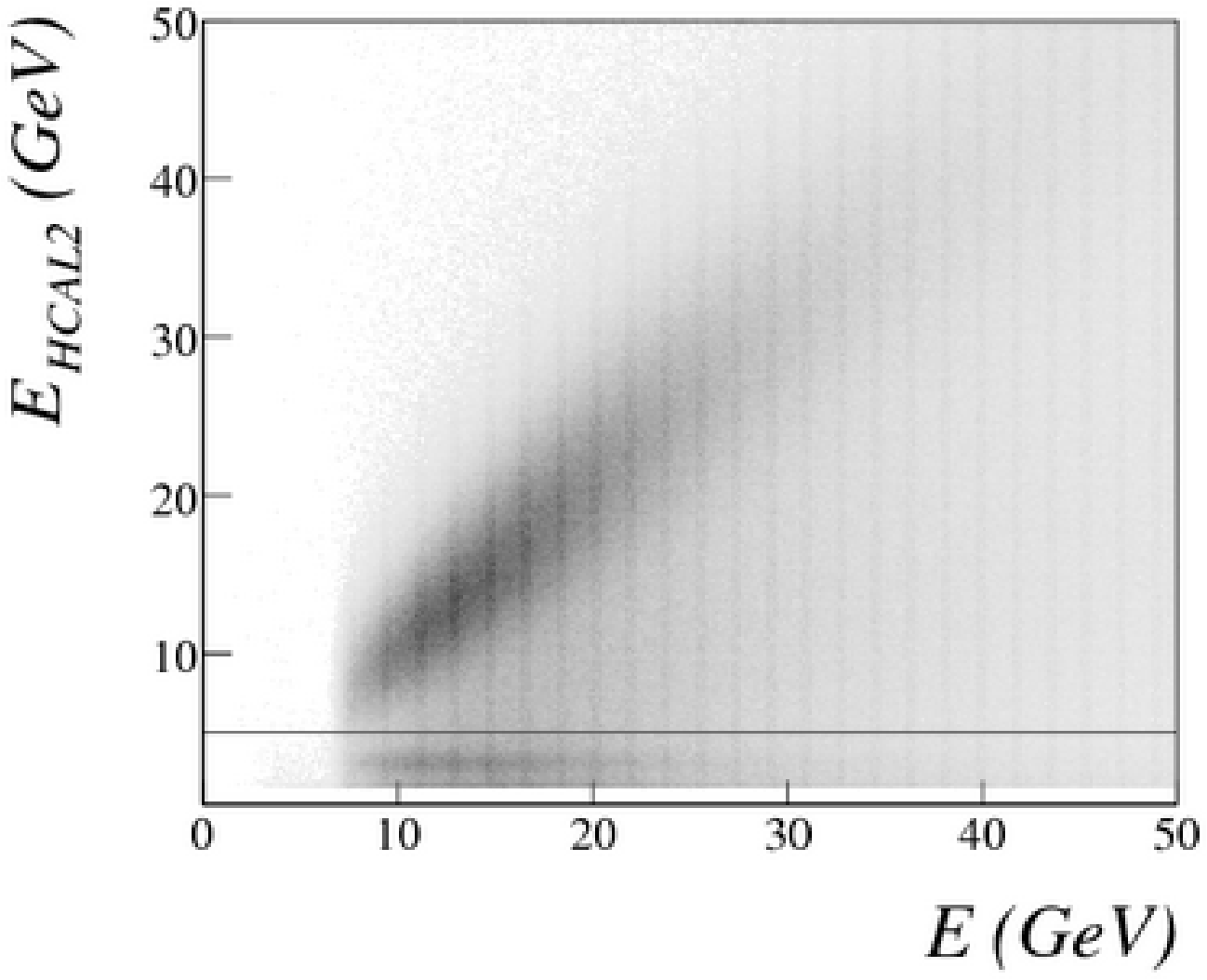} 
\end{center}
\caption{Correlation
between the energy measured in HCAL1 (left) and HCAL2 (right)
and the energy measured by
the spectrometer for the 2004 data.
}
\label{fig:hcal}
\end{figure}
\end{itemize}
Two different asymmetries have been measured.
The ``all hadron'' asymmetries were evaluated using all the particles
identified as hadrons carrying a fraction of available energy $z > 0.2$.
For the ``leading hadron'' analysis, only the events with at least
one hadron with $z > 0.25$ were used, and the hadron with largest $z$
was defined as leading. 
In the case the energy of the reconstructed leading hadron was higher 
than the missing
energy evaluated from all the charged reconstructed hadrons, no further test
was performed.
If this was not the case, the event was accepted only if 
in the hadron calorimeters no particle with energy larger (within the
energy resolution) than that of the leading hadron was detected.

After these cuts, the number of events with a leading hadron amounted 
to 22\% of the number of miniDST events.

\section{Data analysis II - kinematical distributions and asymmetry evaluation}
\subsection{Final samples of events}
In table  \ref{tab:stat}, the final statistics used for the asymmetry 
evaluation is given
for all the periods both for the positively and negatively charged
"leading hadron" and the "all hadron" sample.
\begin{table*}[htb]
\caption{Final statistics  used for the asymmetry evaluation}
\label{tab:stat}
\begin{center}
\vspace*{.5cm}
\begin{tabular}{ l l c c c c }
\hline
Year  & Period & \multicolumn{2}{c}{Leading hadron sample} & \multicolumn{2}{c}{All hadron sample} \\
 & & positive hadrons &  negative hadrons & positive hadrons & negative hadrons\\
\hline
2002 & 1 & 0.48$\, \cdot \, 10^6$ & 0.38$\, \cdot \, 10^6$ & 0.71$\, \cdot \, 10^6$ & 0.59$\, \cdot \, 10^6$ \\
2002 & 2 & 0.32$\, \cdot \, 10^6$ & 0.26$\, \cdot \, 10^6$ & 0.48$\, \cdot \, 10^6$ & 0.40$\, \cdot \, 10^6$ \\
\hline
2003 & 1 & 1.68$\, \cdot \, 10^6$ & 1.33$\, \cdot \, 10^6$ & 2.46$\, \cdot \, 10^6$ & 2.03$\, \cdot \, 10^6$ \\
\hline
2004 & 1 & 1.44$\, \cdot \, 10^6$ & 1.13$\, \cdot \, 10^6$ & 2.12$\, \cdot \, 10^6$ & 1.74$\, \cdot \, 10^6$ \\
2004 & 2 & 1.87$\, \cdot \, 10^6$ & 1.47$\, \cdot \, 10^6$ & 2.75$\, \cdot \, 10^6$ & 2.26$\, \cdot \, 10^6$ \\
\hline
\end{tabular}\\[2pt]
\end{center}
\vspace*{.8cm}
\end{table*}

\subsection{Kinematical distributions}
\begin{figure}[tb]
\includegraphics[width=.49\textwidth]{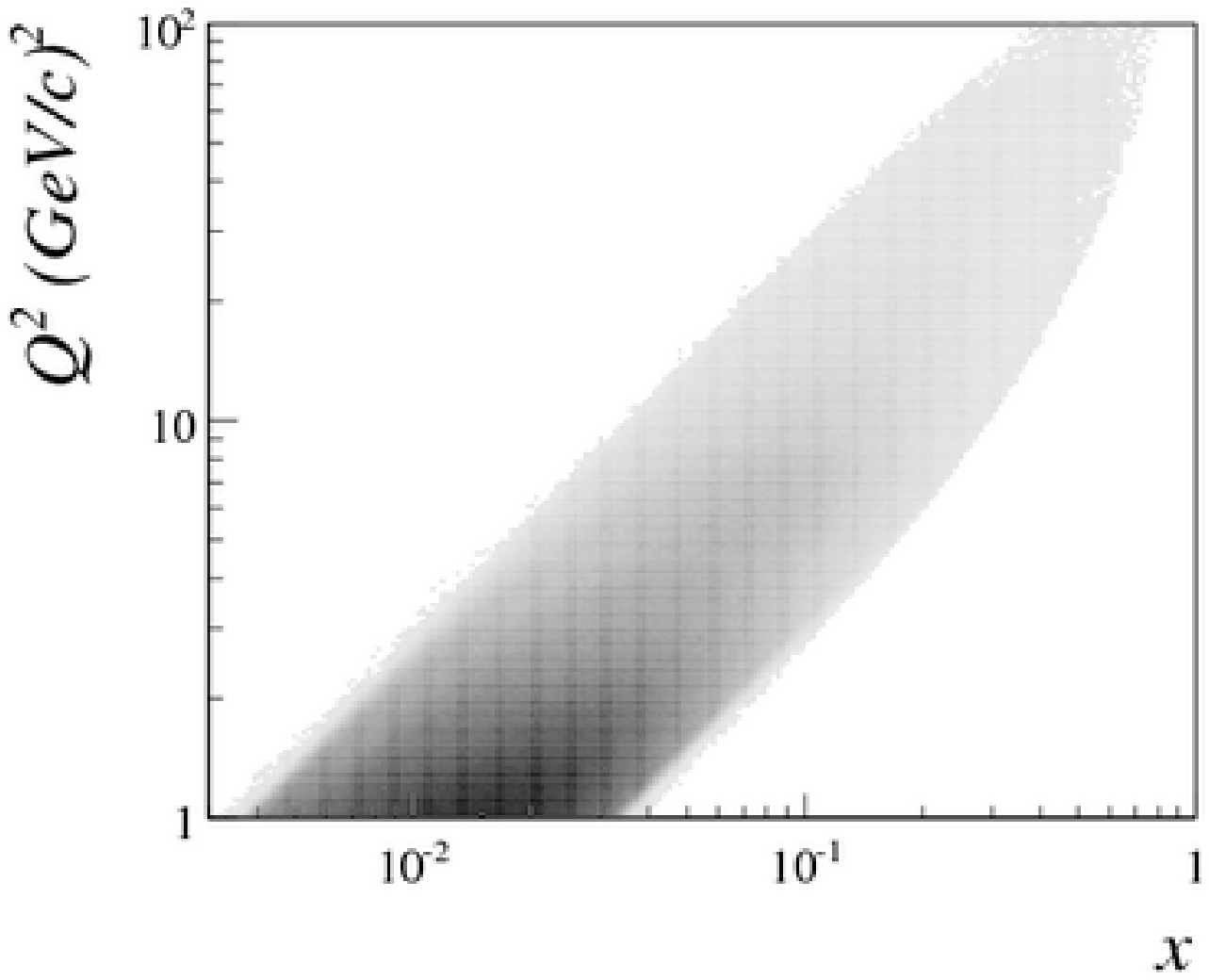} 
\includegraphics[width=.49\textwidth]{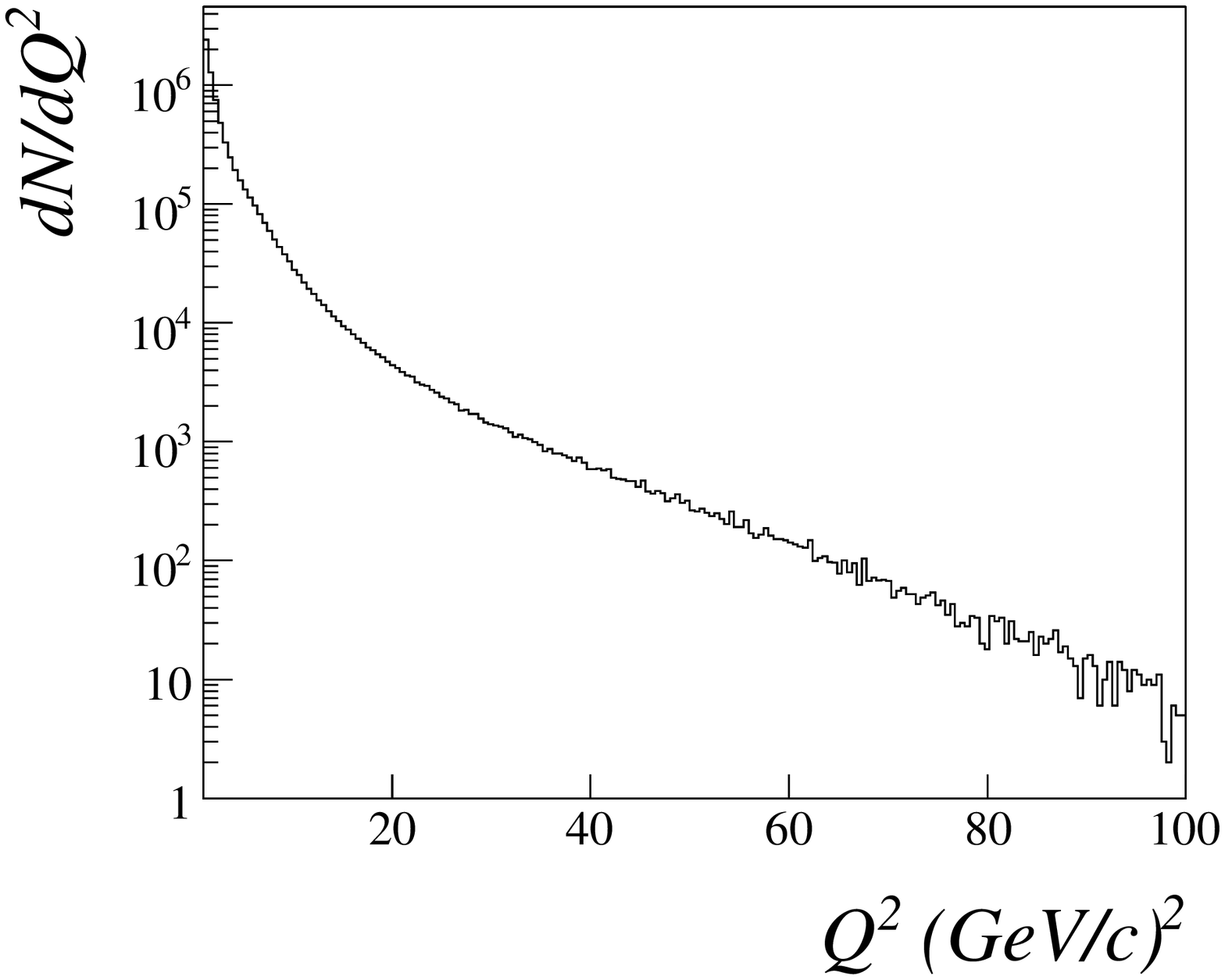} 
\includegraphics[width=.49\textwidth]{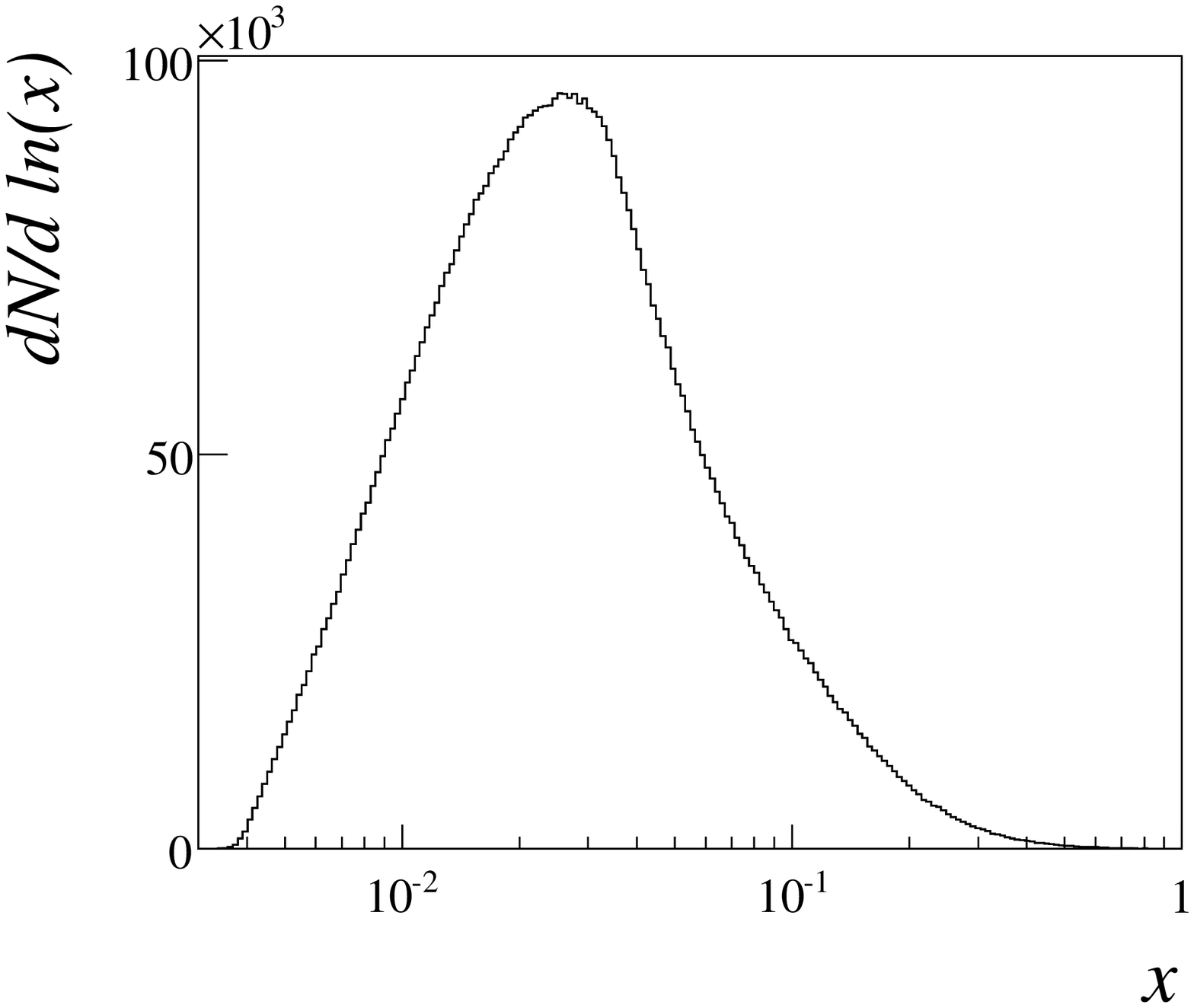} 
\caption{Scatter-plot of $Q^2$ vs $x$, and the corresponding $Q^2$- and 
$x$-distributions
for the 2004 positive plus negative leading hadron sample.
}
\label{fig:q2x}
\end{figure}

\begin{figure}[tb]
\includegraphics[width=.49\textwidth]{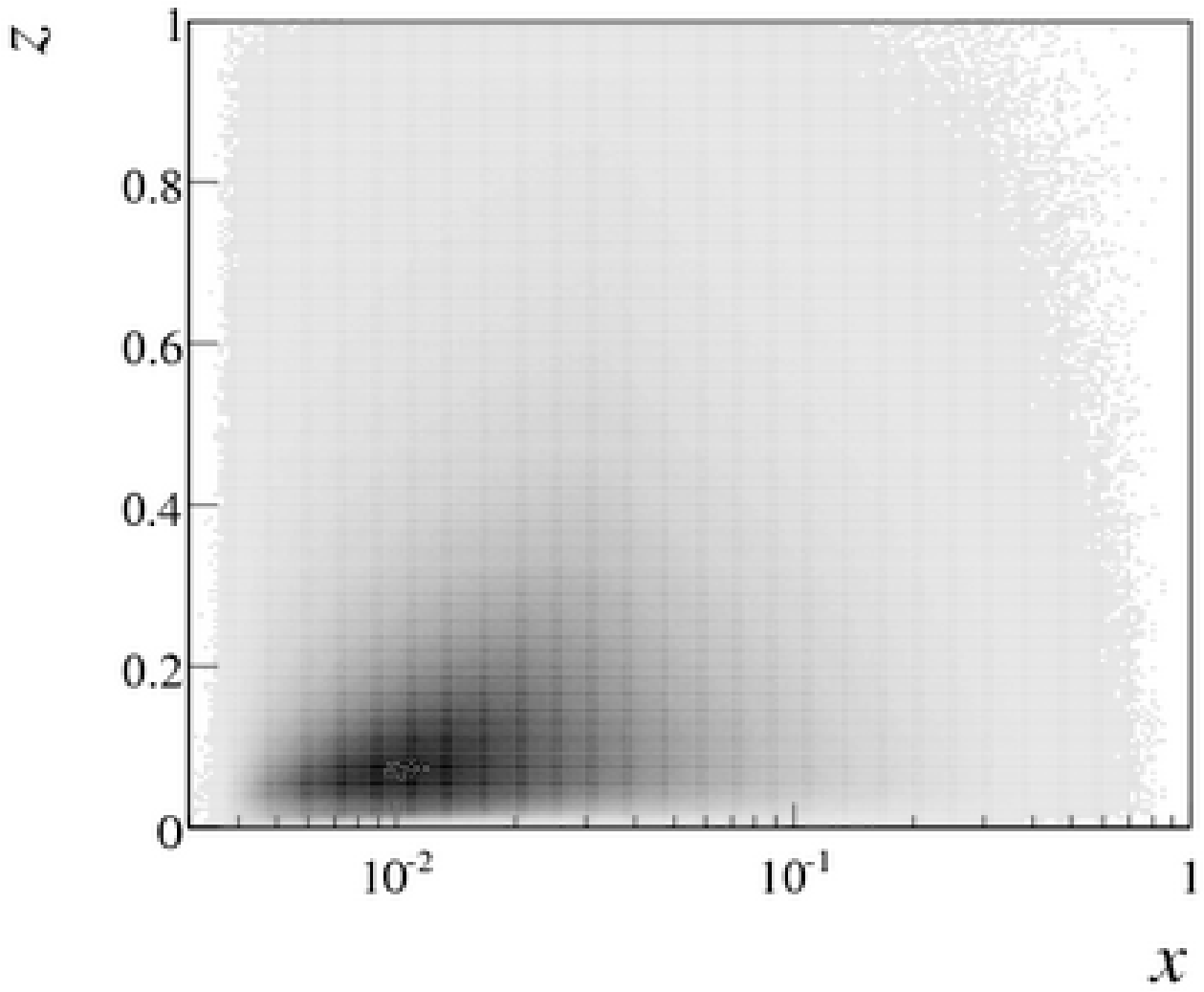} 
\includegraphics[width=.49\textwidth]{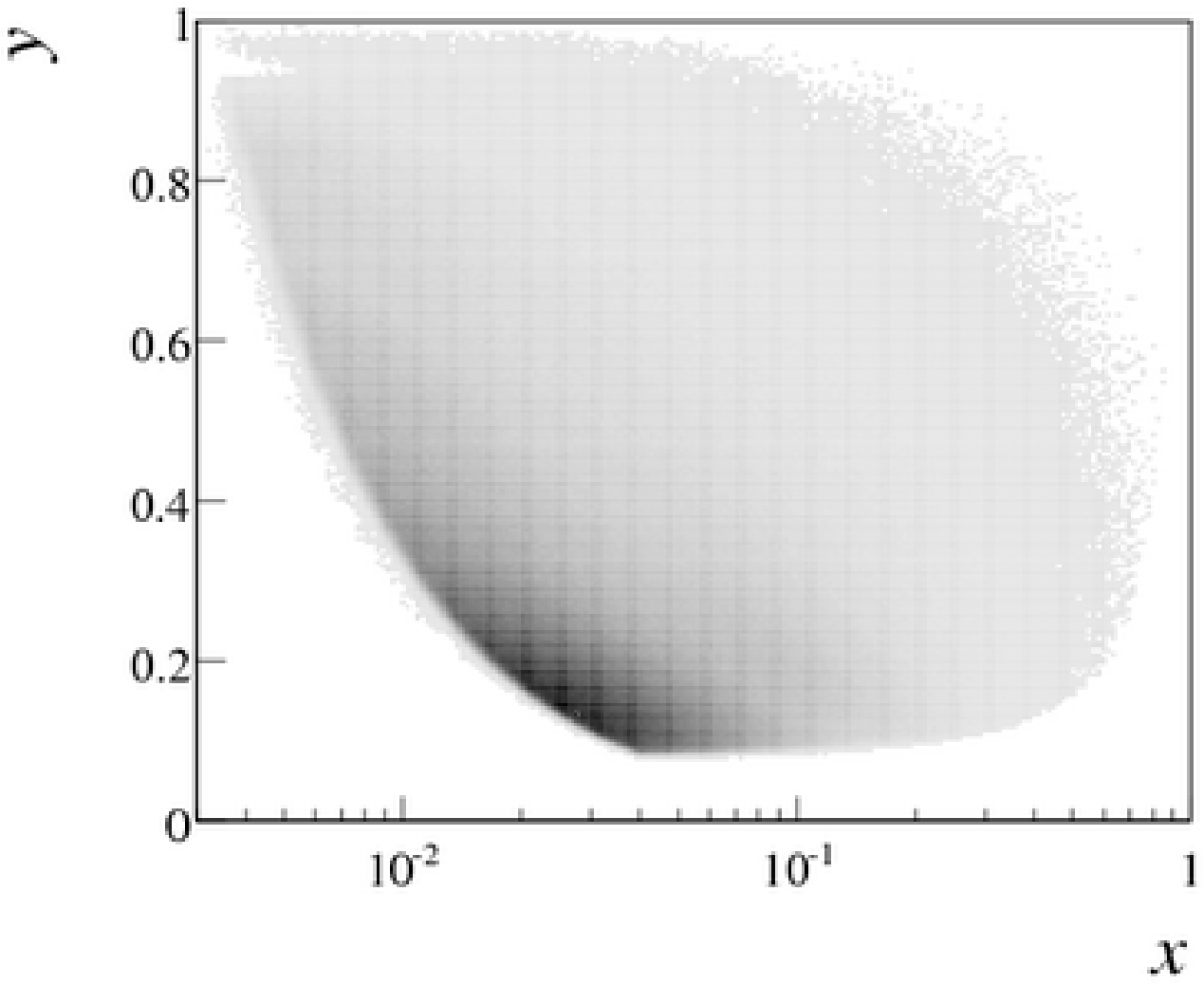} 
\includegraphics[width=.49\textwidth]{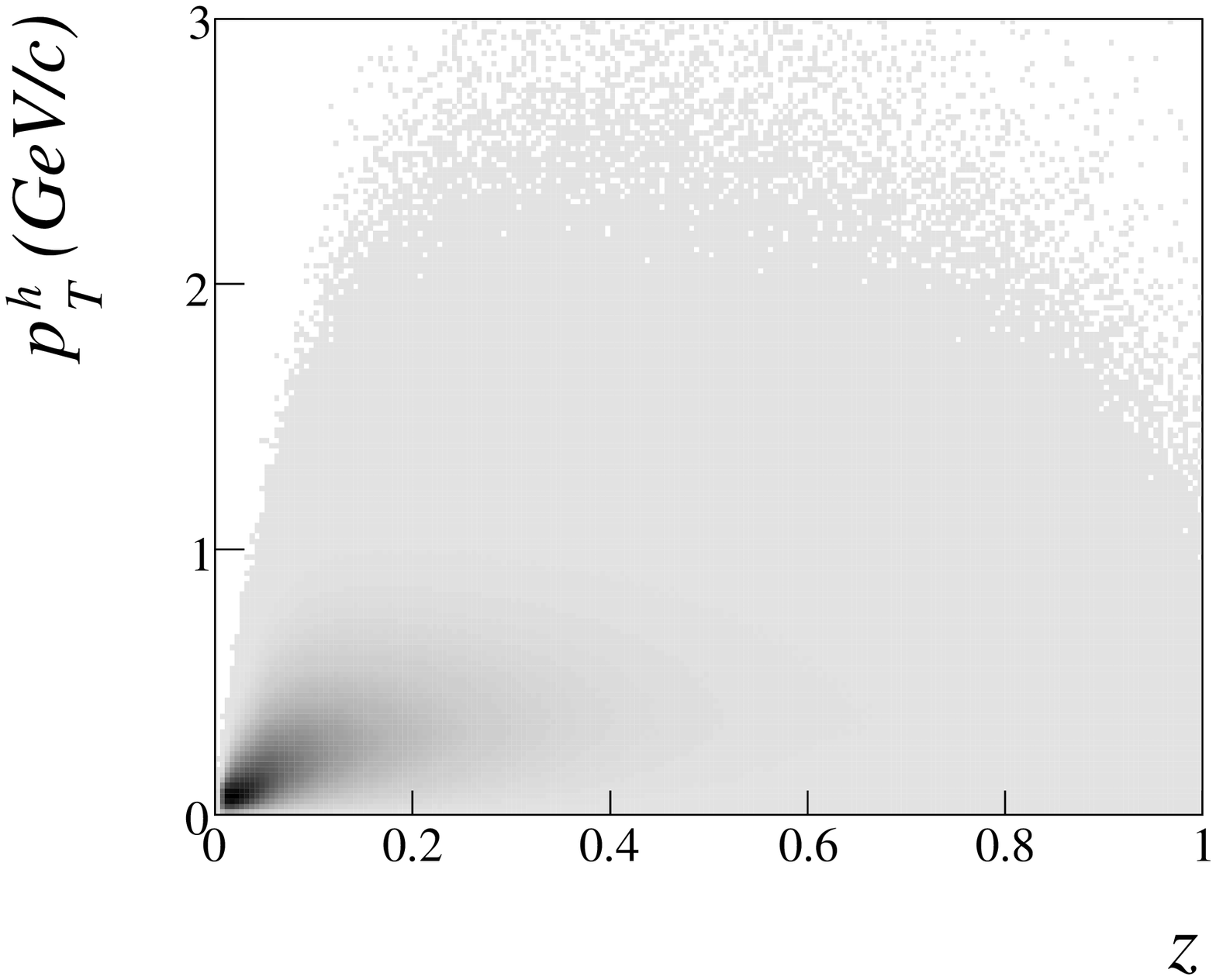} 
\caption{Scatter-plots  of $z$ vs $x$, $y$ vs $x$, and
$z$ vs $p^h_T$
for the 2004 all positive plus negative hadron samples.
}
\label{fig:xzxpt}
\end{figure}

\begin{figure}[tb]
\includegraphics[width=.49\textwidth]{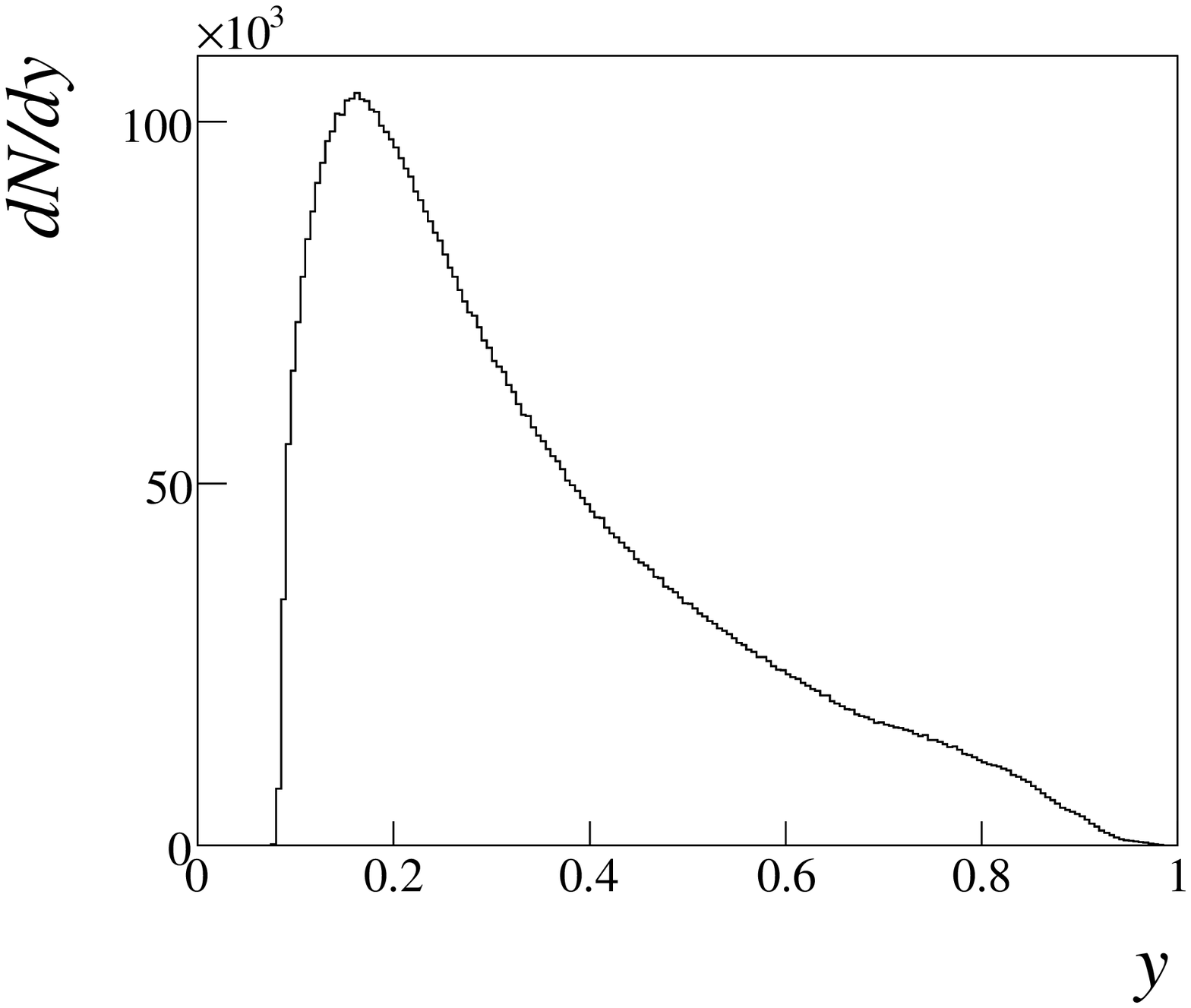} 
\includegraphics[width=.49\textwidth]{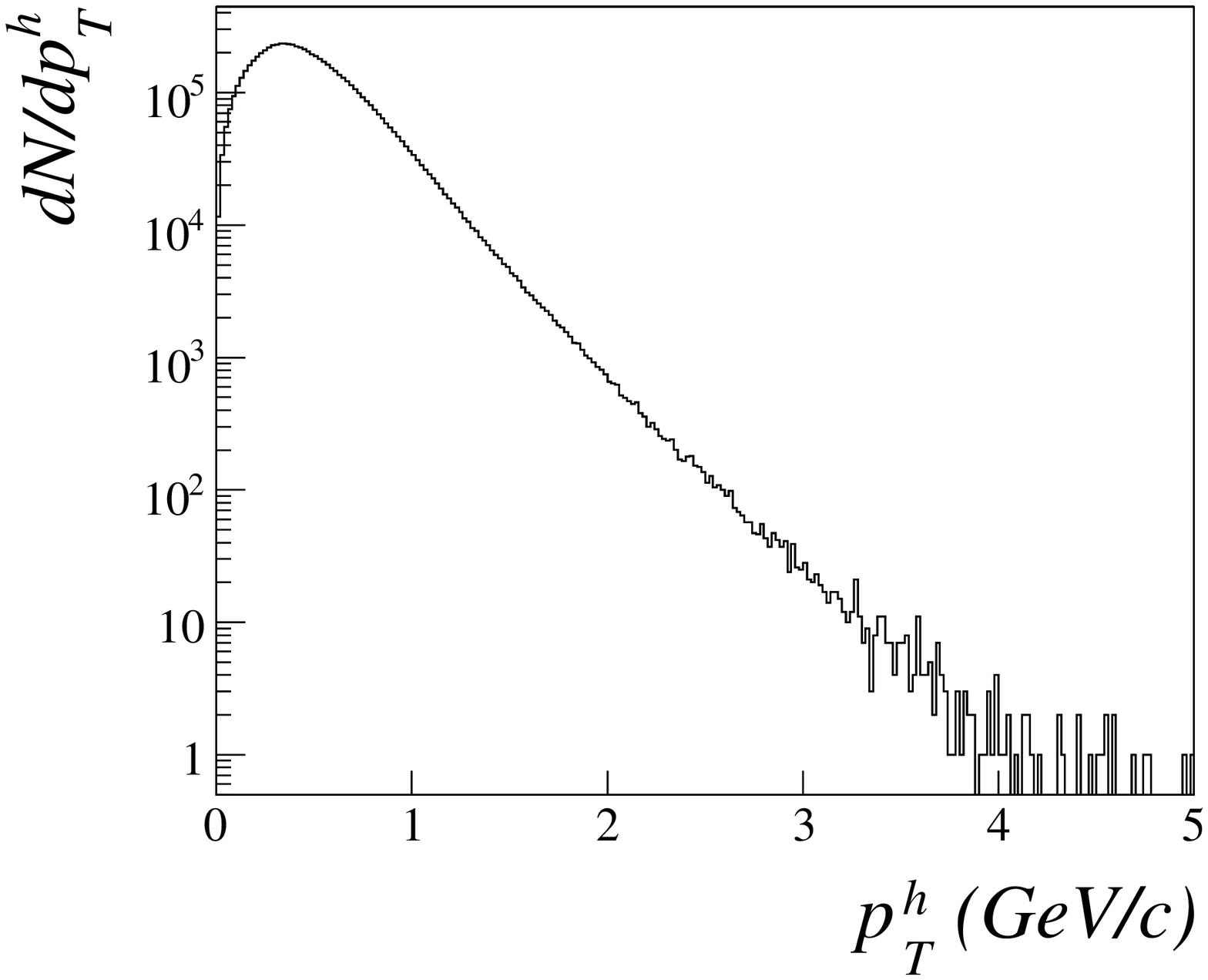} 
\includegraphics[width=.49\textwidth]{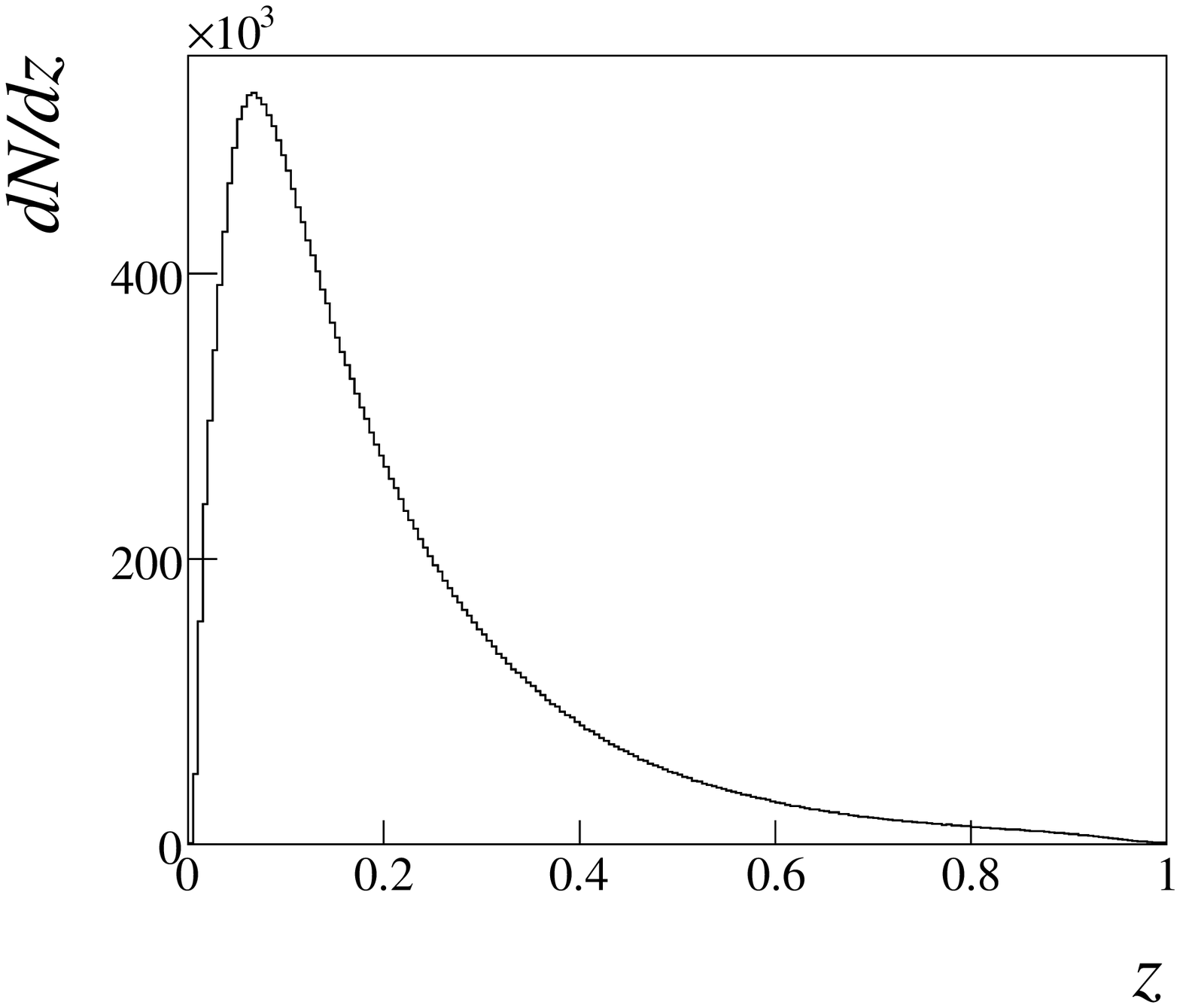} 
\includegraphics[width=.49\textwidth]{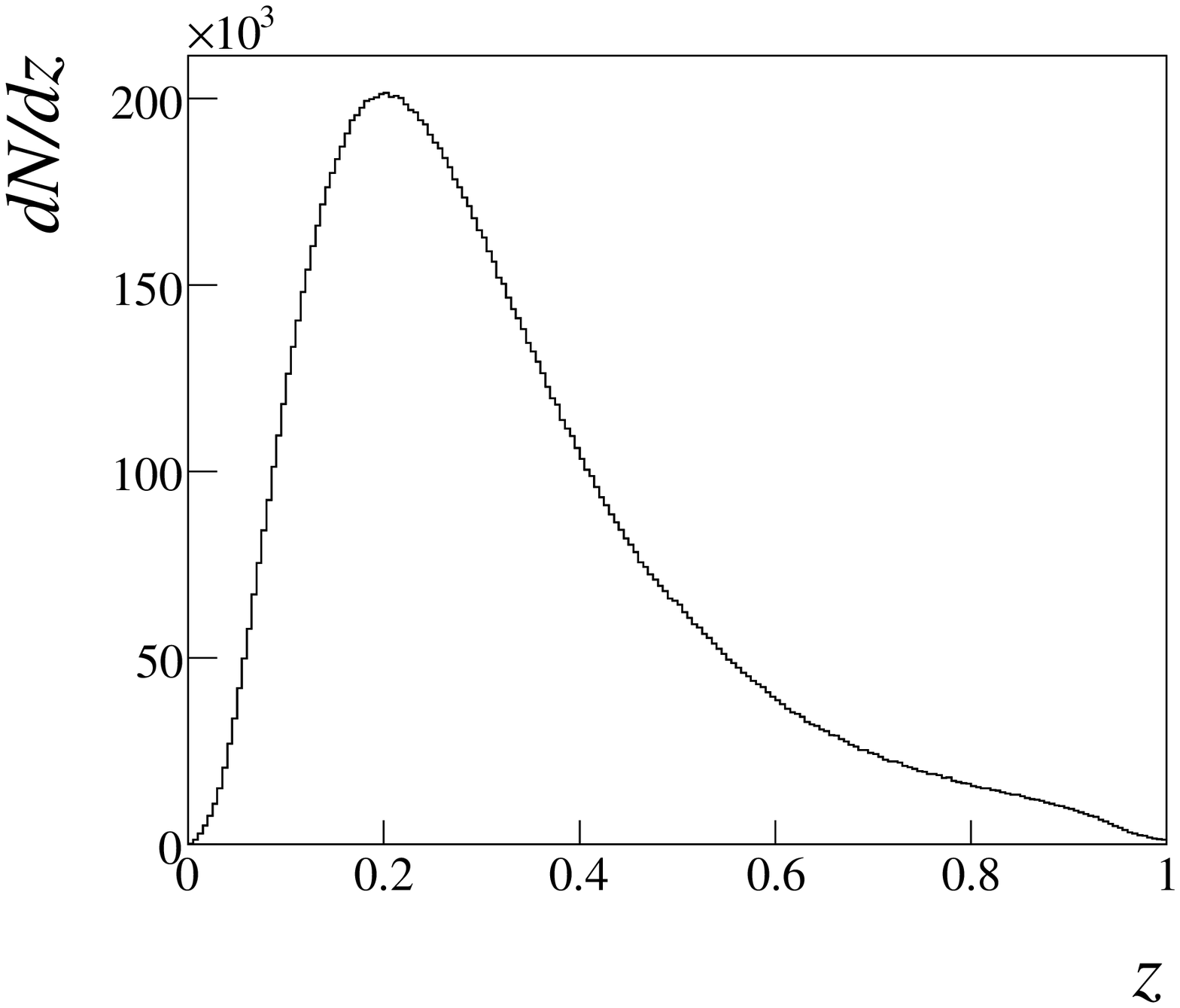} 
\caption{Upper plots: distributions of $y$
and $p^h_T$ (without the $p^h_T$ cut)
for the 2004 all positive plus negative hadron sample.
Lower plots: $z$-distribution (without the $z$ cut)
for the same events (left) and for the 2004 leading 
positive plus negative hadron sample (right).
}
\label{fig:kdist}
\end{figure}

The distributions of some kinematical quantities from
the final 2004  leading hadron events are shown in
Figs.~\ref{fig:q2x} to \ref{fig:kdist}.
The plots have been produced after all the cuts described
in the previous section, if not specified differently.

The $Q^2$ vs $x$ scatter-plot and its projections
are shown in Fig.~\ref{fig:q2x}.

The $x$-$z$ correlation (without the $z$ cut), the $x$-$y$ correlation, 
and the $z$-$p^h_T$ correlation (without the $z$ and the $p^h_T$ cuts)
are shown in Fig.~\ref{fig:xzxpt}, while the $y$, 
$p^h_T$ (without the $p^h_T$ cut) and the $z$ 
(without the $z$ cut) distributions are shown in Fig.~\ref{fig:kdist}.

\subsection{Asymmetry evaluation}
%
\subsubsection{Binning}
%
The Collins and Sivers asymmetries were evaluated as function of
$x$, $p^h_T$, and $z$ dividing the corresponding kinematical range in 
bins (with variable width, in order to have 
a comparable statistics in each of them), and integrating over the other 2
variables.
No attempt has been done to extract the asymmetries in a 2- or 3-dimensional 
grid.
In total, the asymmetries where evaluated in the 9  $x$-bins, 
9  $p^h_T$-bins, and 8 $z$-bins for the 
all hadron samples (7 for the leading hadron, because of the
$z > 0.25$ cut) given in Table~\ref{tab:binning}.

\begin{table}[bt]
\begin{center}
\begin{tabular}[h]{| l | l | l |}
 \hline
0.003 $< x<    $ 0.008 &    0.20 $\leq z <$ 0.25 & 0.10 $ < p^h_T \le $ 0.20 GeV/$c$ \\
0.008 $\leq x< $ 0.013 &    0.25 $\leq z <$ 0.30 & 0.20 $ < p^h_T \le $ 0.30 GeV/$c$ \\
0.013 $\leq x< $ 0.020 &    0.30 $\leq z <$ 0.35 & 0.30 $ < p^h_T \le $ 0.40 GeV/$c$ \\
0.020 $\leq x< $ 0.032 &    0.35 $\leq z <$ 0.40 & 0.40 $ < p^h_T \le $ 0.50 GeV/$c$ \\
0.032 $\leq x< $ 0.050 &    0.40 $\leq z <$ 0.50 & 0.50 $ < p^h_T \le $ 0.60 GeV/$c$ \\
0.050 $\leq x< $ 0.080 &    0.50 $\leq z <$ 0.65 & 0.60 $ < p^h_T \le $ 0.75 GeV/$c$ \\
0.080 $\leq x< $ 0.130 &    0.65 $\leq z <$ 0.80 & 0.75 $ < p^h_T \le $ 0.90 GeV/$c$ \\
0.130 $\leq x< $ 0.210 &    0.80 $\leq z <$ 1.00 & 0.90 $ < p^h_T \le $ 1.30 GeV/$c$ \\
0.210 $\leq x< $ 1.000 &                         & 1.30 $ < p^h_T  $          \\
 \hline
\end{tabular}
\end{center}
\caption{Final binning in $x$, $z$, and $p^h_T$.}
\label{tab:binning}  
\end{table} 

\begin{figure}[bt]
\begin{center}
\includegraphics[width=0.49\textwidth]{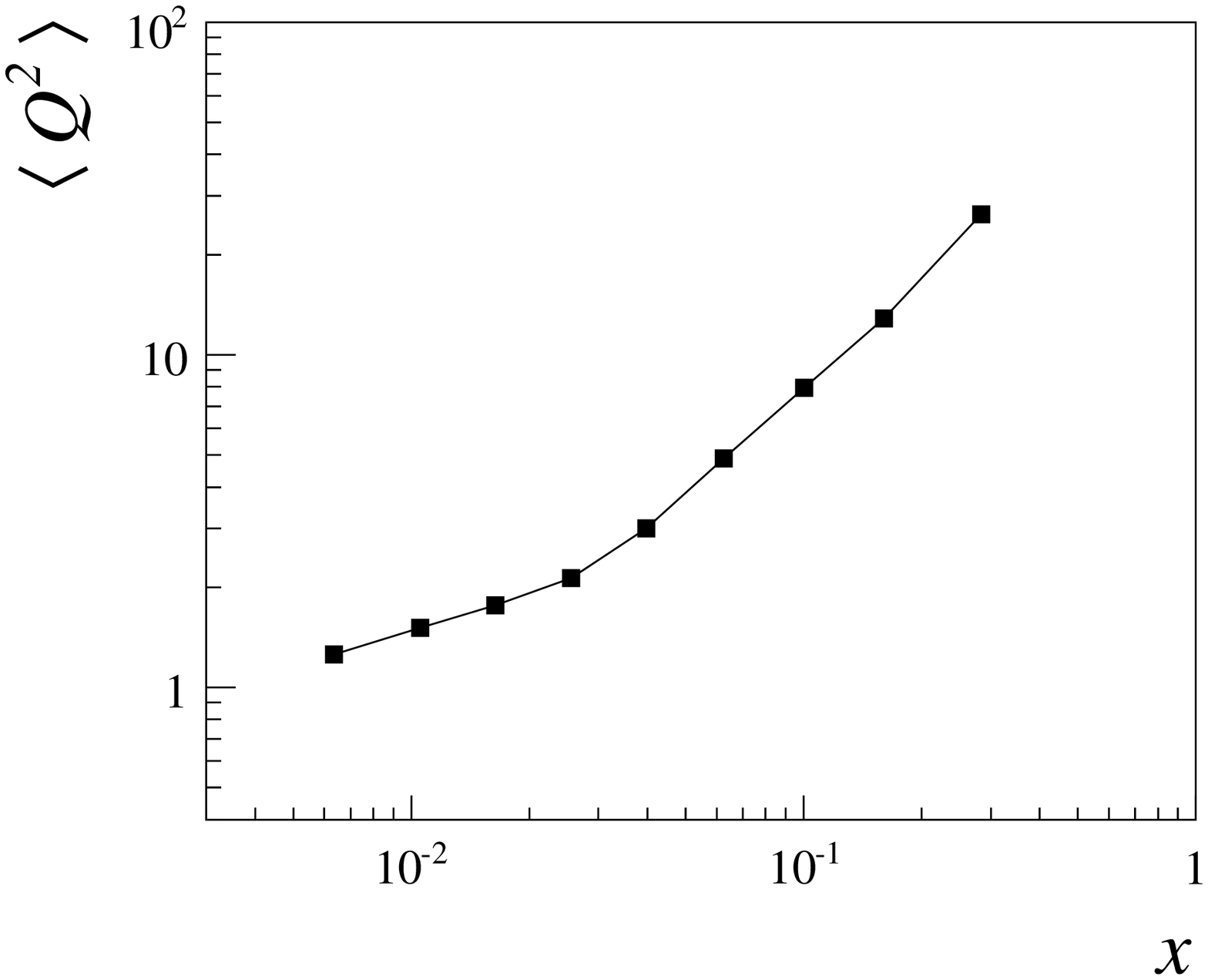}
\includegraphics[width=0.49\textwidth]{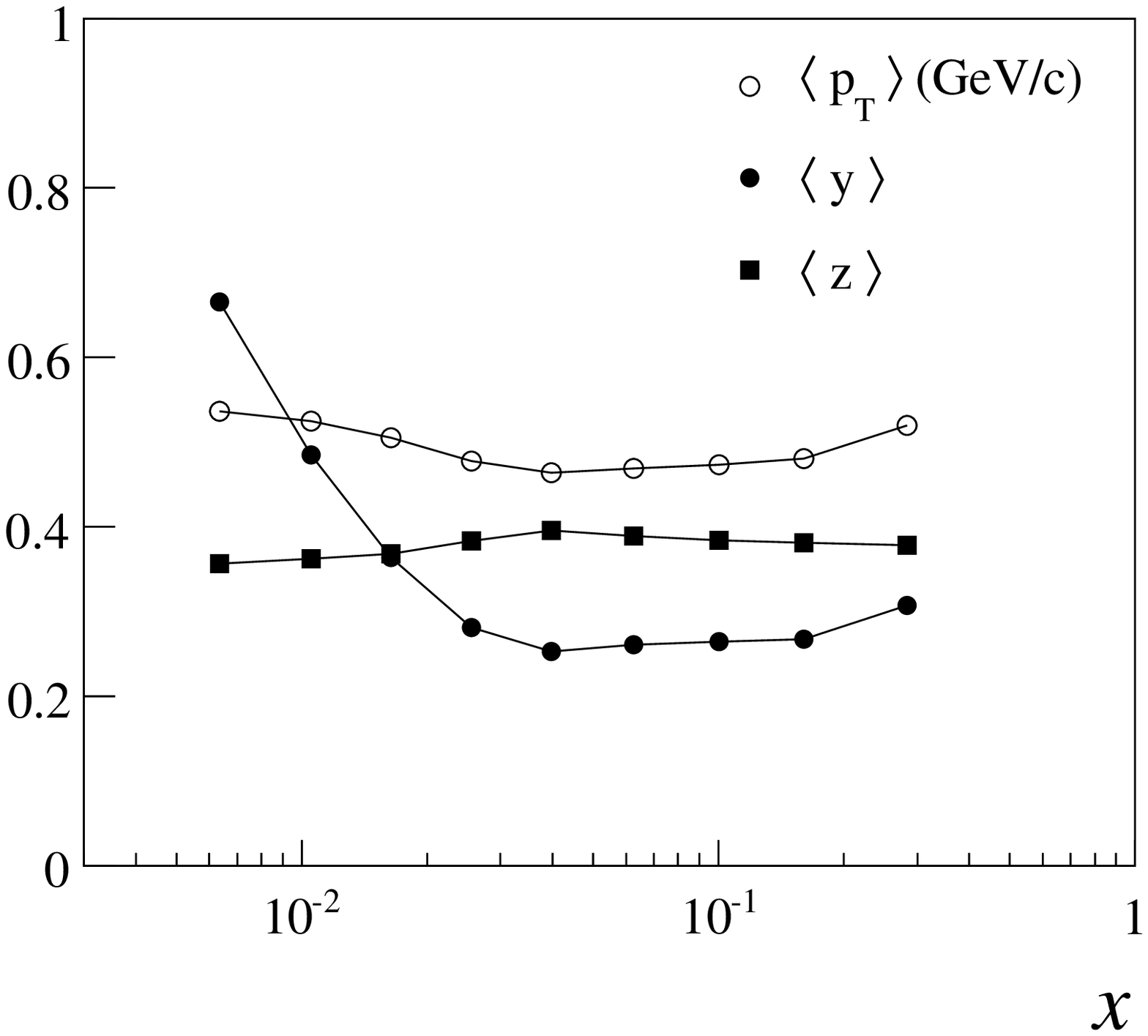}
\end{center}
\caption{Mean values of $Q^2$ 
(left), $z$, $p^h_T$ and $y$ (right) in the different $x$ bins
for all positive hadrons, 2003 and 2004 data.
}
\label{fig:kxbinsp}
\end{figure}

Tables with the mean values of $Q^2$, $z$, $p^h_T$ and $y$
in all the $x$ bins are available on HEPDATA~\cite{hepdata}.
As an example, Fig.~\ref{fig:kxbinsp} shows the mean values of $Q^2$ 
(left), $z$, $p^h_T$ and $y$ (right) in the different $x$ bins
for the all positive  hadron sample
from 2003 and 2004 data.

\subsubsection{Raw asymmetry evaluation}
\label{subs:rpm}
%
The Collins and Sivers asymmetries have been evaluated
separately, in each kinematic bin, for each data taking period,
and for positive and negative hadrons.

Using Eq.~(\ref{eq:sicross}) and (\ref{eq:sicross1}), the $\Phi$ distribution
of the number of events for each cell and for each 
polarisation state can be written as
\begin{equation}
N^{\pm}_{j,k}(\Phi_j)  =  
F^{\pm}_k \, n_k  \, \sigma \cdot a^{\pm}_{j,k}(\Phi_j) 
\cdot (1 \pm \epsilon^{\pm}_{j,k} \, \sin \Phi _j) ,
\label{eq:angdis}
\end{equation}
where $j=C, S$, and $F$ is the muon flux, $n$ the number of target particles,
$\sigma$ the spin averaged cross-section, and $a_{j}$ the product of 
angular acceptance and efficiency of the spectrometer.
The index $k = u, d$ refers to the target cell.
The $\pm$ signs indicate the two target spin orientations: $+$
means spin up, $-$ spin down.
Here the Collins and Sivers angles $\Phi_j$ are evaluated using the 
expressions given
in Sections~\ref{sec:collins} and~\ref{sec:sivers}, always assuming target 
spin up, at variance
with what was done in~\cite{Alexakhin:2005iw}.

The asymmetries $\epsilon^{\pm}_{j,k}$ are
\begin{equation}
\epsilon^{\pm}_{C,k} = c^{\pm}_{C,k} \cdot A_{Coll} \, , \; \; \; \;
\epsilon^{\pm}_{S,k} = c^{\pm}_{S,k} \cdot A_{Siv} \, ,
\end{equation}
where
\begin{equation}
c^{\pm}_{C,k} = f \cdot |P^{\pm}_{T,k}| \cdot D_{NN} \, , \; \; \; \;
c^{\pm}_{S,k} = f \cdot |P^{\pm}_{T,k}| \, .
\end{equation}
The factor $f$ is the polarised target dilution factor, 
$P^{\pm}_{T,k}$ the deuteron polarisation, and
$D_{NN}$ the transverse spin
transfer coefficient from the initial to 
the struck quark given in Eq.~(\ref{eq:dnn}).
The absolute values of the target polarisation $P_{T}$ 
in the two cells and for the two spin orientations in the  same cells differed
up to about 7\%.

In order to estimate $\epsilon^{\pm}_{j,k}$ from the measured number of events
a method, which in the
following will be called ``ratio product method'' (RPM), has been used
which minimises the effect on the estimated asymmetries
of possible acceptance variations.

In each data taking period, the ``ratio product'' 
quantities~\cite{Alexakhin:2005iw}
\begin{equation}
A_j(\Phi_{j}) = \frac{ N_{j,u}^{+}(\Phi_{j})}
                     {N_{j,u}^{-}(\Phi_{j})} \cdot
\frac{ N_{j,d}^{+}(\Phi_{j})} 
                     {N_{j,d}^{-}(\Phi_{j})} , \; \; \; \; \; j=C,S 
\label{eq:ratiop}
\end{equation}
were evaluated from the number of events in the two cells.
For small values of the quantities $\epsilon^{\pm}_{j,k}$ it is:
\begin{eqnarray}
A_j(\Phi_{j}) & = & C_F \cdot C_{a,j} \cdot
    \frac{ (1 + \epsilon^{+}_{j,u} \cdot \sin \Phi _j)
          \cdot(1 + \epsilon^{+}_{j,d} \cdot \sin \Phi _j) }
       { (1 - \epsilon^{-}_{j,u} \cdot \sin \Phi _j)
          \cdot(1 - \epsilon^{-}_{j,d} \cdot \sin \Phi _j) }
\nonumber \\
 & \simeq & C_F \cdot C_{a,j} \cdot (1 + A^m_{j} \cdot \sin \Phi _j) 
\label{eq:rpcalc}
\end{eqnarray}
where
\begin{equation}
A^m_{j}=\epsilon^{+}_{j,u}+\epsilon^{+}_{j,d}+\epsilon^{-}_{j,u}
           +\epsilon^{-}_{j,d} = 4 \cdot < \epsilon_j > \, , \\
\label{eq:rpcalc2}
\end{equation}
\begin{equation}
C_F  = 
  \frac{ F^{+}_u \cdot F^{+}_d }{ F^{-}_u \cdot F^{-}_d } \, , \; \; \; \;
C_{a,j}  = 
  \frac{ a^{+}_{j,u}(\Phi_j) \cdot a^{+}_{j,d}(\Phi_j) }
       { a^{-}_{j,u}(\Phi_j) \cdot a^{-}_{j,d}(\Phi_j) } \,  .
\end{equation}
The raw asymmetries $A^m_{C}$ and $A^m_{S}$
were evaluated by fitting respectively the quantities $A_C(\Phi_{C})$ 
and  $A_S(\Phi_{S})$ with the functions
$p_0 \cdot(1 + A^m_{C} \cdot \sin \Phi_{C})$ and 
$p_0 \cdot(1 + A^m_S \cdot \sin \Phi_{S})$,
where $p_0$ is a free parameter.

In this method
the only requirement to avoid systematic errors due to
acceptance effects is that the factor $C_{a,j}$ does not depend on
$\Phi_j$.
This is surely true, in particular, under the 
reasonable assumption that the ratio
$a_{j,u}^{+}(\Phi_{j})/$ $a_{j,d}^{-}(\Phi_{j})$ 
before the polarisation reversal be equal to the corresponding ratio
$a_{j,u}^{-}(\Phi_{j})/$ $a_{j,d}^{+}(\Phi_{j})$
after the reversal in each $\Phi_{j}$ bin.
In this case, $C_{a,j}$ is equal to 1 for all $\Phi_{j}$ values.
Also, with the cuts applied on the incoming beam it is
expected that $C_F = 1$ and indeed the fits to $A^m_{j}$ give $p_0$ values 
always compatible with 1.

In addition to the fact that this estimator has ``soft'' requirements
on the acceptance stability, it is independent of the relative luminosity
and combines all the data from the two target cells.
A further advantage of the use of the RPM,
is that at first order (for small values of the involved asymmetries)
all spin-independent effect, e.g. Cahn asymmetry, are factored out.

Concerning the $\Phi _j$ binning, the interval $(0,2\pi)$ was 
divided in 16 bins.
Monte Carlo studies indicated that the angular resolution (rms) is 
0.07 rad, much smaller than the bin size.

\subsubsection{Collins and Sivers asymmetries evaluation}
To extract the Collins and Sivers asymmetries,
the target polarisation values given in table~\ref{tab:polvalues}
have been used.
\begin{table}[bt]
\begin{center}
\begin{tabular}[h]{c l  r r }\hline
Year  & Period/Subperiod & \multicolumn{2}{c}{Polarisation} \\ 
 & &  upstream & downstream \\ \hline
2003 & 1 / 1   & -49.7 & 52.8 \\
 & 2 / 2 (1$^{st}$ part) &  49.4 & -42.6 \\
 & 2 / 2 (2$^{nd}$ part) &  51.3 & -44.6 \\ 
\hline
2004 & 1 / 1 &  50.70 & -43.52 \\
 & 1  / 2 (1$^{st}$ part) & -44.8 &  46.0\\
 & 1 / 2 (2$^{nd}$ part) & -38.6 &  40.4\\
 & 2 / 1 (1$^{st}$ part)  & -46.1 &  47.4\\
 & 2 / 1 (2$^{nd}$ part)  & -46.4 &  47.4\\
 & 2 / 2   &  49.9 & -42.8\\
\hline
\end{tabular}
\end{center}
\caption{Target
  polarisation values for the 2003 and 2004 data taking.}
\label{tab:polvalues}  
\end{table} 

The dilution factor $f$ has been taken constant and equal to 0.38 for all data 
taking periods.

Using the event kinematics, the mean values of the quantities
$c^{\pm}_{C,k} = f \cdot |P^{\pm}_{T,k}| \cdot D_{NN}$ and 
$c^{\pm}_{S,k} = f \cdot |P^{\pm}_{T,k}|$ have been evaluated for 
each bin and each period of
data taking.
These values have than been used to calculate the Collins and Sivers 
asymmetries from the corresponding raw asymmetries $A^m_j$.

\section{Data analysis III - Systematic studies}
%
Given the high statistics of our measurements, a number of systematic studies
have been performed in order to determine the size of possible
systematic errors.

Extensive tests both to measure false asymmetries and to investigate 
the stability of the physics results were done, for each measured asymmetry,
and in each
data taking period. They are briefly described in the following.

\subsection{False asymmetry studies}
%
Two different kinds of asymmetries expected to be zero were
measured using the final event samples, for the Collins and
Sivers angles, positive and negative hadrons, leading and all hadrons, 
and in each $x, \, z, \, p^h_T$ bin.

The first one was built by splitting the two
target cells in two parts and by combining
the data from the same cell. 
The mean values of all the resulting asymmetries were found to
be compatible with zero, as it should. The distributions of 
the pulls of these asymmetries with respect to 
zero, i.e. the values of the resulting asymmetries divided 
by the corresponding standard deviation,
 were well compatible with Standard Normal distributions,
giving no indication of the presence of systematic effects.

The second asymmetry was measured after scrambling the data
collected in each period and measuring the asymmetry in the standard way. 
The runs of the two sub-periods before and after the polarisation reversal
were mixed by labelling as ``spin up'' 
  every 1st, 3rd, 5th, ... run of the two sub-periods and as
``spin down'' all the others.
This selection was truncated in a way, that each fake period had the
  same amount of runs from each real sub-period to ensure
  roughly the same amount of events. 
As in the previous case, the asymmetries are expected to be zero,
and the measured asymmetries were compatible with zero.
The refined statistical analysis of pulls  with respect to 
zero did not give any hint for the presence of systematic effects.

\subsection{Stability of physics asymmetries vs acceptance and time}
\label{sec:pulls}
Possible acceptance effects
on the physics results from each data taking period
were tested by checking the compatibility of the asymmetries evaluated 
after splitting  the data according to: the location inside the target;
 the region of the spectrometer
in which the scattered muon was measured;
 the trigger of each event.
This work was done for both the Collins and
Sivers asymmetries  of positive and negative hadrons, leading and all,
and in each $x, \, z, \, p^h_T$ bin.

The compatibility test consisted in comparing the distribution of
the pulls
\begin{equation}
P_{i,k} = \frac{A^m_{i,k} - A^m_k }{\sqrt{\sigma_{A^m_{i,k}}^2 - \sigma_{A^m_k}^2}}
\end{equation}
with the normal standard distribution.
Here $A^m_k$ are the raw asymmetries evaluated from the whole data sample of 
one specific period
$k$.
They are evaluated for positive and negative
hadrons, Collins and Sivers asymmetries in each $x$, $p^h_T$ and $z$ bin,
for a total of $2 \times 2 \times ( 9 + 9 + 7) = 100$ values
in each period.
The corresponding variance is $\sigma_{A^m_k}^2$.
The asymmetries $A^m_{i,k}$ are the corresponding ones evaluated for all the 
sub-samples $i$ in which the original data set was divided.
The  variances of $A^m_{i,k}$ are $\sigma_{A^m_{i,k}}^2$, and
the use of the difference with $\sigma_{A^m_k}^2$ takes into
account almost completely all the correlations between $A^m_{i,k}$ and
$A^m_k$.
The compatibility with the normal standard distribution was checked also for
some specific group of asymmetries, like positive and negative hadrons,
and Collins and Sivers asymmetries.

The effect of the different acceptances for events with the primary vertex
in the two target cells was investigated, evaluating the physics 
asymmetries separately
for the events with the primary vertex in the inner half and in the outer
half of the two target cells. 
Combining events from only the outer halves (upstream part of cell \textit{u}
and downstream part of cell \textit{d}) as well as from the inner halves
only (downstream part of cell \textit{u} and upstream part of cell \textit{d})
gave results compatible to each other and to the ones using the
full sample.

The asymmetries were also evaluated dividing the data samples accordingly 
to the azimuthal angle $\Phi_{\mu^\prime }$ of the 
scattered muon in the laboratory system, namely in the regions 
$0  < \Phi_{\mu^\prime } <  \pi /2$, 
$ \pi /2 < \Phi_{\mu^\prime } <   \pi$,
$ \pi    < \Phi_{\mu^\prime } < 3\pi /2$, and
$3 \pi /2< \Phi_{\mu^\prime }  <2p$.

As in the previous case, the pull distributions turned out  to be centred 
around zero, and in very good agreement with the standard distribution.
The RMS were statistically compatible with 1, as expected in the 
absence of systematic effects.

The physics asymmetries have been evaluated splitting the data in different
samples according to the different triggers. The results turned out to
be statistically compatible, giving once more no evidence for
systematic effect.

A further test was done to investigate the stability in time of 
the physics asymmetries.
To do this, each of the two sub-periods entering the standard  asymmetry
calculation was split in these 9 groups of subsequent runs (each corresponding
to $\approx$ 10 hours  of data taking).
The asymmetries were then calculated using the full data
set of one sub-period and all the 9 groups of  runs of the other 
sub-period separately.
In total 54 asymmetries were evaluated from the 2003 and 2004 data.
This method is known to be quite powerful to single out significant 
time dependencies within single periods.
As for the previous test, the pulls relative to the mean asymmetry 
for each time slot
region were calculated, but no evidence for systematic errors 
was found in their distributions.

A further test was done to check the possible effect of $z$ and $p^h_T$ 
acceptance.
The asymmetries were
evaluated also in 4 complementary $z$ and $p^h_T$ bins:
($z<0.5$, $p^h_T<0.5$), ($z<0.5$, $p^h_T>0.5$), ($z>0.5$, $p^h_T<0.5$),
($z>0.5$, $p^h_T>0.5$) and their weighted means were compared with the 
standard results.
The differences between the two sets of measurements are
essentially invisible, much much smaller than the statistical errors, 
again as expected in absence of systematic effects.

\subsection{Stability of the acceptance in the Collins and Sivers angles}
%
A stringent test
on the $\Phi_j$ ($j=C,s$) dependence of the acceptance ratio,
already used for the previously published data,
consists in checking the $\Phi_j$ dependence of the ratios:
\begin{eqnarray} 
R_j(\Phi_j) &=&  \frac{N_{j,u}^{+}(\Phi_j) \cdot N_{j,d}^{-}(\Phi_j)}
{N_{j,u}^{-}(\Phi_j) \cdot N_{j,d}^{+}(\Phi_j)}
\end{eqnarray}
At the first order, assuming the absolute value of the target polarisation
to be the same in each cell before and after reversal, it is
\begin{eqnarray} 
R_j(\Phi_j) &\simeq&  \frac{ F^{+}_u \cdot F^{-}_d }{ F^{-}_u \cdot F^{+}_d }
\cdot   \frac{ a^{+}_{j,u}(\Phi_j) \cdot a^{-}_{j,d}(\Phi_j) }
       { a^{-}_{j,u}(\Phi_j) \cdot a^{+}_{j,d}(\Phi_j) } \, .
\end{eqnarray}
In the very likely case in which
$a_{j,u}^{+}(\Phi_{j}) / a_{j,d}^{-}(\Phi_{j}) =
a_{j,u}^{-}(\Phi_{j}) / a_{j,d}^{+}(\Phi_{j})$ it is
\begin{eqnarray} 
R_j(\Phi_j) &\simeq&  \frac{ F^{+}_u \cdot F^{-}_d }{ F^{-}_u \cdot F^{+}_d }
\cdot   \left( \frac{ a^{+}_{j,u}(\Phi_j) }
       { a^{-}_{j,u}(\Phi_j) } \right)^2 \nonumber \\
 &\simeq&  \frac{ F^{+}_u \cdot F^{-}_d }{ F^{-}_u \cdot F^{+}_d }
\cdot   \left( \frac{ a^{-}_{j,d}(\Phi_j) }
       { a^{+}_{j,d}(\Phi_j) } \right)^2 \, ,
\end{eqnarray}
thus, the constancy in $\Phi_j$ of $R_j(\Phi_j)$ implies
for each cell the ratio of the acceptances before and
after the reversal to be constant in $\Phi_j$.
It must be noted that this constancy is not required
to have unbiased estimators in the case the asymmetries are
evaluated with the RPM.
Still, this test is quite convincing on the stability of
the apparatus.

\begin{figure}[bt]
\includegraphics[width=\textwidth]{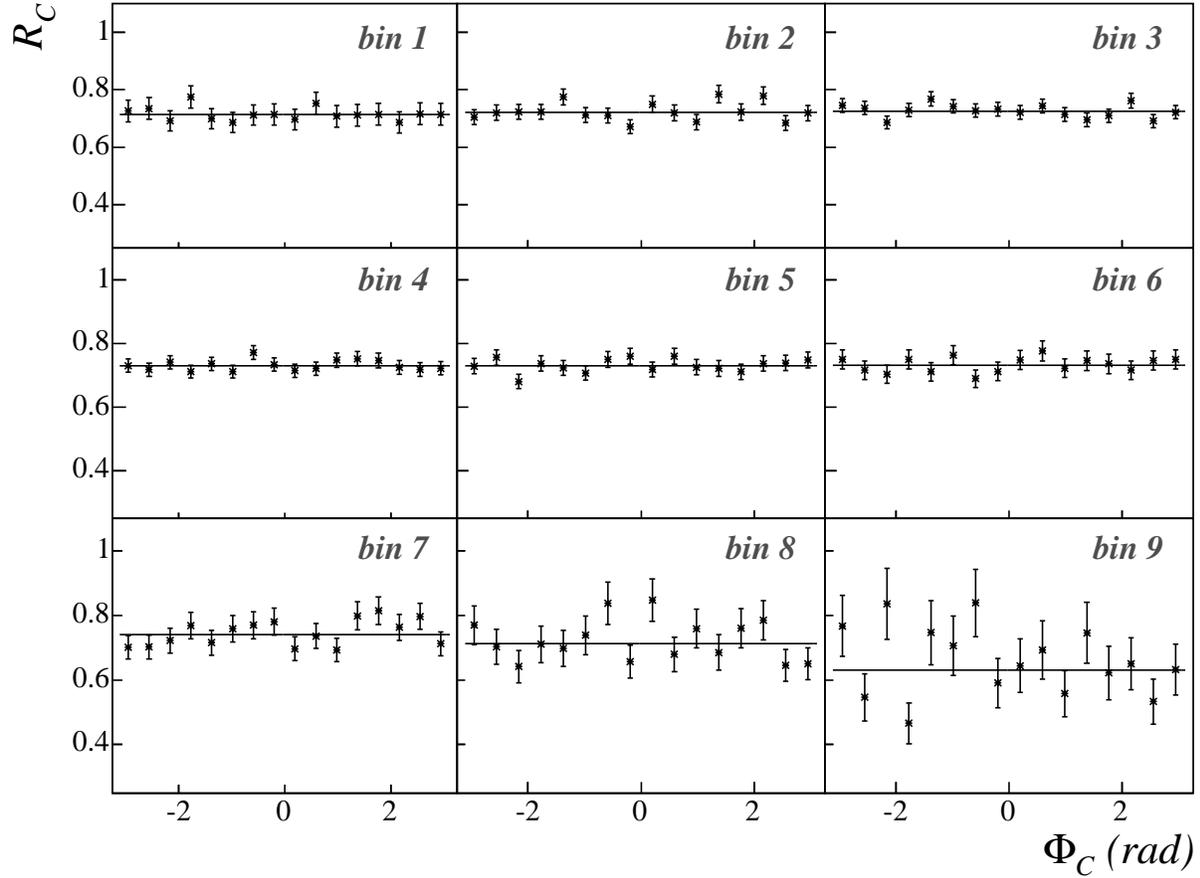}
\caption{Distribution of the $R_C$ values
for the Collins leading hadron sample vs. $x$ for the 
first period of data taking in 2004.
The line shows the result the fit with a constant value.
}
\label{fig:exa_r_01}
\end{figure}

\begin{figure*}[bt]
\begin{center}
\includegraphics[width=.6\textwidth]{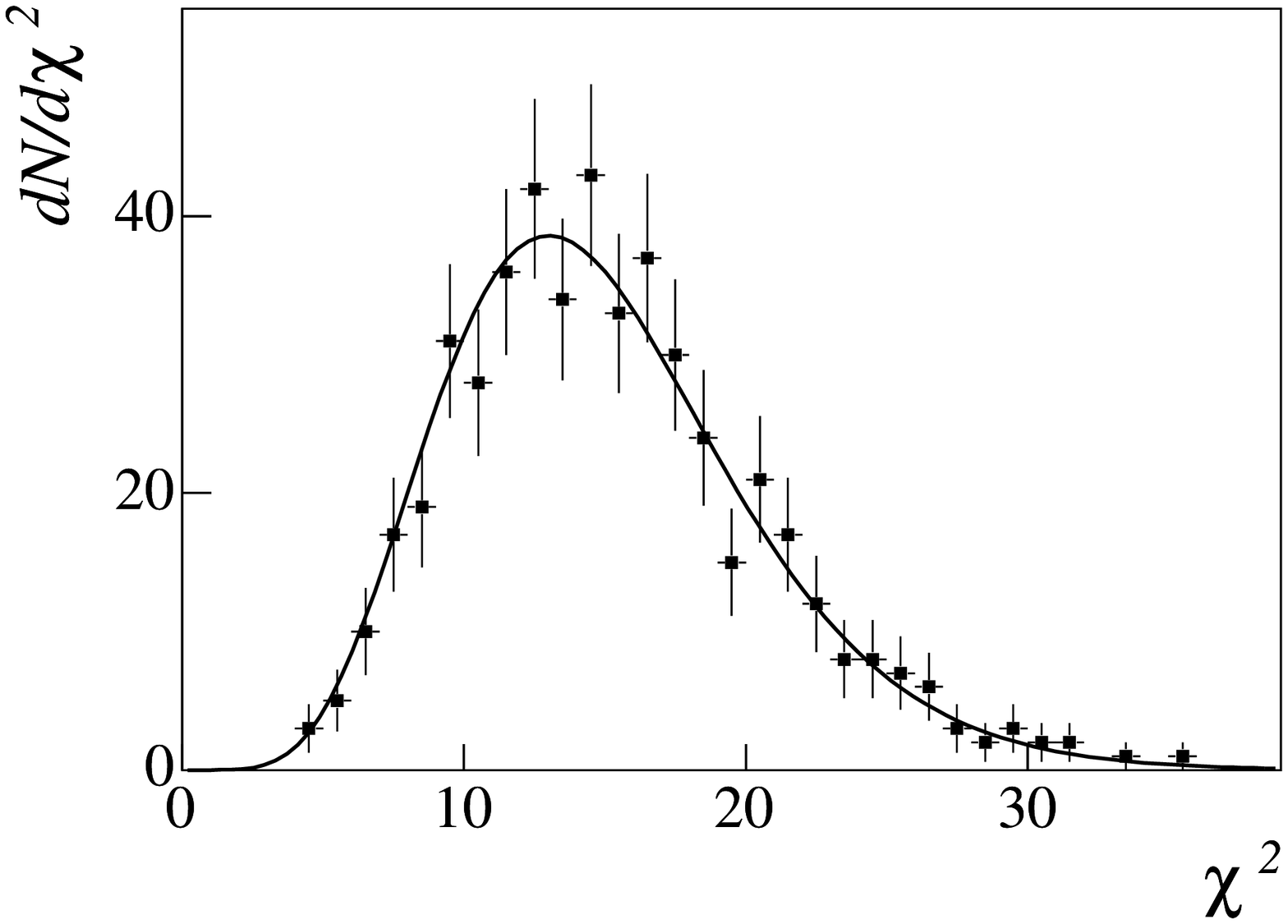}
\end{center}
\caption{The $\chi ^2$ distribution of a constant fit on $R(\Phi)$ 
for the leading hadron sample compared to the normalised
$\chi ^2$ distribution for $ndf = 15$.
}
\label{fig:res_r_01}
\end{figure*}

The ratios $R_j$ were calculated in each bin of $x, \, z, \, p^h_T$,
for the Collins and 
Sivers angles, for leading and all hadrons and for both charges.
In each bin the $\Phi_j$ distribution was fitted with a constant.
Fig. \ref{fig:exa_r_01} shows an example of the $R_C$ values
for the Collins leading hadron sample in the 9 $x$-bins
for the first period of data taking in 2004.
The lines are the results of the fit.
The quality of these fits are very good, as can be seen from 
Fig.~\ref{fig:res_r_01}
where the distribution of the $\chi ^2$ of all the fits is
 compared with the expected $\chi ^2$ distribution for $\nu = 15$ degrees of
freedom normalised to the number of entries.
To conclude, this test gave an independent indication on the stability of the
acceptances.

Summarising, all the test we performed on the effects of acceptance and
on fluctuation in time of the measured asymmetries gave results statistically
compatible with what expected in the absence of systematic effects.

\subsection{Further tests on the fit of the ratio product quantities}

\begin{figure*}[bt]
\begin{center}
\includegraphics[width=.6\textwidth]{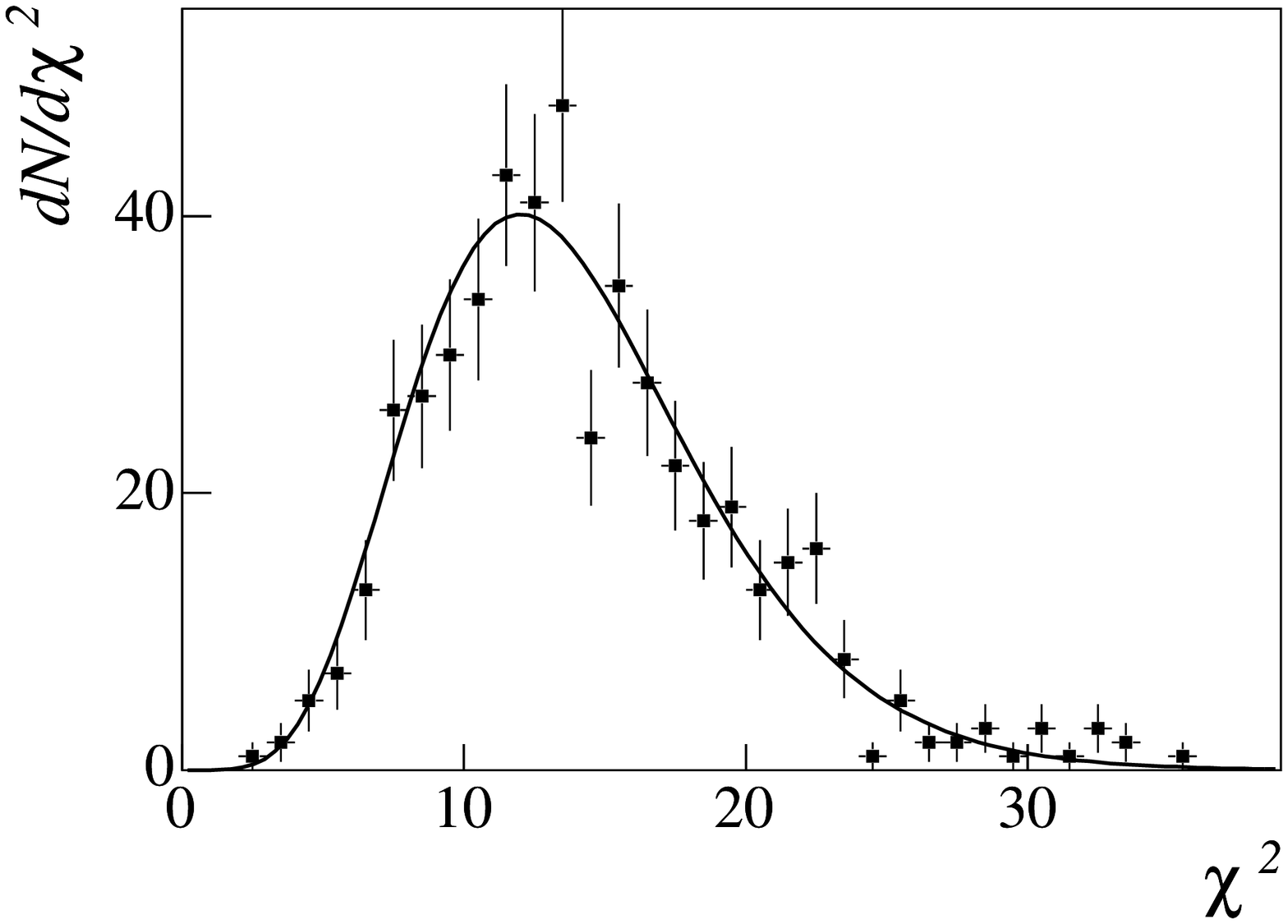}
\end{center}
\caption{$\chi ^2$ distribution of the two parameter fit 
used to extract the asymmetries $A^m_{j}$ compared to the normalised
$\chi ^2$ distribution for $ndf = 14$.
}
\label{fig:res_r_02}
\end{figure*}

The quality of the fit of the double ratio quantities $A^m_{j}$
with the function
$p_0 \cdot(1. + A^m_{j} \cdot \sin \Phi_{j})$ 
has been checked looking at the $\chi ^2$ distribution.
Fig.~\ref{fig:res_r_02} shows the $\chi ^2$ distribution 
for Collins and Sivers asymmetries,
in each bin of $x, \, z, \, p^h_T$ for leading hadrons of both charges
and for all the 5 data taking periods in 2002, 2003, and 2004. 
The curve is
the  expected $\chi ^2$ distribution for 14 degrees of freedom
normalised to the number of entries. A perfect agreement
can be observed.

The effect of using two other different fitting functions, namely
$(1 + A_{m,j} \cdot \sin \Phi_{j})$ and
$(p_0 +  A_{m,j} \cdot  \sin \Phi_{j})$,
as well as changing the $\Phi_{j}$ bin size (8 bins instead of 16)
have also been investigated.
In both cases the effects turned out to be negligible, of the order of a
few percents of the statistical error.

\subsection{Different estimators}

To estimate the size of possible systematic effects, 
the asymmetries have also been evaluated using two other estimators
which rely on different assumptions
of the acceptance variations, namely the ``difference method'' (DM, in the
following) and the ``geometric mean method'' (GMM), which are commonly
used in evaluating asymmetries.

The DM was used to extract the final values of the asymmetries published
in~\cite {Alexakhin:2005iw}.
The evaluation of the asymmetries is performed separately for the 
two target cells and, after checking their compatibility, they are  
combined by taking weighted averages.
The asymmetries $\epsilon_{C,k}$ and $\epsilon_{S,k}$,
where $k= u, d$ indicates the target cell,
were evaluated from the number of events with
the two target spin orientations
(+ spin up, and - spin down)
by fitting the quantities
\begin{eqnarray}
D_{j,k}^m(\Phi_{j}) = \frac{N_{j,k}^{+}(\Phi_{j}) 
                      - r_k \cdot N_{j,k}^{-}(\Phi_{j})}
{N_{j,k}^{+}(\Phi_{j})
                       + r_k \cdot N_{j,k}^{-}(\Phi_{j})} \, , \; \;
j=C, S
\label{eq:dasim}
\end{eqnarray}
with the functions
$\epsilon_{C,k} \cdot \sin \Phi_{C}$ and 
$\epsilon_{S,k} \cdot \sin \Phi_{S}$.
The normalisation factor $r_k$ was  taken equal to the ratio
of the total number
of detected events in the two orientations of the target polarisation:
$r = N^{+}_{h, tot, k}/N^{-}_{h, tot, k}$.
This procedure is correct if the difference of the absolute
value of the target polarisation before and after
the spin reversal is negligible. 
For what concerns the acceptances, they cancel in Eq.~(\ref{eq:dasim}) as long as the ratio
$a_{j,k}^{+}(\Phi_{j}) / a_{j,k}^{-}(\Phi_{j})$
is constant in $\Phi_{j}$.

In the GMM, again the asymmetries are evaluated separately for the 
two target cells and then combined by taking weighted averages.
The method consists in building the measured quantities
\begin{eqnarray} 
G_{j,k}^m(\Phi_{j}) & = & \frac{
\sqrt{N_{j,k}^{+}(\Phi_{j}) \cdot N_{j,k}^{-}(\Phi_{j} +\pi )} -
\sqrt{N_{j,k}^{-}(\Phi_{j}) \cdot N_{j,k}^{+}(\Phi_{j} +\pi )}
}{
\sqrt{N_{j,k}^{+}(\Phi_{j}) \cdot N_{j,k}^{-}(\Phi_{j} +\pi )} +
\sqrt{N_{j,k}^{-}(\Phi_{j}) \cdot N_{j,k}^{+}(\Phi_{j} +\pi )}  \, .
}
\end{eqnarray}
Assuming:\\
- a negligible difference in the absolute value of the target polarisation
before and after the spin reversal (as for the DM), and\\
- that the acceptances satisfy the relation
\begin{eqnarray} 
\frac{a_{j,k}^{+}(\Phi_{j}) }{ a_{j,k}^{+}(\Phi_{j} +\pi )} =
\frac{a_{j,k}^{-}(\Phi_{j}) }{ a_{j,k}^{-}(\Phi_{j} +\pi )}
\end{eqnarray}
the asymmetries are evaluated by fitting the quantities $G_{j,k}^m(\Phi_{j})$
with the functions
$\epsilon_{C,k} \cdot \sin \Phi_{C}$ and 
$\epsilon_{S,k} \cdot \sin \Phi_{S}$
in the range $0 \leq \Phi_{C,S} < \pi $.

The advantage of this method is that the luminosity cancels.
Still, the requirement on the acceptance stability is more demanding
than in the case of the ratio product method, which was thus used to 
evaluate the final asymmetries.

The fitted asymmetry values from the three methods were very close.
Given the advantages of the RPM, 
the evaluation of the asymmetries with the GMM and DM has been considered 
only a cross-check of the result, and the results of the experiment 
given in the
following are those obtained using the RPM method.

\subsection{2-D fits}
In an entirely different approach, we have estimated the Collins and 
Sivers asymmetries using the standard linear least square method (LSM)
and fitted in each kinematics bin the measured asymmetries
\begin{equation}
A_j(\phi_{h},\phi_{S}) = \frac{ N_{j,u}^{+}(\phi_{h},\phi_{S})}
                     {N_{j,u}^{-}(\phi_{h},\phi_{S})} \cdot
\frac{ N_{j,d}^{+}(\phi_{h},\phi_{S})} 
                     {N_{j,d}^{-}(\phi_{h},\phi_{S})}
\label{eq:2dfit}
\end{equation}
with the function 
\begin{eqnarray}
H(\phi_{h},\phi_{S}) & = & a_1 + a_2\sin(\phi_{h}+\phi_{S}-\pi)
                 + a_3\sin(3\phi_{h}-\phi_{S}) +
\nonumber \\
&&
                 + a_4\sin(\phi_{h}-\phi_{S})
                 + a_5\cos(\phi_{h}-\phi_{S}) \, .
\end{eqnarray}

 The first parameter should be one. The second, third and fourth terms
arise from the transverse component of the target nucleon spin when the 
lepton beam is unpolarised, while the last term originates from the 
interaction between a longitudinally polarised lepton beam and a 
transversely polarised target (see ref~\cite{eannalen}).
The second term is the Collins term. The fourth term is the Sivers 
term. The last term has also physics interest of its own, it is related
to the $g_{1T}$ transverse momentum distribution function. 

  In order to have enough statistics in each $(\phi_{h},\phi_{S})$ bin, 
we have plotted the data in 8 bins of $\phi_h$ and 8 bins of $\phi_S$. 
Each kinematic bin is thus split in 64 $(\phi_{h},\phi_{S})$ bins. 
We have performed both 5-parameter fits and 
4-parameter fits, in this second case fixing a1 to its expected value of 1. 
Also, we have done two independent fits, one minimising a $\chi^2$ with 
a linear LSM, the second using MINUIT~\cite{minuit}. 
The results of the four fits are 
in excellent agreement, and both the fitted Collins and Sivers 
asymmetries and their errors turn out to be essentially identical to the 
values given by the one dimensional fits, as expected from the 
orthogonality of the different terms. 
The Collins and Sivers asymmetries as given by the 2-D fit turn out to be 
slightly correlated (the correlation coefficient ranges from -0.25 to 0.25),
a known effect due to the  considerably non uniform 
population of the 64 $(\phi_{h},\phi_{S})$ bins.

A full discussion of the procedure 
and of the results will be the subject of a separate paper.

\subsection{Systematic errors}
As spelled out in the previous sections,
all tests performed in the different data taking periods
did not give any evidence for the presence of systematic
effects.

The conclusive test was to look at the compatibility
of the physics results obtained separately for all the
data taking periods.
As already mentioned, from 2002 to 2004 data were collected
in 5 periods and the compatibility test is significant.
It has to be stressed that the 2002 data from which the results
have already been published, have been reanalysed using the
slightly different event selection and analysis described here.
The published data turned out to be very close to the new ones, 
with differences of the order of half of the systematic
errors only in the less populated bins, essentially because of the
different method used to evaluate the asymmetries (RPM
 instead of the DM used previously).

The test was performed separately for all hadron and leading hadrons asymmetries, following the
procedure described in Section~\ref{sec:pulls}. 
The Collins and Sivers asymmetries in each $x$, $p^h_T$ and $z$ bin,
for positive and negative hadrons, obtained in the 5
periods were compared with their weighted mean.
The pulls are defined as
\begin{equation}
P_{k,j} = \frac{A^m_{k,j} - A^m_j }{\sqrt{\sigma_{A^m_{k,j}}^2 - \sigma_{A^m_j}^2}} \, , \; \; \; \; k=1,5 \,  
\end{equation}
where $A^m_j$ are the weighted means of the asymmetries,
and are expected to follow a normal standard distribution.
In total there are 500 $P_{k,j}$ values for the leading hadron
asymmetries and 520 for the all hadron asymmetries.

\begin{figure}[bt]
\includegraphics[width=0.48\textwidth]{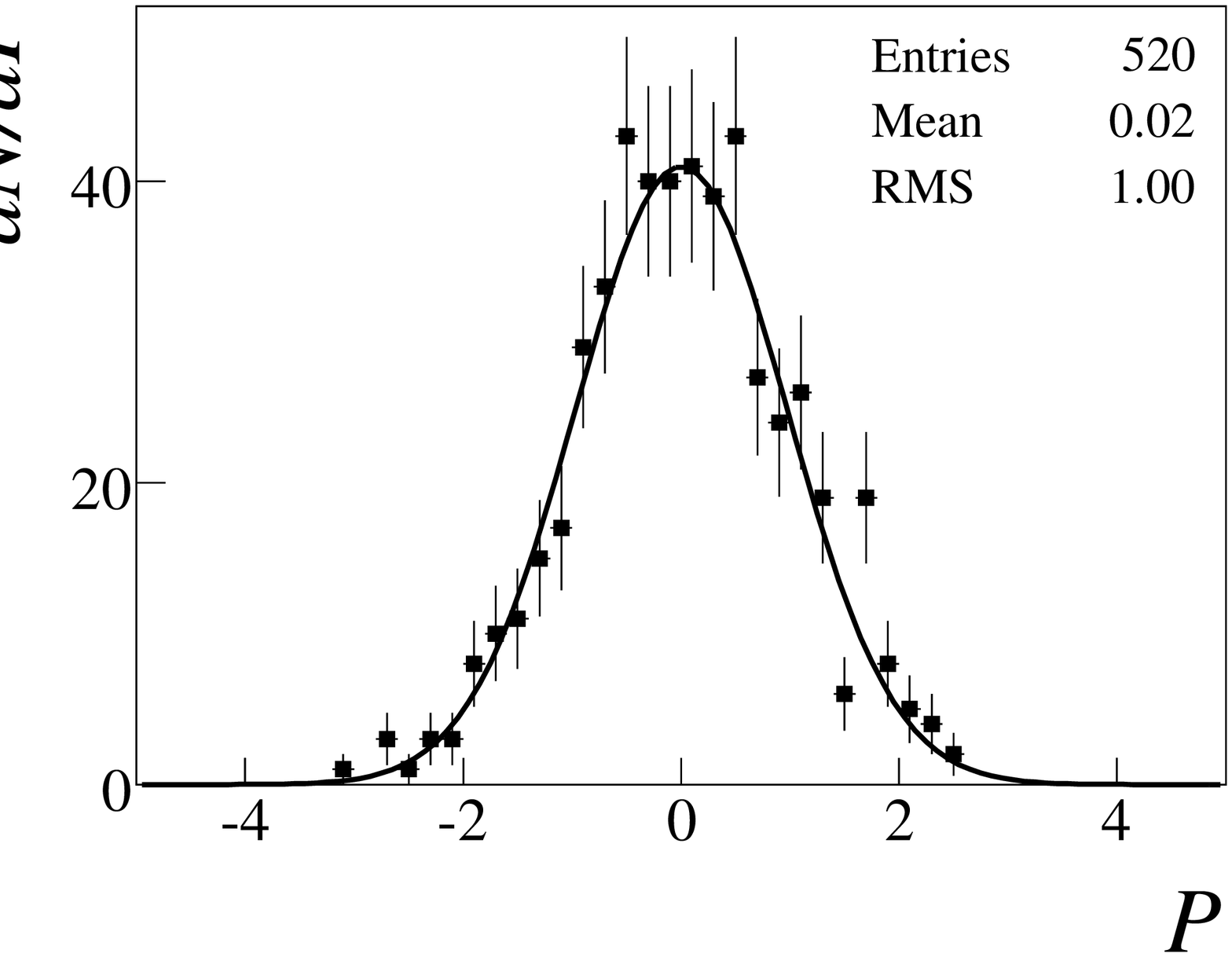} \hfill
\includegraphics[width=0.48\textwidth]{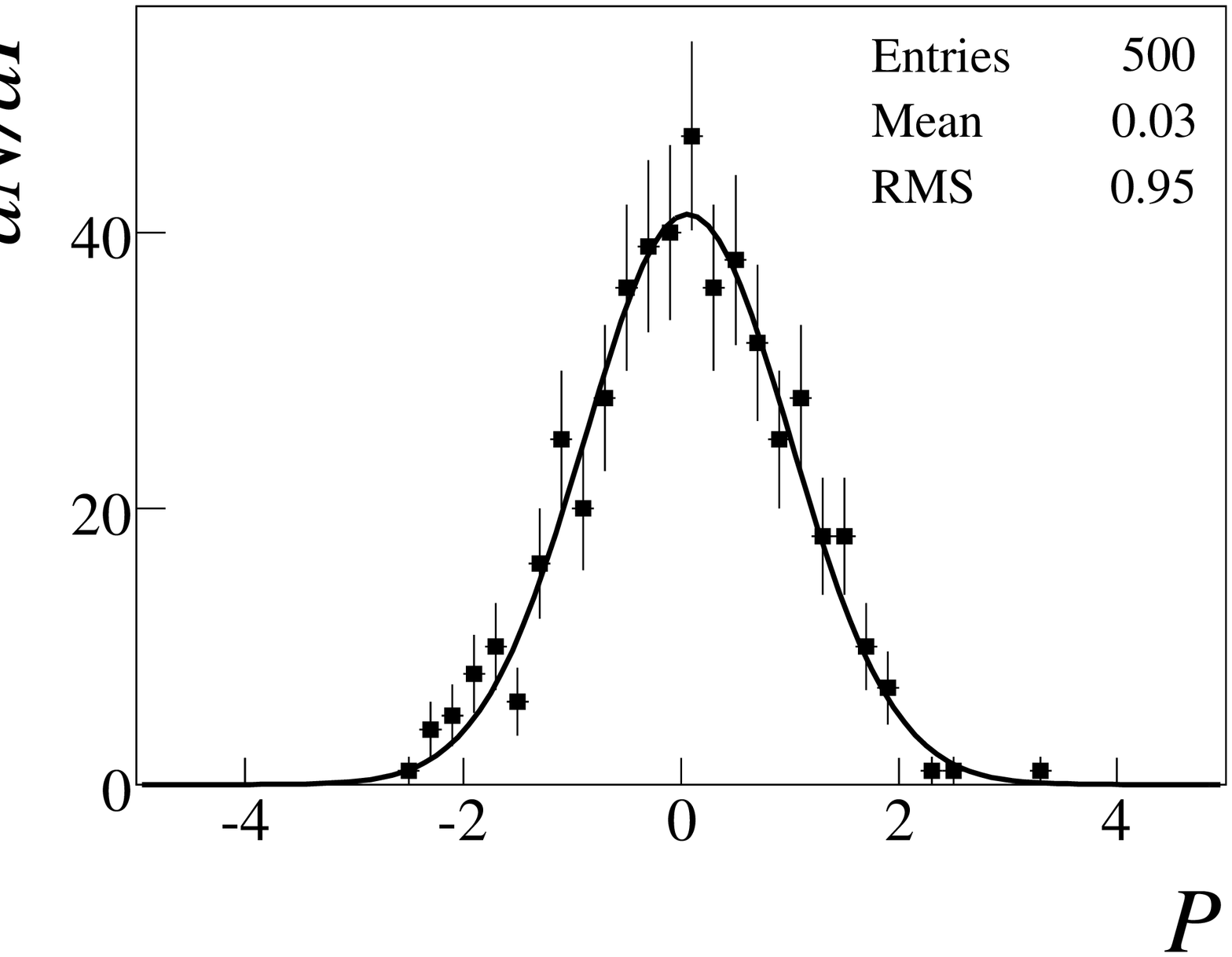}
\caption{Distributions of $P_{k,j}$ (see text) for the leading hadron
(left) and all hadron (right) asymmetries.
}
\label{fig:da_pulls}
\end{figure}

The distribution of the $P_{k,j}$ values for all hadrons and for leading hadrons
are shown in figure~\ref{fig:da_pulls}.
As expected, these pulls follow a
normal distribution.
A very good agreement with purely statistical fluctuations has been seen for 
all the various subsamples of the asymmetries; the RMS of the 
different distributions are given for completeness in table~\ref{tab:rms}. 
\begin{table}
\caption{RMS of the pull distributions for the different samples of
asymmetries}
\begin{center}
\begin{tabular}{|c|c|c|cc|}
\hline
\multicolumn{3}{|c}{Asymmetries} & \multicolumn{2}{|c|}{RMS}  \\ \hline
\multirow{4}{*}{positive hadrons} & \multirow{2}{*}{leading} & Collins & 0.882  &  $\pm 0.056$ \\   
                                  &                          & Sivers  & 0.895  &  $\pm 0.057$ \\  
  \cline{2-2}
                                  & \multirow{2}{*}{all    } & Collins & 0.919  &  $\pm 0.057$ \\   
                                  &                          & Sivers  & 0.974  &  $\pm 0.060$ \\   
\hline
\multirow{4}{*}{negative hadrons} & \multirow{2}{*}{leading} & Collins & 1.043  &  $\pm 0.066$ \\   
                                  &                          & Sivers  & 0.971  &  $\pm 0.061$ \\  
  \cline{2-2}
                                  & \multirow{2}{*}{all}     & Collins & 1.092  &  $\pm 0.068$ \\   
                                  &                          & Sivers  & 0.991  &  $\pm 0.061$ \\   
\hline\hline
                                    \multicolumn{3}{|c|}{leading}      & 0.950  &  $\pm 0.030$ \\   
                                    \multicolumn{3}{|c|}{all}          & 0.996  &  $\pm 0.031$ \\ 
\hline
                                    \multicolumn{3}{|c|}{positive}     & 0.919  &  $\pm 0.029$ \\   
                                    \multicolumn{3}{|c|}{negative}     & 1.026  &  $\pm 0.032$ \\  
\hline\hline
\end{tabular}
\end{center}
\label{tab:rms} 
\end{table}

Since even in this last test we could not observe any indication
for systematic effects, we concluded that  the systematic 
errors due to acceptance and efficiency effects
are considerably smaller than the statistical errors.

The asymmetry scale uncertainty due to the uncertainties
on $P_T$ is of 5\%.
The error on the dilution factor $f$,
which takes into account the uncertainty on the  target composition, is
of the order of 6\%. 
When combined in quadrature, these errors give a global scale uncertainty of
8\%.

\section{Results and comparison with models}

\subsection{Measured asymmetries}
\label{sec:meas}

The results from the different data taking periods have been combined
by making the standard weighted mean.

\begin{figure*}[bt] %
\begin{center}
\includegraphics[width=1\textwidth]{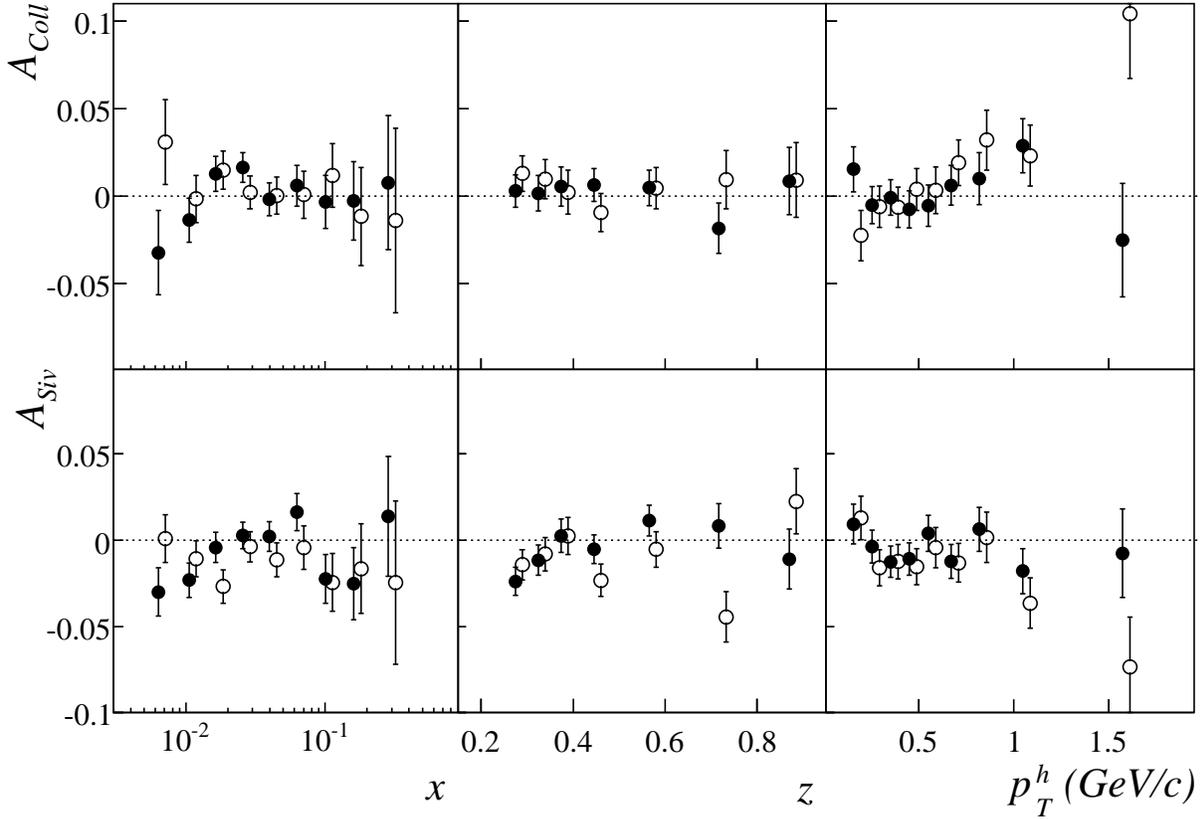}
\end{center}
\caption{Collins asymmetry (top) and Sivers asymmetry 
(bottom) against $x$, $z$ and 
$p_T^h$ for positive (full circles) and negative 
leading hadrons (open circles) from 2003--2004 data.
Error bars are statistical only.
In all the plots the open circles are slightly shifted 
horizontally with respect to the measured value.
}
\label{fig:r20034l} 
\end{figure*}

\begin{figure*}[bt] %
\begin{center}
\includegraphics[width=1\textwidth]{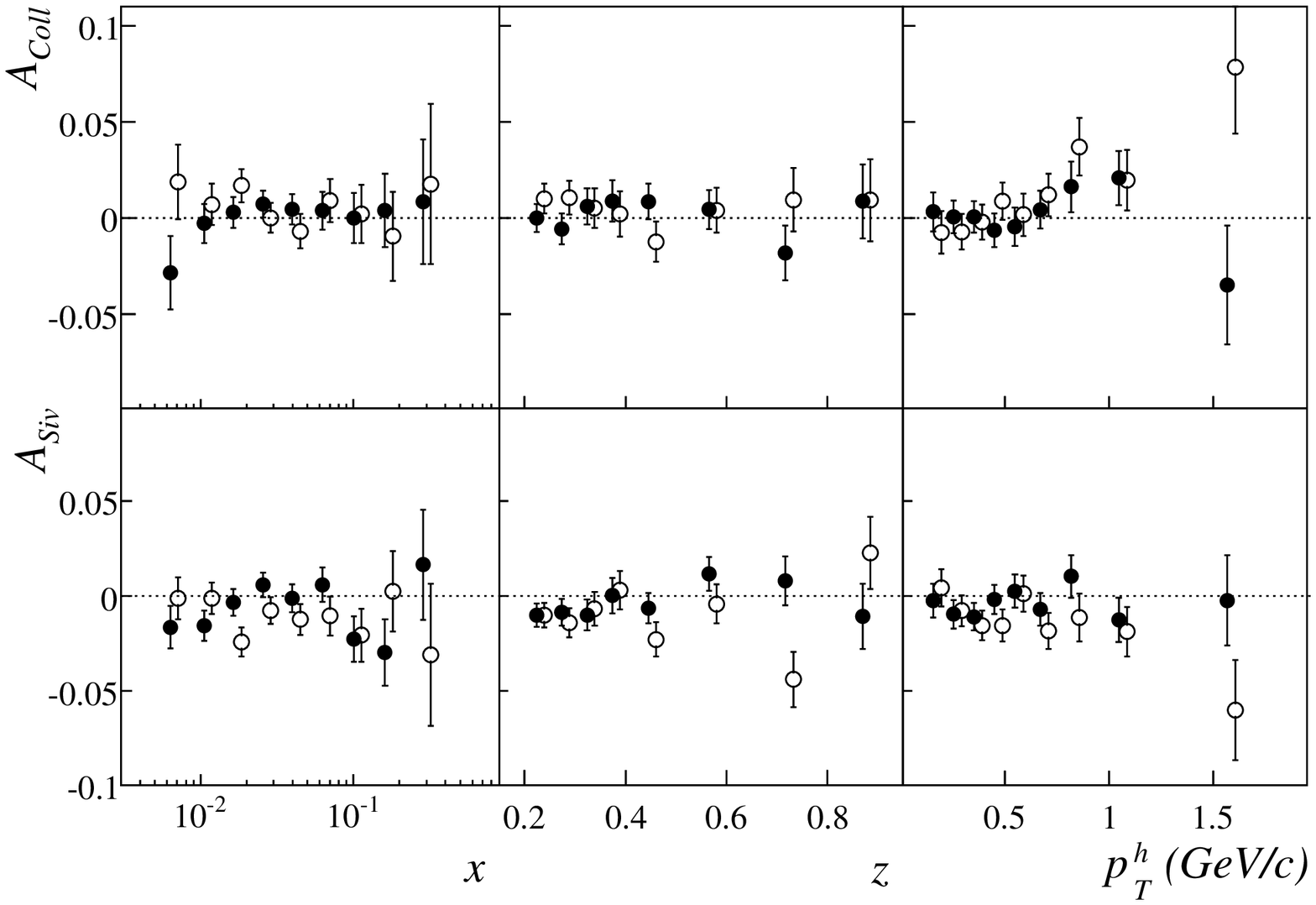}
\end{center}
\caption{Collins asymmetry (top) and Sivers asymmetry 
(bottom) against $x$, $z$ and 
$p_T^h$ for positive (full circles) and negative 
 hadrons (open circles) from 2003--2004 data.
Error bars are statistical only.
In all the plots the open circles are slightly shifted 
horizontally with respect to the measured value.
}
\label{fig:r20034a} 
\end{figure*}

Plots of the measured values of 
 the Collins and Sivers asymmetries 
$A_{Coll}$ and $A_{Siv}$ for the 2003--2004 leading hadron sample
against the three kinematic
variables $x$, $z$ and $p_T^h$ are given in Fig.~\ref{fig:r20034l}.
Full points and open points refer to   positive and negative hadrons 
respectively.
The errors shown in the figure are only statistical. 
The same asymmetries are shown in Fig.~\ref{fig:r20034a} 
for the 2003--2004 all hadron sample.
Again, the errors shown in the figure are only statistical. 

\begin{figure*}[bt] %
\begin{center}
\includegraphics[width=.8\textwidth]{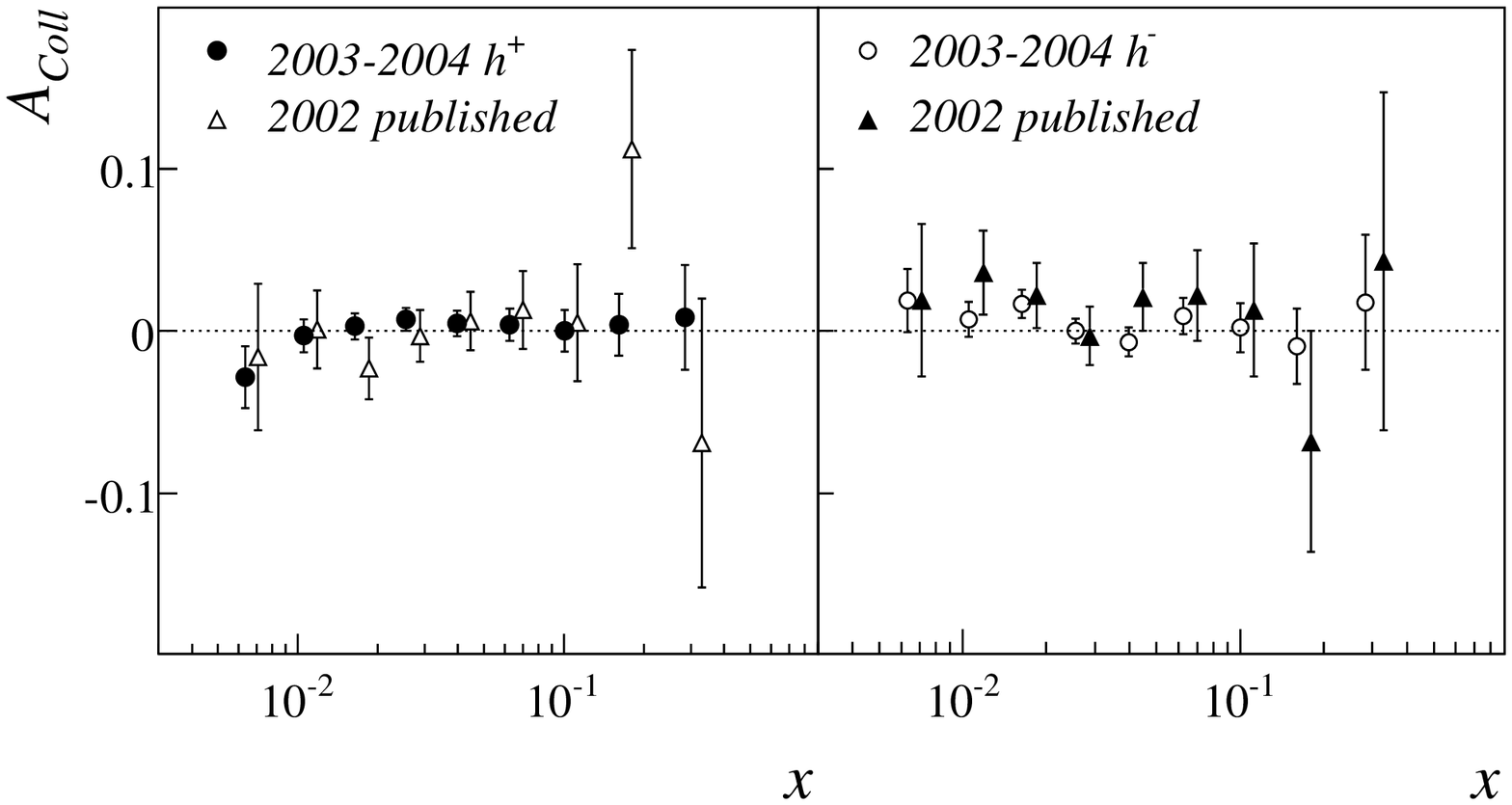}
\end{center}
\caption{Collins asymmetry against $x$ for all positive (left) and all 
negative (right) hadrons from 2002 data~\cite{Alexakhin:2005iw} 
(triangles) and  from 2003--2004 data (circles).
The 2002 values are slightly shifted 
horizontally with respect to the measured value.
}
\label{fig:cfr0234} 
\end{figure*}

The improvement in statistics with respect to the 2002 published
result is clearly visible in Fig.~\ref{fig:cfr0234}, where, as an example,
the published Collins asymmetry data for all positive (left) and all negative
(right) hadrons are compared with the 2003--2004 results.

\begin{figure*}[bt] %
\begin{center}
\includegraphics[width=.8\textwidth]{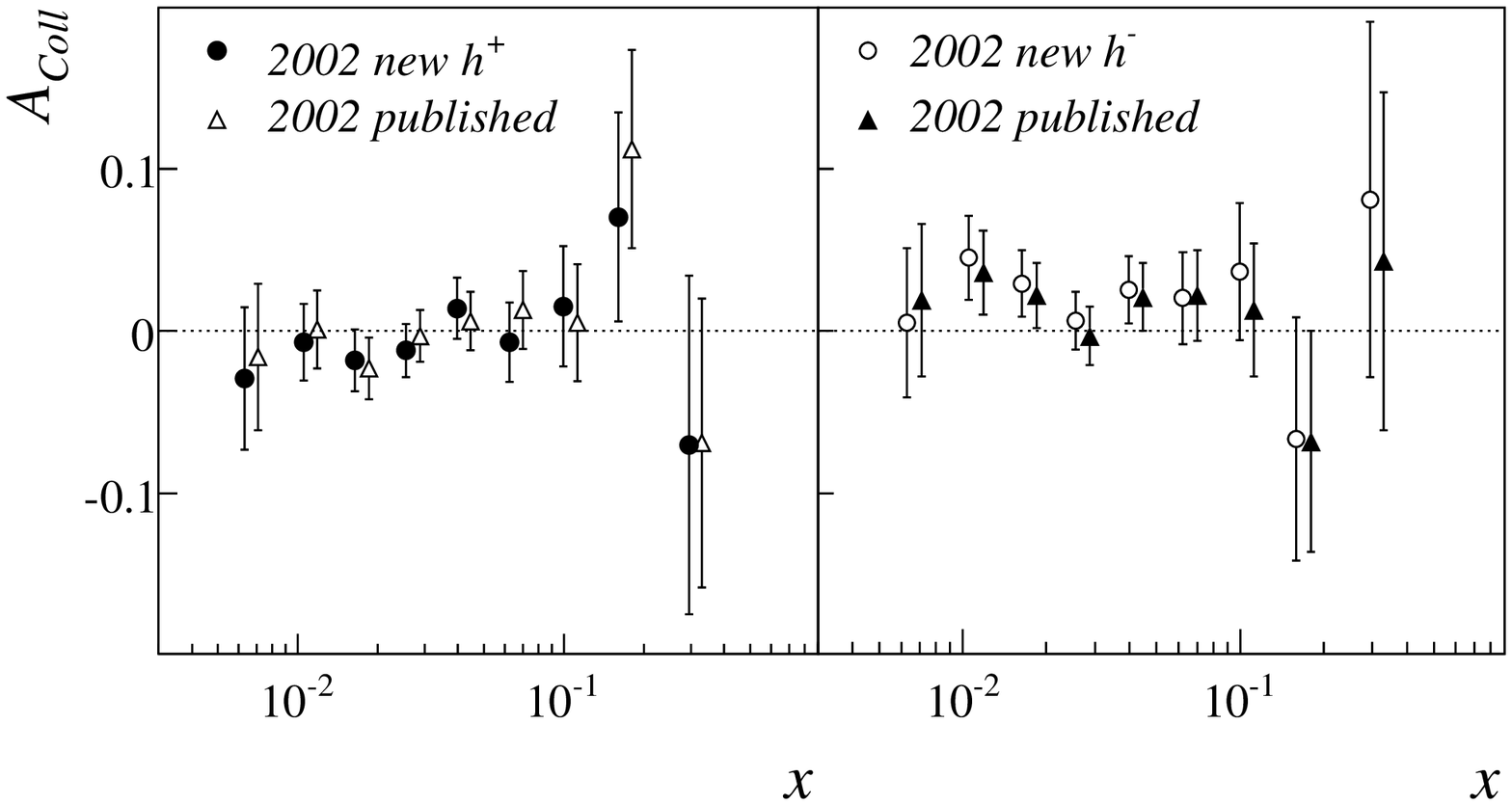}
\end{center}
\caption{Collins asymmetry against $x$ for positive (left) and negative (right)
 hadrons from 2002 data.
The triangles are the published results~\cite{Alexakhin:2005iw} and the circles
the results of the new analysis.
The published values are slightly shifted 
horizontally with respect to the measured value.
}
\label{fig:cfr0202} 
\end{figure*}

As already mentioned, the 2002 data have been reanalysed using the
event selection and the analysis described in this paper.
The effect on the measured asymmetries is very small, and
is mainly due to use of the RPM in evaluating the raw asymmetries.
As an example, Fig.~\ref{fig:cfr0202} shows the comparison between the 
published and the new results for the Collins asymmetry.
The new values from the 2002 data have been combined with the
results from the 2003--2004 data to evaluate the final asymmetries.

\begin{figure*}[bt] %
\begin{center}
\includegraphics[width=1\textwidth]{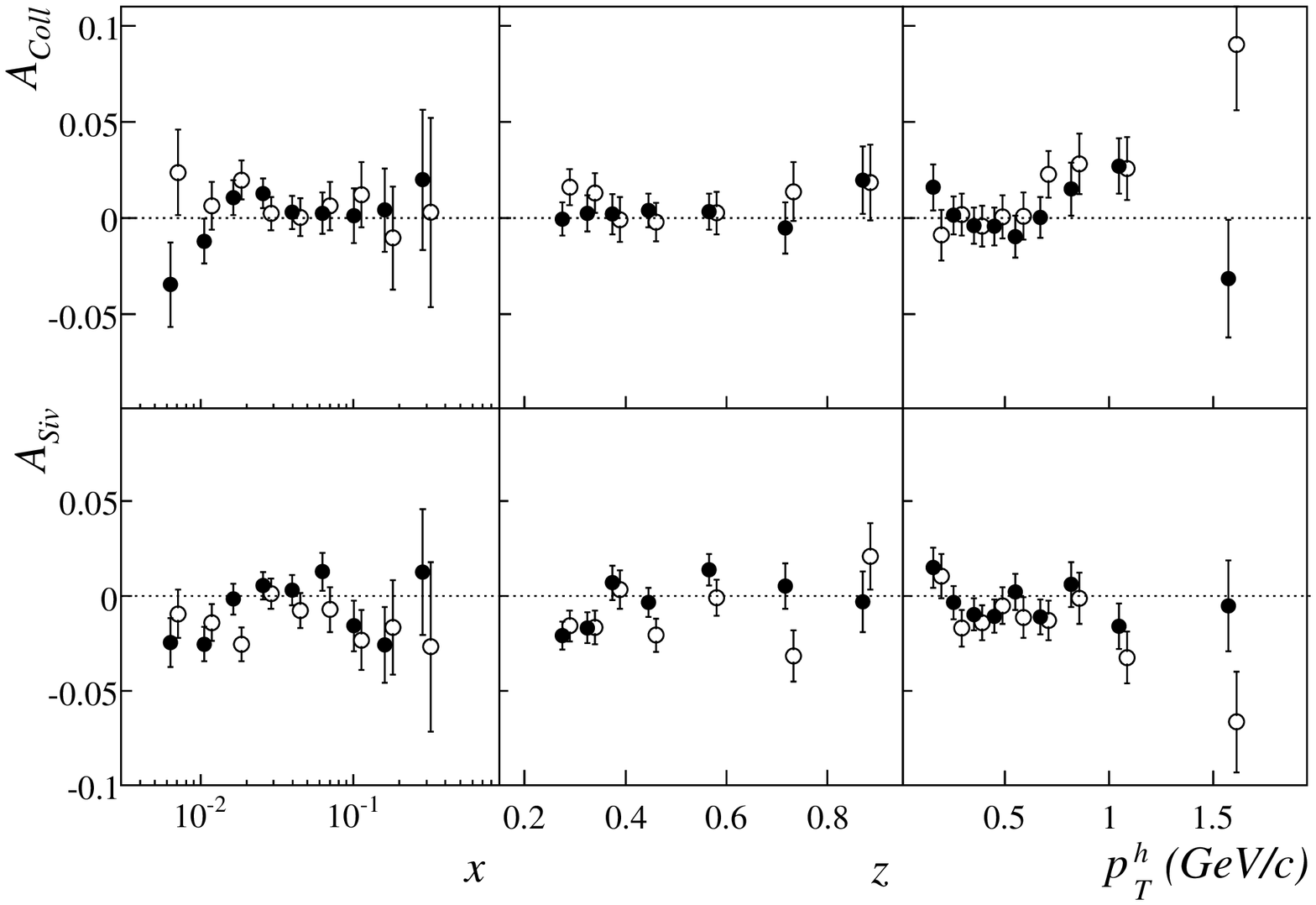}
\end{center}
\caption{Overall results for Collins asymmetry (top) and Sivers asymmetry 
(bottom) against $x$, $z$ and 
$p_T^h$ for positive (full circles) and negative 
leading hadrons (open circles) from 2002, 2003, and 2004 data.
Error bars are statistical only.
In all the plots the open circles are slightly shifted 
horizontally with respect to the measured value.
}
\label{fig:r0234l}
\end{figure*}

\begin{figure*}[h!] %
\begin{center}
\includegraphics[width=1\textwidth]{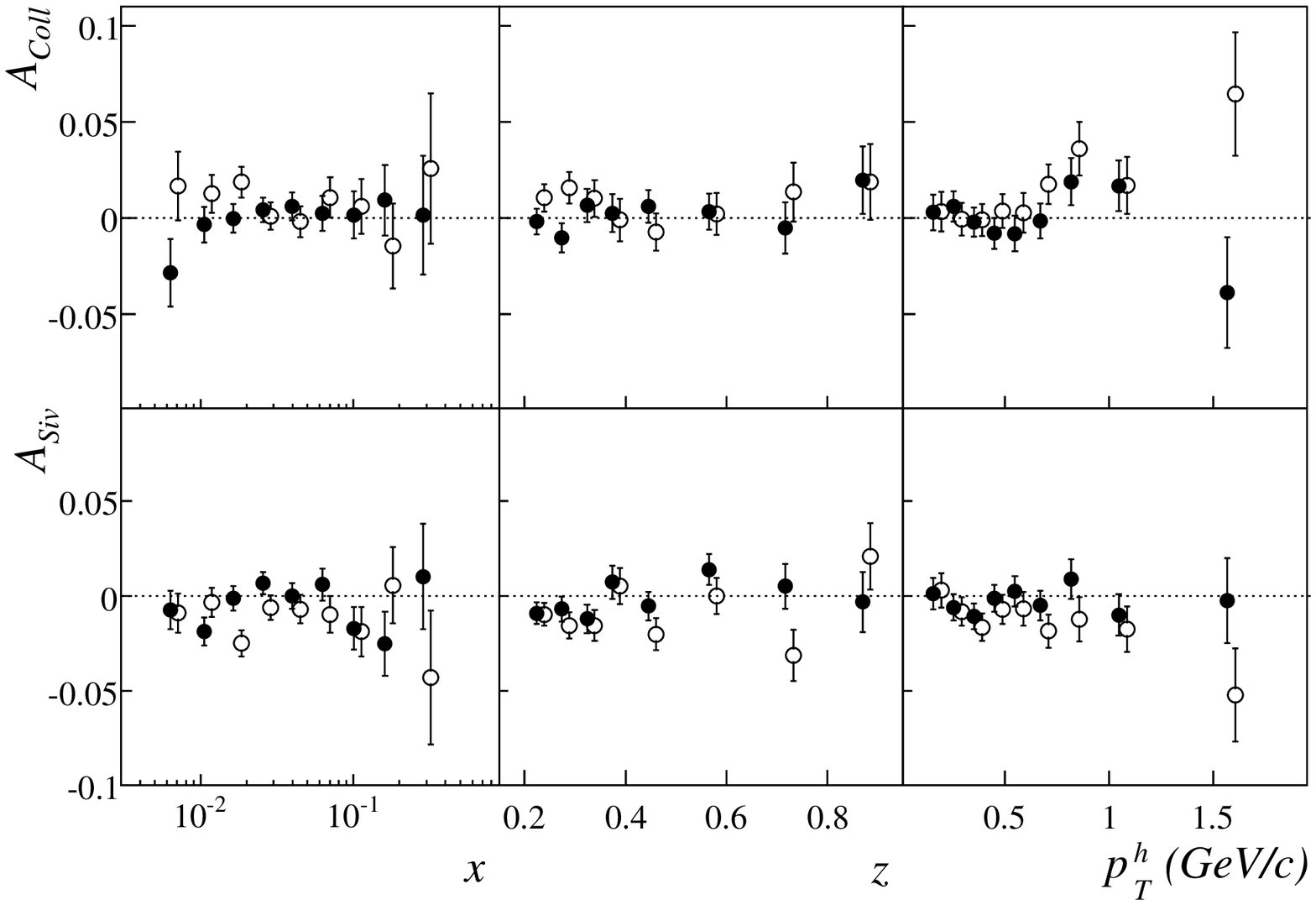}
\end{center}
\caption{Overall results for Collins asymmetry (top) and Sivers asymmetry 
(bottom) against $x$, $z$ and 
$p_T^h$ for all positive (full circles) and all negative 
hadrons (open circles) from 2002, 2003, and 2004 data.
Error bars are statistical only.
In all the plots the open circles are slightly shifted 
horizontally with respect to the measured value.
}
\label{fig:r0234a} 
\end{figure*}

The overall results from 2002--2004 deuteron target 
for the leading hadron sample and for the all hadron sample are given in
Fig.~\ref{fig:r0234l} and~\ref{fig:r0234a} respectively.
All these measured asymmetries
are available on HEPDATA~\cite{hepdata}.
%

\subsection{Comments and comparison with models}

As apparent from Fig.~\ref{fig:r0234l} and \ref{fig:r0234a}, 
all the measured asymmetries are small, 
if any, and compatible with zero. 
This trend already characterised the published data of the 2002 run, 
and is confirmed by the new data with considerably improved precision. 
Small asymmetries is not a surprise.
From the very beginning it was predicted that transverse spin effects be small
 in the deuteron due to the opposite sign which was expected for the $u$ 
and $d$ distributions, causing cancellations on the asymmetries 
of an isoscalar target, very much like in the helicity case.
Still, it was not obvious that they would have been so small.

An analysis of the results on the deuteron can be done only in conjunction 
with corresponding proton data, which up to now have been measured only 
by the HERMES Collaboration~\cite{Hermest}. 
The measured non zero Collins asymmetry on the proton
 has provided convincing evidence that both the transversity distribution 
$\Delta _T u(x)$ and
 the Collins mechanism $\Delta_T ^0 D_u^h (z)$ are not zero. 
Independent evidence that the Collins mechanism is a real 
measurable effect has come from the recent analysis of the
BELLE Collaboration~\cite{Abe:2005zx,Ogawa:2006bm}. 
Furthermore, the HERMES data on a proton target have provided
convincing evidence that the Sivers  mechanism is also at work.

It is fair to say that the accuracy of the present HERMES data has allowed
 to extract only the leading contribution to the proton transverse asymmetry, 
that is the u quark contribution. 
Also, the interpretation of the HERMES data has led to surprising 
assumptions about the relative size of the favoured and unfavoured 
spin dependent fragmentation functions.
 A global analysis using all  the available deuteron and  proton data 
should allow now to provide first estimates of 
both the u and the d quark contribution, and clearly constitutes 
the next step in this work.
 
In the following, first naive expectations, based on the simple
formulas of Sections~\ref{sec:collins} and \ref{sec:sivers},
are given for the deuteron asymmetries, 
then the new deuteron data are compared to a few existing model calculations.

\subsubsection{Collins asymmetry }
Although the measured deuteron asymmetries refer to unidentified hadrons, 
in the following it will be assumed that the hadrons be pions (actually more
 than 80\% are pions), so that the algebra considerably simplifies.
Further simplification can be obtained by neglecting 
the sea contribution (i.e. $\Delta_{T}\bar{q} = \Delta_{T}s = 0$
and $\bar{q} = s = 0$) and considering only the range $0.1 < x < 0.3$.
This is justified by the fact that the 
PDFs are expected to be considerably different from zero in the valence
region, and the HERMES data show non-zero values in 
the range $0.05 < x < 0.3$.
Assuming
\begin{eqnarray}
D_u^{\pi^+} = D_d^{\pi^-} = D_1 \, , \;
D_d^{\pi^+} = D_u^{\pi^-} = D_2 \, , \\
\Delta_{T}^0 D_u^{\pi^+} = \Delta_{T}^0 D_d^{\pi^-} = \Delta_{T}^0 D_1 \, , \;
\Delta_{T}^0 D_d^{\pi^+} = \Delta_{T}^0 D_u^{\pi^-} = \Delta_{T}^0 D_2 \, ,
\end{eqnarray}
and using Eq.~(\ref{eq:collass}), one gets for $\pi^+$ on a proton target:
\begin{equation}
A_{Coll}^{p, \pi^+}  \simeq 
\frac{4 \Delta_{T} u_v \Delta_{T}^0 D_1 + \Delta_{T} d_v \Delta_{T}^0 D_2}
     {4  u_v D_1 + d_v D_2}
\label{eq:colpip1}
\end{equation}
and for $\pi^-$:
\begin{equation}
A_{Coll}^{p, \pi^-}  \simeq 
\frac{4 \Delta_{T} u_v \Delta_{T}^0 D_2 + \Delta_{T} d_v \Delta_{T}^0 D_1}
     {4  u_v D_2 + d_v D_1} \, .
\label{eq:colpim1}
\end{equation}
If, as suggested from the HERMES data, $\Delta_{T}^0 D_1 = - \Delta_{T}^0 D_2$,
and taking $D_2 \simeq 0.5 D_1, \, d_v \simeq 0.5 u_v$, the previous expressions become
\begin{equation}
A_{Coll}^{p, \pi^+}  \simeq 
\frac{\Delta_{T} u_v }{u_v} \frac{\Delta_{T}^0 D_1}{D_1}
\label{eq:colpip2}
\end{equation}
and
\begin{equation}
A_{Coll}^{p, \pi^-}  \simeq 
- \frac{4}{2.5} \frac{ \Delta_{T} u_v }{u_v} \frac{\Delta_{T}^0 D_1}{D_1}
\label{eq:colpim2}
\end{equation}
respectively. As already stressed, the u-quark contribution is dominant.

For a deuteron target equations (\ref{eq:colpip1}), (\ref{eq:colpim1}), 
 (\ref{eq:colpip2}), and (\ref{eq:colpim2}) become respectively
\begin{equation}
A_{Coll}^{d, \pi^+}  \simeq  
\frac{\Delta_{T} u_v  + \Delta_{T} d_v}{u_v + d_v }
\frac{4 \Delta_{T}^0 D_1 + \Delta_{T}^0 D_2}
     {4 D_1 + D_2}
\label{eq:colpid1a}
\end{equation}
\begin{equation}
A_{Coll}^{d, \pi^-}  \simeq 
\frac{\Delta_{T} u_v  + \Delta_{T} d_v}{u_v + d_v }
\frac{\Delta_{T}^0 D_1 + 4 \Delta_{T}^0 D_2}{D_1 + 4 D_2}
\label{eq:colpid1b}
\end{equation}
and
\begin{equation}
A_{Coll}^{d, \pi^+}  \simeq  
\frac{3}{7} \frac{\Delta_{T} u_v  + \Delta_{T} d_v}{u_v} 
             \frac{\Delta_{T}^0 D_1}{D_1}
\label{eq:colpid2a}
\end{equation}
\begin{equation}
A_{Coll}^{d, \pi^-}  \simeq 
- \frac{3}{4.5} \frac{\Delta_{T} u_v  + \Delta_{T} d_v}{u_v} 
          \frac{\Delta_{T}^0 D_1}{D_1}
\label{eq:colpid2b}
\end{equation}

Both the $\pi^+$ and the  $\pi^-$ Collins asymmetries on the deuteron are 
proportional to $\Delta _T u_v(x) + \Delta _T d_v(x)$, therefore cancellation 
is expected to reduce considerably the effect which has been measured 
on the proton.
As a matter of fact, assuming $\Delta_{T} d_v = 0$ (no limit on the size of 
$\Delta_{T} d_v$ is provided by the HERMES data) one derives
relations between the Collins asymmetry measured by HERMES and COMPASS
which are only marginally satisfied.
Thus, the present precise data on  $A_{Coll}^{d, \pi}$ 
should allow to extract $\Delta _T d_v$.

\begin{figure*}[tb] %
\begin{center}
\includegraphics[width=.8\textwidth]{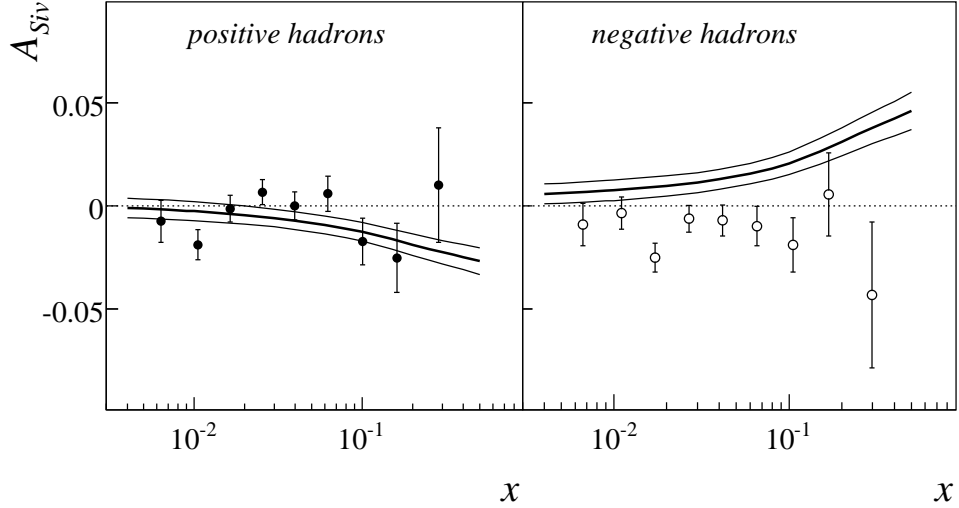}
\end{center}
\caption{Comparison between the present results for positive 
(left) and negative (right)
hadrons with the calculations of Ref.~\cite{vy} (scenario 1).
}
\label{fig:collvog} 
\end{figure*}

This was not the case in so far. Three global analyses have been 
performed with the published data,
 trying to derive bounds on the transversity distributions
and the Collins fragmentation functions.
In Ref.~\cite{vy} the Soffer bound $|\Delta _T q| = (q+\Delta  q)/2 $
was used, 
a fit of the HERMES data set was performed, and the Collins 
 functions were extracted.
  Two different scenarios for favoured and unfavoured Collins 
fragmentation functions were considered, 
but the fits always favoured a relation 
$ \Delta _T ^0 D_1 \sim - \Delta _T ^0 D_2 $.
The comparison of the fit results with the COMPASS data shows 
a fair agreement, as apparent
from Fig.~\ref{fig:collvog}, although the data do not exhibit
the trend with $x$ which is suggested by the model. 
The upper and lower curves in the figures correspond to the 1-$\sigma$ errors
of the fitted parameters.

\begin{figure*}[tb] %
\begin{center}
\includegraphics[width=.8\textwidth]{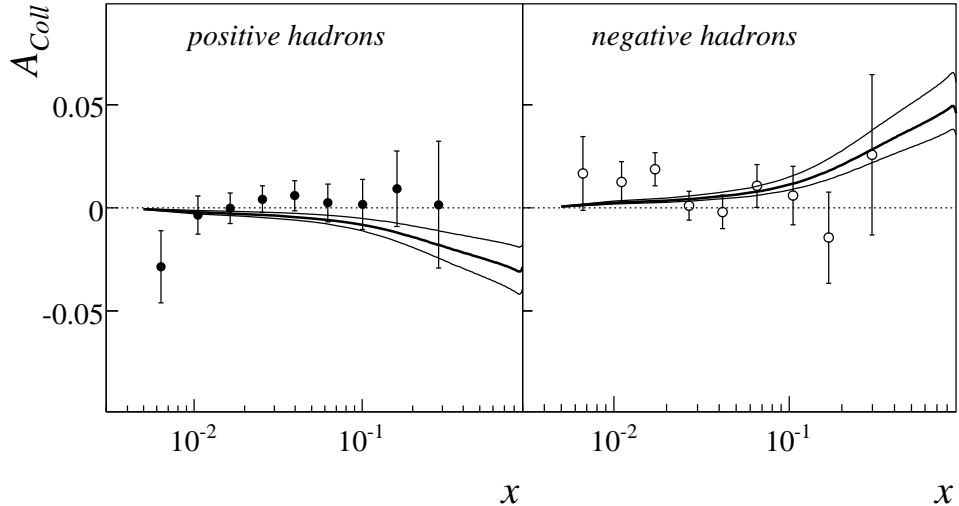}
\end{center}
\caption{Comparison between the present results for positive 
(left) and negative (right)
hadrons with the calculations of Ref.~\cite{efre}.
}
\label{fig:collefr} 
\end{figure*}

In Ref.~\cite{efre} a chiral quark-soliton model
 was used for the transversity distributions, and the 
Collins fragmentation function 
was derived from a fit to the HERMES data, which do not 
constrain the $\Delta _T d$ distribution.
 A comparison with the present COMPASS results shows again 
a fair agreement (Fig.~\ref{fig:collefr}).
The upper and lower curves in the figures correspond to the uncertainty
in the Collins fragmentation functions as obtained from the fit.
Independent extraction of the Collins function was performed by 
fitting the BELLE data. 
The result was found to be compatible with the one obtained 
fitting the HERMES data.

\begin{figure*}[tb] %
\begin{center}
\includegraphics[width=.8\textwidth]{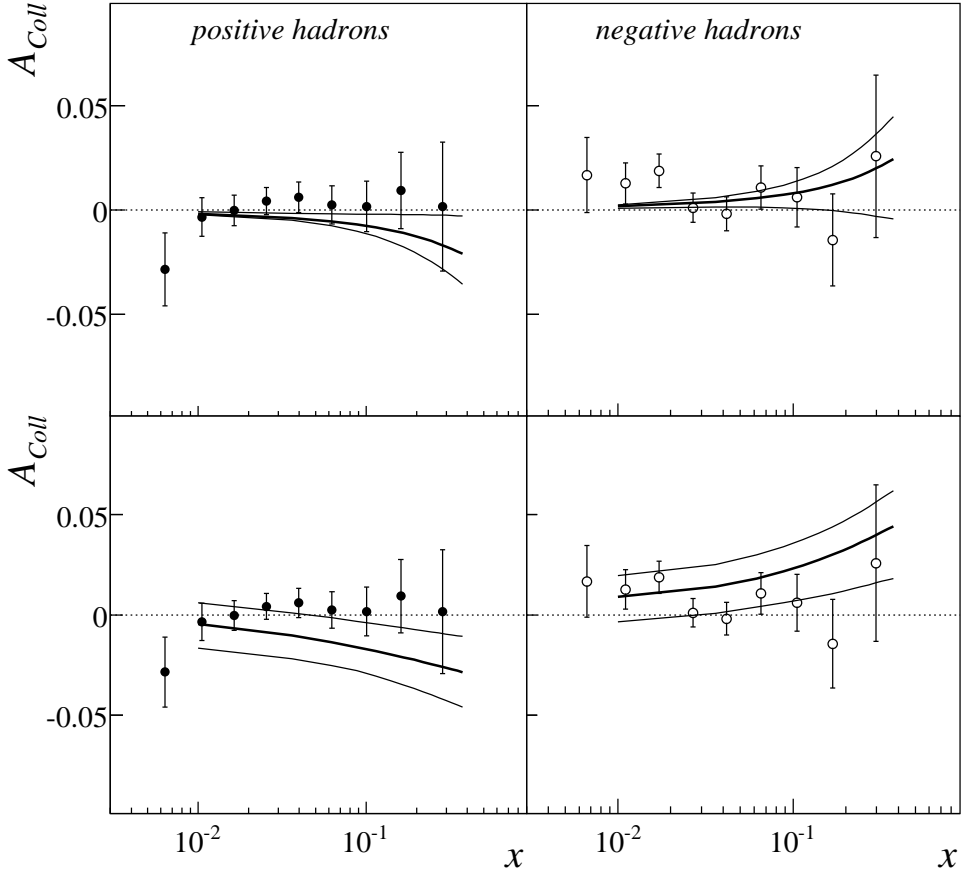}
\end{center}
\caption{Comparison between the present results for positive (left) 
and negative (right) hadrons with the calculations of Ref.~\cite{anse},
where two scenarios were tested for transversity:
$\Delta_T q=\Delta q$ was assumed in one scenario (upper figures),
while $\Delta _T q = (q+\Delta  q)/2 $ was assumed for the calculations
of the lower figures.
}
\label{fig:collans} 
\end{figure*}

Similar results were obtained in Ref.~\cite{anse}. 
Two different scenarios were used for transversity, either 
$\Delta _T q = \Delta  q $, or 
the Soffer bound, and the Collins fragmentation functions
 were extracted from a fit to the HERMES data. 
The fits were very good in both cases. 
The extracted Collins functions were then used to reproduce the 
published 2002 COMPASS data. 
The agreement is acceptable for both scenarios, as apparent from 
Fig.~\ref{fig:collans}, although also in this case the expected increase 
of $A_{Coll}$ with increasing $x$ is not manifested by the COMPASS
data.
The upper and lower curves in all the figures correspond to 1-$\sigma$ 
deviations of the parameters.
Also in this case the BELLE data are reproduced, and one can conclude that 
the Collins mechanism 
in SIDIS and in $e^+ e^- \rightarrow$ hadrons are the same.

To summarise, the new data are compatible with the uncertainty bands given 
by the 
phenomenological calculations, but the trend of the data with $x$ is not the 
one suggested
by the central value of the calculations: this is a clear indication
that the COMPASS data will surely help in constraining the parameters of the 
models.

\subsubsection{Sivers asymmetry}
Also in this case it is useful to consider the expressions one obtains 
for $A_{Siv}$ in the hypothesis that all hadrons are pions.
Again, the simplified analysis is restricted to the valence region.

Neglecting the sea contribution (i.e. $\Delta^{T}_0\bar{q} = \Delta_0^{T}s = 0$
and $\bar{q} = s = 0$ at all $x$) and assuming 
$D_u^{\pi^+} = D_d^{\pi^-} = D_1$ and
$D_d^{\pi^+} = D_u^{\pi^-} = D_2$,
on a proton target, from Eq.~(\ref{eq:sivass}) one gets for $\pi^+$:
\begin{equation}
A_{Siv}^{p, \pi^+}  \simeq 
\frac{4 \Delta_0^{T} u_v D_1 + \Delta_0^{T} d_v D_2}
     {4  u_v D_1 + d_v D_2}
\label{eq:sivpip1}
\end{equation}
and for $\pi^-$:
\begin{equation}
A_{Siv}^{p, \pi^-}  \simeq 
\frac{4 \Delta_0^{T} u_v D_2 + \Delta_0^{T} d_v D_1}
     {4  u_v D_2 + d_v D_1} \, .
\label{eq:sivpim1}
\end{equation}
Assuming $D_2 \simeq 0.5 D_1, \, d_v \simeq 0.5 u_v$, 
the previous expressions become
\begin{equation}
A_{Siv}^{p, \pi^+}  \simeq 
\frac{\Delta_0^{T} u_v }{u_v} 
\label{eq:sivpip2}
\end{equation}
and
\begin{equation}
A_{Siv}^{p, \pi^-}  \simeq 
\frac{2 \Delta_0^{T} u_v + \Delta_0^{T} d_v}{2.5 u_v}
\label{eq:sivpim2}
\end{equation}
respectively. Since the  Sivers asymmetries for $\pi^-$ 
as measured by HERMES is about zero, in this very simplified 
treatment it follows that
\begin{equation}
\Delta_0^{T} d_v = - 2 \Delta_0^{T} u_v \, .
\label{eq:sivfh}
\end{equation}

For a deuteron target the Sivers asymmetries can be written as
\begin{equation}
A_{Siv}^{d, \pi^+}  \simeq   
\frac{\Delta_0^{T} u_v  + \Delta_0^{T} d_v}{u_v + d_v}  
\label{eq:sivpid1a}
\end{equation}
and
\begin{equation}
A_{Siv}^{d, \pi^-}  \simeq 
\frac{\Delta_0^{T} u_v  + \Delta_0^{T} d_v}{u_v + d_v}
\label{eq:sivpid1b}
\end{equation}
which implies $A_{Siv}^{d, \pi^+}  \simeq   A_{Siv}^{d, \pi^-}$.
The approximatively zero Sivers asymmetries for positive and negative
hadrons observed in COMPASS require
\begin{equation}
\Delta_0^{T} d_v \simeq -  \Delta_0^{T} u_v \, ,
\end{equation}
a relation which is also obtained in some models.

From Eq.(\ref{eq:sivpip2}), (\ref{eq:sivpim2}), and  (\ref{eq:sivfh}),
a relation between the Sivers asymmetry measured by COMPASS and those measured 
by HERMES can be derived
\begin{equation}
A_{Siv}^{d, \pi^+}  \simeq   A_{Siv}^{d, \pi^-}  \simeq 
-\frac{A_{Siv}^{p, \pi^+}}{1.5} 
\label{eq:sivpid1}
\end{equation}
which is only marginally satisfied by the data.

\begin{figure*}[tb] %
\begin{center}
\includegraphics[width=.8\textwidth]{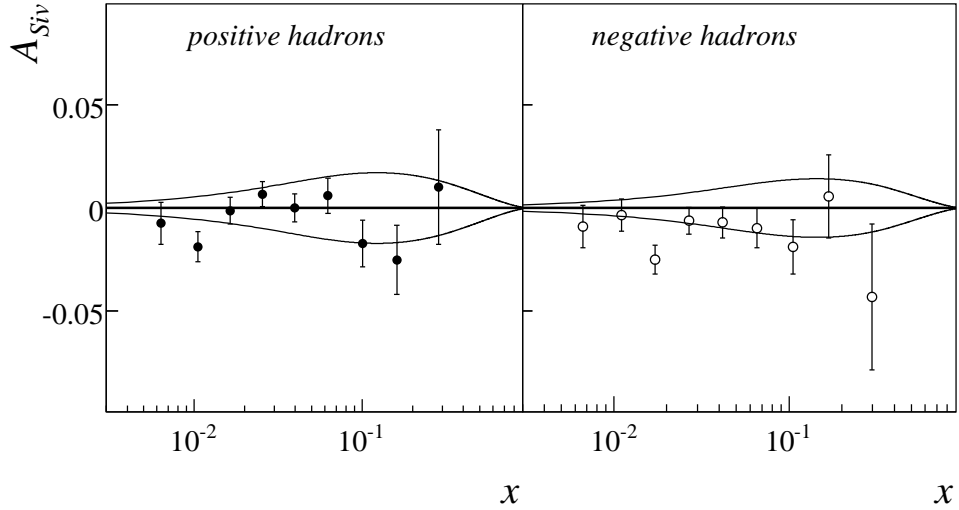}
\end{center}
\caption{Comparison between the present results for positive (left) 
and negative (right)
hadrons with the calculations of Ref.~\cite{Collins:2005ie}.
}
\label{fig:sivefr} 
\end{figure*}

Also the Sivers data have been looked upon independently by three different
groups. In Ref.~\cite{Collins:2005ie} a fit of the HERMES data was performed, 
on the assumption that 
$\Delta_0^T d(x) = - \Delta_0^T u(x)$. A good agreement with the HERMES data
was obtained, and a zero asymmetry in case of a deuterium target was 
predicted, as shown in Fig.~\ref{fig:sivefr}.
The curves in the figure indicate the expected size of the effect 
on $A_{Siv}$ of the $1/N_c$-corrections.
The COMPASS data fall well within the band resulting from the model.

\begin{figure*}[tb] %
\begin{center}
\includegraphics[width=.8\textwidth]{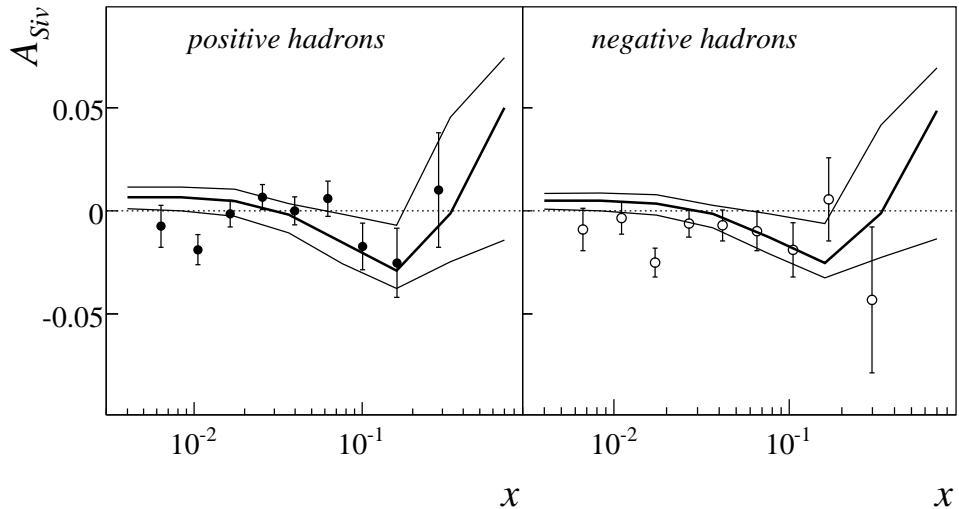}
\end{center}
\caption{Comparison between the present results for positive 
(left) and negative (right)
hadrons with the calculations of Ref.~\cite{sanse}.
}
\label{fig:sivans} 
\end{figure*}

The authors of Ref.~\cite{sanse} could estimate the Sivers functions
by fitting both the HERMES and the published COMPASS data
getting $\Delta_0^{T} d_v \simeq -  \Delta_0^{T} u_v$. 
Leading Order MRST01 sets of unpolarised
distribution functions~\cite{Martin:2002dr} were used, 
together with Kretzer's set of 
fragmentation functions~\cite{Kretzer:2000yf}.
A very good agreement with the experimental data was reached,  
as apparent from Fig.~\ref{fig:sivans}, where the upper and lower curves  
correspond to 1-$\sigma$ deviation of the parameters.
In this model the $z$ and $p_T^h$ dependence of the
single spin asymmetries are also well described.

\begin{figure*}[tb] %
\begin{center}
\includegraphics[width=.8\textwidth]{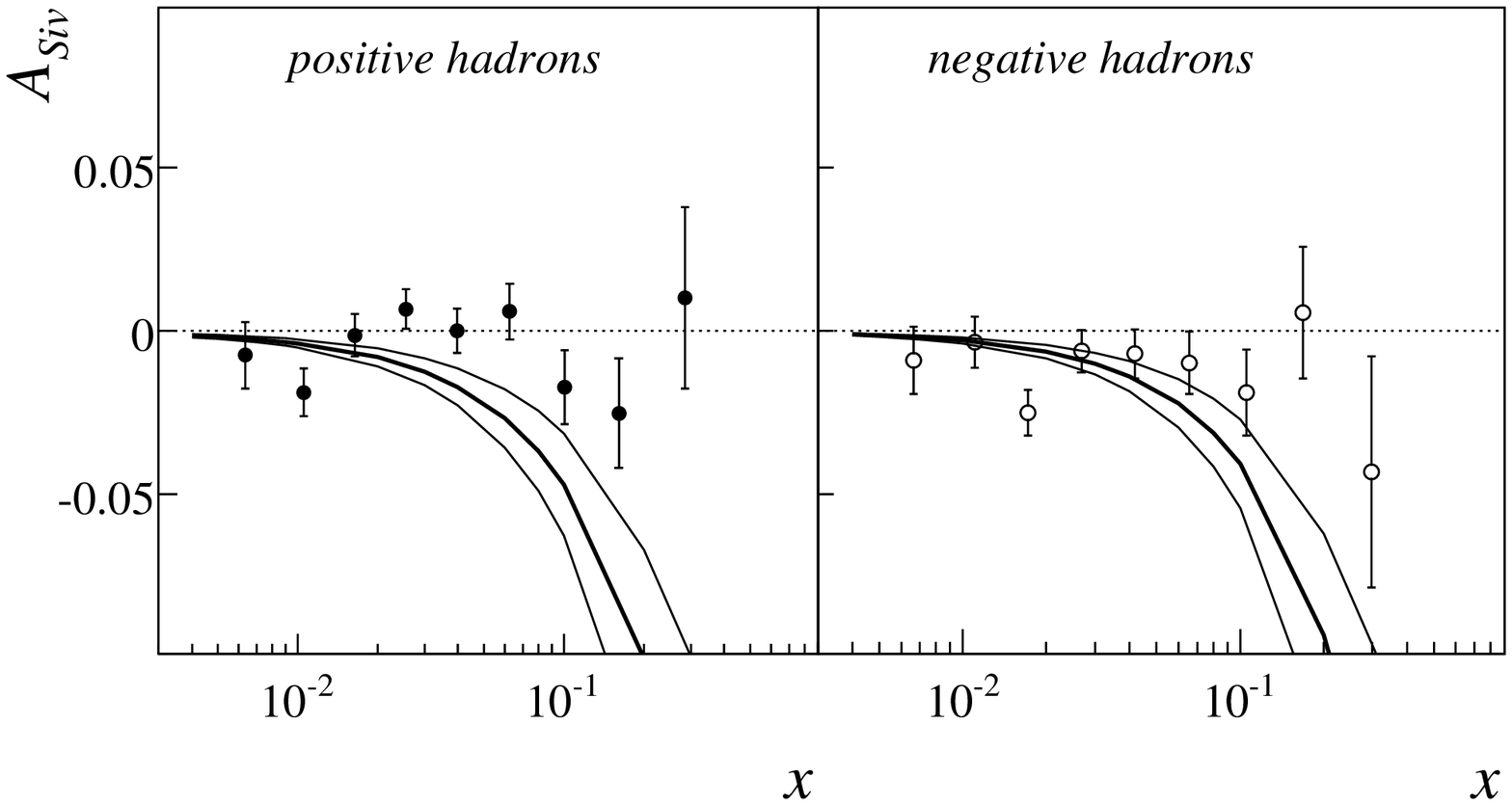}
\end{center}
\caption{Comparison between the present results for positive (left) 
and negative (right)
hadrons with the calculations of Ref.~\cite{vy}.
}
\label{fig:sivvog} 
\end{figure*}

In Ref.~\cite{vy} it was assumed that the final hadron 
transverse momentum is the transverse
 momentum in the Sivers function, i.e. a collinear fragmentation was assumed. 
GRV98 leading order distribution functions~\cite{Gluck:1994uf} were used, 
along with Kretzer's  
fragmentation functions.
The fit of the HERMES data is very good, 
and gave as a result $\Delta_0^{T} d_v \simeq -  \Delta_0^{T} u_v$,
but the prediction for 
COMPASS are
not in agreement with the new data, as shown in Fig.~\ref{fig:sivvog}.
The upper and lower curves in the figures correspond to  1-$\sigma$ errors
in the fitted parameters.
Here again, a new global fit using the deuteron data should be able to 
extract the 
Sivers function for the d-quark and improve the agreement with the data. 

Very recently, the smallness of the Sivers asymmetry for positive and 
negative hadrons on the deuteron has been interpreted as evidence for
the absence of gluon orbital angular momentum in the 
nucleon~\cite{Brodsky:2006ha}.

\section{Conclusions}
After providing the very first SIDIS data on transverse spin asymmetries on a 
transversely polarised deuteron target, 
COMPASS now publishes the overall results from the data collected
 in 2002, 2003, and 2004, increasing the statistics as compared to 
the published 2002 data by a factor of $\sim$7,
so that even at large $x$ the errors on the measured asymmetries are only 
a few percent. 

All the measured asymmetries are small, mostly compatible with zero within 
the measurement errors. 
Presently, the most likely interpretation, taking into account the 
corresponding measurements 
of the HERMES collaboration on a proton target, is that in the COMPASS 
isoscalar target 
there is a cancellation between the proton and the neutron asymmetries.
Also, the independent evidence provided by the BELLE data that the 
Collins effect is a real physical mechanism 
guarantees that the transversity distributions $\Delta_T q(x)$ can be extracted
 from the single spin asymmetries measured in SIDIS.
A global analysis of all the presently available data is now mandatory, and
the inclusion of the new precise deuteron data from COMPASS will surely 
allow to provide first estimates of the down quark
transversity distribution $\Delta _T d$ and of the Collins
fragmentation functions $\Delta_{T}^0 D$.
Already now the COMPASS data for the Sivers asymmetry provide convincing evidence
on the cancellation of the u and d quark Sivers distribution 
functions.

It has to be stressed, however, that the measured effects are rather small, 
of the order of a few per cent at most, and that flavour separation 
requires much larger 
statistics than presently available.
Also, within the present errors, the physics interpretation is not 
straightforward 
and has led to some surprises. 
The present data are of fundamental importance
because they have opened up the road to transverse momentum dependent 
distribution and fragmentation functions, 
but they will by no means suffice to determine these new functions.
New data are needed, and particularly, new proton data. HERMES is 
still finalising
 the analysis of their 2005 proton run, which will double the 
statistics of their analysed data.
The BELLE Collaboration is producing more and more accurate data 
on the Collins fragmentation function. 
In the near future,
COMPASS plans proton runs, which should result in particularly 
reduced error bars at large $x$, 
thanks to the much improved geometrical acceptance of the spectrometer
which is obtained using the new COMPASS polarised target magnet. 
A global analysis will then again be necessary,
and from all those  measurements it should be able to provide the 
u and d quark transversity distributions.

\section{Acknowledgement}
We thank M. Anselmino, X. Artru, V. Barone, and A. Prokudin for many 
interesting discussions.
We gratefully acknowledge the support of the CERN management
and staff, as well as
the skills and efforts of the technicians of the collaborating
institutes. 


\end{document}